\documentclass[aip,
jmp,
 amsmath,amssymb,
reprint, onecolumn 
]{revtex4-1}

\usepackage{graphicx}
\usepackage{dcolumn}
\usepackage{bm}

\usepackage[utf8]{inputenc}
\usepackage[T1]{fontenc}
\usepackage{mathptmx}
\usepackage{etoolbox}

\usepackage{multirow}

\makeatletter
\def\@email#1#2{%
 \endgroup
 \patchcmd{\titleblock@produce}
  {\frontmatter@RRAPformat}
  {\frontmatter@RRAPformat{\produce@RRAP{*#1\href{mailto:#2}{#2}}}\frontmatter@RRAPformat}
  {}{}
}%
\makeatother

\begin{document}

\allowdisplaybreaks


\title{Vacuum electromagnetic field correlations between two moving points}
\author{Michael Vaz}

\email{michael.vaz@ens-paris-saclay.fr}

\author{Hervé Bercegol}

\affiliation{%
SPEC, CEA, CNRS, Université Paris-Saclay, F-91191, Gif-sur-Yvette, France
}%

\date{\today}

\begin{abstract}
A renewed experimental interest in quantum vacuum fluctuations brings back the need to extend the study of electromagnetic vacuum correlations. 
Quantum or semi-classical models developed to understand various configurations should combine the effects of the zero-point fluctuations with those of blackbody radiation. 
In this paper, after a brief historical introduction and a rapid study of the electric field correlations in time domain, we propose exact and approximate expressions for the vacuum field correlations in Fourier space seen by moving points. We first present an exact computation of the electric field correlations, expressed in frequency space, between two points moving with opposite constant velocities on parallel trajectories. We also consider the electric field self-correlations, i.e. on the same moving point but at different frequencies, and comment the results related to special relativity.
Then, we compute the exact main symmetrized quadratic electromagnetic field correlations between two points diametrically opposed on the same circular trajectory, with diameter $r$, covered at constant angular velocity $\Omega$. We derive the expressions for the electromagnetic field correlations with itself and with its spatial derivatives, still at the locations of the moving points. Since the points we consider are accelerating, both the zero-point fluctuations and the blackbody spectrum give non-trivial results, for two-point correlations as well as for self-correlations. 
In both cases, results are shown at any vacuum temperature. For practical uses, we provide the first-order approximations in the small parameter $\Omega r/c$ with $c$ being the speed of light. 
\end{abstract}

\maketitle

\section{Introduction}

Fluctuations of the vacuum electromagnetic fields are one of the foremost characteristic of quantum electrodynamics, "the best theory we have"\cite{milonni1993quantum}. Only recently, these fluctuations were directly measured, through the interaction of femtosecond laser pulses in vacuum \cite{Riek-Seletskiy} or in a non linear crystal \cite{benea2019electric}. This new interest in vacuum fluctuations revive an older research area focused in the consequences at diverse scale of those fluctuations, at a microscopic \cite{Bethe} or macroscopic level \cite{casimir1948attraction}. While static properties are still an active research field \cite{menezes2015vacuum,Pozas-Kerstjens}, dynamic interaction \cite{dodonov,reiche_2022} leading to dissipative phenomena remain a puzzling theoretical conundrum. To aid the exploration of dynamic effects of the vacuum field, this paper presents the computation of field correlations in the frequency domain, between two different spatial points, taking into account their relative motion. 

The standard Fermi problem of interacting atoms \cite{Fermi1932}, as comprehensively described by Lindel et al. \cite{Lindel_PRR}, shows that the correlations to be considered depend on the operator ordering chosen in the quantum equations, although the final physical prediction should not depend on it. A symmetric ordering, as recommended by Dalibard, Dupont-Roc and Cohen-Tannoudji \cite{dalibard1982vacuum}, gives the vacuum field the main role in atomic interaction problems \cite{buhmann2013dispersion}. 
Thus, in this paper, we will be interested only in the vacuum fluctuations. The full atom+field configuration is considered in companion papers: attraction \cite{Vaz1} and friction \cite{Vaz2} forces between two revolving atoms can be obtained using these new field correlations.

The first theory of quantum mechanics left the electromagnetic field out of the game. It even hypothesized a violation of the Larmor formula \cite{larmor1897lxiii,jackson1998classical} which states that an accelerating charge should radiate, which was not described by the quantum models of Hydrogen, in particular for orbitals with angular momentum \cite{bohr1913constitution,schrodinger1926undulatory}. However, already in 1900, Planck's blackbody radiation involved quanta of energy for light \cite{planck1901law}, even if Planck himself decided to describe it through an equilibrium with quantized material oscillators. This phenomenon will later inspire Einstein for its description of the photoelectric effect \cite{einstein1905heuristic}, who showed that the electromagnetic field also needed to be quantized. This was then proposed by Dirac in 1927 \cite{dirac1927quantum}, who reestablished the possibility for Maxwell's equations to hold at microscopic scale \cite{dalibard1982vacuum}. In the same process, Dirac also introduced the concept of vacuum state and its associated zero point fluctuations, object of many questions and investigations throughout the twentieth century. 

Quantum electrodynamics (QED), in particular the quantum properties of the vacuum, leads to accurate predictions; the Lamb shift \cite{Bethe} and the Casimir effect \cite{casimir1948attraction} being among the most famous ones. Very precise experimental measurements \cite{lamb, lamoreaux97} have confirmed the existence of such a quantum electromagnetic vacuum. Nevertheless, in accordance with classical theory, no energy can directly be extracted from the vacuum, so measurements may seem impossible at first glance. Indeed, the vacuum expectation values (VEV) of the fields are vanishing, so that direct field measurements always lead to trivial results. However, quadratic forms in the field can give non-zero outcomes, which would not be possible for a deterministic classical electromagnetic field. From this model, the vacuum is no longer electromagnetically inert: in particular, the fields at two points in space and time are generally correlated, meaning that the fluctuations are not independent. This is usually understood in quantum field theory through propagation of virtual or thermal photons \cite{milonni1993quantum}, which can lead to various effects, such as the dispersion forces between atoms or non polar molecules, introduced by London \cite{london1937general} and pursued by Casimir, Polder \cite{CasimirPolder} and others \cite{McLachlan}. This explains the interest in calculating and measuring vacuum field correlations. Recently, experimentalists have been able to measure such correlations through different ways, using material probes, highlighting their importance in modern physics \cite{Riek-Seletskiy,benea2019electric}. 

Back in the years of "the old quantum theory", Einstein and Hopf treated the case of individual atomic friction on blackbody radiation \cite{einstein-hopf}, a subject which was revived recently \cite{mkrtchian2003universal} and continues to be studied to this day \cite{lach2012enhancement,milton_guo,Sinha_2022}. Decades after Casimir's work, people began to consider the dissipative properties arising from the zero temperature vacuum correlations, gathered under the name of dynamical Casimir effects \cite{dodonov,milton_guo}. Following a stream of research that attributes the origin of the second law of thermodynamics to basic quantum features \cite{popescu2006entanglement}, studying vacuum friction could also bring a possible quantitative appraisal of
interatomic dissipative forces, potential microscopic contributions to entropy growth. Hence, when working on two atoms or more, it becomes important to know how to compute quadratic forms in the vacuum fields for two moving points, hence accounting for dynamical effects in the correlations. 

In the following, we first recall how correlations have been defined, calculated and used in the past. Then, we calculate in general electric field correlations in the time domain. We Fourier transform these, first in the static case which is a known result that poses no particular difficulty. Problems appear when considering points in relative motion. We solve here two generic cases: the collision of two points moving on parallel rectilinear trajectories; the circular motion of two diametrically opposed points.
\subsection*{State of the art in field correlations}
The literature expresses these correlations in two main ways. The first one, through the use of the fluctuation-dissipation theorem (FDT) \cite{Kubo_1966,callen_welton}, is based on the relation between the dissipation of some matter response and the driving vacuum field. However, the vacuum fluctuations should be thought of as a property of the vacuum electromagnetic field only, without the need for any material. From this point of view, a commonly practiced way of dealing with these vacuum correlations is to consider matter as an intermediary that disappears in the end. The idea is to make use of virtual dipolar field to propagate the vacuum one, allowing us to rewrite the field correlations with the use of the dyadic Green function, the elementary solution of the non-homogeneous Maxwell equations \cite{buhmann2013dispersion}. But these two common ways of presenting the correlations can only be applied to the case of two points which are not moving through space. However, QED offers the possibility of obtaining these correlations without taking a material response as an intermediary. Hence, here we present correlations between two points in physical space using only the quantum field operators, as we can find inside some integrals in Refs \cite{milonni1993quantum,CasimirPolder} for instance. This method can be used to obtain correlations beyond FDT, as we will show, in particular when one is looking for Fourier transforms with respect to the time of the vacuum electromagnetic field seen by a moving point.

According to quantum field theory, the electromagnetic field is described by operators; namely, the electric and magnetic field in vacuum can be written as \cite{glauber1963coherent}:
\begin{equation}\label{quantum_EB}
    \begin{array}{c}
        \begin{aligned}
        \hat{\mathbf{E}}[\Hat{\mathbf{x}},t] = i \sum_{\mathbf{k},\lambda} \sqrt{\frac{\hbar \omega_k}{2 \epsilon_0 V}} \left( \hat{a}_{\mathbf{k},\lambda} e^{i(\mathbf{k}\cdot \Hat{\mathbf{x}} - \omega_k t)} \mathbf{e}_{\mathbf{k},\lambda} - \hat{a}^{\dagger}_{\mathbf{k},\lambda} e^{-i(\mathbf{k}\cdot \Hat{\mathbf{x}} - \omega_k t)} \mathbf{e}_{\mathbf{k},\lambda}^* \right),
        \end{aligned}
    
        \vspace{0.2cm} \\

        \begin{aligned}
        \hat{\mathbf{B}}[\Hat{\mathbf{x}},t] = i \sum_{\mathbf{k},\lambda} \sqrt{\frac{\hbar}{2 \epsilon_0 \omega_k V}} \left( \hat{a}_{\mathbf{k},\lambda} e^{i(\mathbf{k}\cdot \Hat{\mathbf{x}} - \omega_k t)} (\mathbf{k} \times \mathbf{e}_{\mathbf{k},\lambda}) - \hat{a}^{\dagger}_{\mathbf{k},\lambda} e^{-i(\mathbf{k}\cdot \Hat{\mathbf{x}} - \omega_k t)} (\mathbf{k} \times \mathbf{e}_{\mathbf{k},\lambda}^*) \right)
        \end{aligned}
    \end{array}
\end{equation}
where $\hat{a}_{\mathbf{k},\lambda}$ and $\hat{a}_{\mathbf{k},\lambda}^\dagger$ are the well known annihilation and creation operators of the photon mode $(\mathbf{k},\lambda)$ of energy $\hbar \omega_k = \hbar k c$, $c$ being the speed of light in vacuum, $V$ is a quantization volume that will cancel out later in the calculations and $\mathbf{e}_{\mathbf{k},\lambda}$ are the unit polarization vectors. These vectors are chosen to be complex to describe the polarization as compatible with the helicity of the photons but the future expressions could also be obtained by imposing a real nature of the polarization vectors (meaning linear polarization), just as done in Ref. \cite{milonni1993quantum} (see sections 2.5 and 2.6). In the previous expression and in the following ones, the $\times$ symbol between vectors is referring to the usual vector product in three dimensions. The quantum rules satisfied by the creation and annihilation operators are
\begin{equation}\label{quantum_rules}
    \begin{array}{cc}
        \langle \varnothing \vert \hat{a}_{\mathbf{k},\lambda} \hat{a}_{\mathbf{k}',\lambda'} \vert \varnothing \rangle = 0, & \langle \varnothing \vert \hat{a}^\dagger_{\mathbf{k},\lambda} \hat{a}^\dagger_{\mathbf{k}',\lambda'} \vert \varnothing \rangle = 0, 
        \vspace{0.2cm}\\
        \langle \varnothing \vert \hat{a}^\dagger_{\mathbf{k},\lambda} \hat{a}_{\mathbf{k}',\lambda'} \vert \varnothing \rangle = N_k, & 
        \langle \varnothing \vert \hat{a}_{\mathbf{k},\lambda} \hat{a}^\dagger_{\mathbf{k}',\lambda'} \vert \varnothing \rangle = 1+N_k, 
        \vspace{0.2cm}\\
        \multicolumn{2}{c}{N_k = N[\omega_k]= \frac{1}{e^{\hbar \omega_k/k_B T}-1}}
    \end{array}
\end{equation}
with $k_B$ being the Boltzmann constant, $T$ the vacuum temperature and $\vert \varnothing \rangle$ is the vacuum state at temperature $T$ (see for instance Section 2.10 of Ref. \cite{milonni1993quantum}). Generally, one should use the density matrix formalism to account for thermalization \cite{von2018mathematical}. In our case, Eqs. (\ref{quantum_rules}) are sufficient because the vacuum state only appears in computations of thermalized VEV (TVEV). At zero temperature, the VEV of an operator $\Hat{O}$ is simply defined by the quantum average over the zero photon state: $\langle 0 \vert \Hat{O} \vert 0 \rangle$. The TVEV is then the average over the radiation field state, which counts the zero point fluctuations together with the statistical mixture of thermal photons: $\langle \Hat{O} \rangle$. It is important to note that the vacuum is encoded in both the field operators, which fulfill Maxwell's equation \textit{in vacuum}, in the classical sense, and in the vacuum state. It is defined, at zero temperature, as the Fock state \cite{fock1932konfigurationsraum,cohen2019quantum} with zero photons in every mode, and more technically when taking into account the thermal photons \cite{von2018mathematical}.


In this paper, we will consider two points $A$ and $B$ precisely localized respectively at positions $\mathbf{r}^A[t]$ and $\mathbf{r}^B[t]$, which have explicit time dependence to account for their motion. Hence, the position operators in the expressions ($\ref{quantum_EB}$) will be replaced by classical trajectories so that we have, for instance for $A$:
\begin{equation}\label{electric_A}
    \begin{aligned}
        \hat{\mathbf{E}}^A[t] = i \sum_{\mathbf{k},\lambda} \sqrt{\frac{\hbar \omega_k}{2 \epsilon_0 V}} \left( \hat{a}_{\mathbf{k},\lambda} e^{i(\mathbf{k}\cdot \mathbf{r}^A - \omega_k t)} \mathbf{e}_{\mathbf{k},\lambda} - \hat{a}^{\dagger}_{\mathbf{k},\lambda} e^{-i(\mathbf{k}\cdot \mathbf{r}^A - \omega_k t)} \mathbf{e}_{\mathbf{k},\lambda}^* \right)
        \end{aligned}.
\end{equation}

After some general considerations, we will consider the case of two points with opposite constant velocities along straight trajectories, for which we will derive the electric-electric correlations. Then, we will focus our calculations on the system of two revolving points at constant angular velocity around their midpoint and show the expressions of the correlations of the electromagnetic field with itself and with its spatial derivatives. 

\section{Generalities about the electromagnetic vacuum correlations}

Usually in the literature, one can find the electric correlations expressed through the fluctuation dissipation theorem, and thus requiring a \textit{source} field and a \textit{response}, in general being material \cite{Kubo_1966}. However, this is not mandatory; the vacuum correlations can be seen as a feature of the vacuum only, without considering an external object, the presence of which could have, by the way, modified the vacuum field.

A way to perform the calculation for two points $A$ and $B$ at respective time and positions $(t,\mathbf{r}^A[t])$ and $(t',\mathbf{r}^B[t'])$ is to consider only the vacuum field expressions from quantum electrodynamics. For the calculation of self-correlations, we take $A = B$. One can then compute, at any times $t$ and $t'$, at any vacuum temperature, the field correlations, also called Wightman functions \cite{wightman1956quantum,streater2000pct}:
\begin{equation}\label{Wij}
    \begin{aligned}
        \langle \Hat{E}_i^A[t] \Hat{E}_j^B [t'] \rangle = \frac{\hbar c}{16 \pi^3 \epsilon_0} \int k^3 \, \mathrm{d}k \, \sin[\theta] \, \mathrm{d}\theta \, \mathrm{d}\phi \, & \bigg( e^{i\left( ks \cos[\theta] - kc (t-t') \right)} \bigg( \frac{1}{e^{\hbar k c / k_B T}-1} +1 \bigg) \\
        & + e^{-i\left( ks \cos[\theta] - kc (t-t') \right)} \frac{1}{e^{\hbar k c / k_B T}-1} \bigg) \bigg( \delta_{ij} - \frac{k_i k_j}{k^2} \bigg)
    \end{aligned} 
\end{equation}
where $i$ and $j$ refers to two projections on an orthonormal basis, $\theta$ is the polar angle with reference $\mathbf{r}^A - \mathbf{r}^B$, $\phi$ is the azimuthal angle and $s = \vert \mathbf{r}^A - \mathbf{r}^B \vert$ is the distance between the two points. In the above expression and in the following ones, the brackets will refer to the TVEV. We have eluded some steps in the derivation of the above expression starting from the electric field expression ($\ref{electric_A}$) since they will be explained in following sections.

To simplify the expression ($\ref{Wij}$), one should set a particular basis, adapted to the vector $\mathbf{r}^A - \mathbf{r}^B$ since this is the reference for the angle $\theta$ above and work directly on projections. However, all of them will involve the same integrals, namely
\begin{equation}
    \begin{array}{c}
        \begin{aligned}
            \mathcal{C}_0 = \int k^3 \, \mathrm{d}k \, \sin[\theta] \, \mathrm{d}\theta \, & \bigg( e^{i\left( ks \cos[\theta] - kc (t-t') \right)} \bigg(\frac{1}{e^{\hbar k c / k_B T}-1} +1 \bigg) \\
            & + e^{-i\left( ks \cos[\theta] - kc (t-t') \right)} \frac{1}{e^{\hbar k c / k_B T}-1} \bigg), 
        \end{aligned} 
    
        \vspace{0.2cm} \\

        \begin{aligned}
            \mathcal{C}_2 = \int k^3 \, \mathrm{d}k \, \sin[\theta] \, \mathrm{d}\theta \, & \bigg( e^{i\left( ks \cos[\theta] - kc (t-t') \right)} \bigg(\frac{1}{e^{\hbar k c / k_B T}-1} +1 \bigg) \\
            & + e^{-i\left( ks \cos[\theta] - kc (t-t') \right)} \frac{1}{e^{\hbar k c / k_B T}-1} \bigg) \cos^2[\theta].
        \end{aligned} 
    \end{array}
\end{equation}

In both cases, the angular integral can be handled as a first step, leading to
\begin{equation}\label{integrals_C}
    \begin{array}{c}
        \displaystyle \mathcal{C}_0 = 2 \int_0^\infty k^3 \, \mathrm{d}k \, \bigg( e^{-ikc(t-t')} \bigg(\frac{1}{e^{\hbar k c / k_B T}-1} +1 \bigg) + e^{ikc(t-t')} \frac{1}{e^{\hbar k c / k_B T}-1} \bigg) \frac{\sin[ks]}{ks},
    
        \vspace{0.2cm} \\

        \begin{aligned}
            \mathcal{C}_2 = -2 \int_0^\infty k^3 \, \mathrm{d}k \, & \bigg( e^{-ikc(t-t')} \bigg(\frac{1}{e^{\hbar k c / k_B T}-1} +1 \bigg) \\
            & + e^{ikc(t-t')} \frac{1}{e^{\hbar k c / k_B T}-1} \bigg) \frac{( 2 - (ks)^2 )\sin[ks] - 2 ks \cos[ks]}{(ks)^3}.
        \end{aligned} 
    \end{array}
\end{equation}

We can then perform the remaining integral using the digamma function $\psi$ defined by $\psi[z]=\Gamma'[z]/\Gamma[z]$ where $\Gamma$ is the Euler Gamma function and the prime stands for the complex derivative (see Appendix \ref{digamma}). It gives 
\begin{equation}
    \begin{array}{c}
        \begin{aligned}
            \mathcal{C}_0 = 4 \frac{3c^2 (t-t')^2+s^2}{(c^2(t-t')^2-s^2)^3} + 2 \left( \frac{k_B T}{\hbar c} \right)^3 \frac{1}{s} \Im & \bigg[ \psi'' \left[ 1 + i \frac{k_B T}{\hbar c} (c(t-t')+s) \right] \\
            & - \psi'' \left[ 1 + i \frac{k_B T}{\hbar c} (c(t-t')-s) \right] \bigg],
        \end{aligned}
    
        \vspace{0.2cm} \\

        \begin{aligned}
            \mathcal{C}_2 = 4 \frac{c^2 (t-t')^2 + 3 s^2}{(c^2(t-t')^2-s^2)^3} - 2 \frac{k_B T}{\hbar c} \frac{1}{s^3} & \Im \bigg[ 2\bigg( \psi \left[ 1 + i \frac{k_B T}{\hbar c} (c(t-t')+s) \right] \\
            & ~~~~~~~ - \psi \left[ 1 + i \frac{k_B T}{\hbar c} (c(t-t')-s) \right] \bigg) \\
            & - 2 i \frac{k_B T}{\hbar c} s \bigg( \psi' \left[ 1 + i \frac{k_B T}{\hbar c} (c(t-t')+s) \right] \\
            & ~~~~~~~~~~~~~~~~~ + \psi' \left[ 1 + i \frac{k_B T}{\hbar c} (c(t-t')-s) \right] \bigg) \\
            & - \left( \frac{k_B T}{\hbar c} s \right)^2 \bigg( \psi'' \left[ 1 + i \frac{k_B T}{\hbar c} (c(t-t')+s) \right] \\
            & ~~~~~~~~~~~~~~~~~~~~~~ - \psi'' \left[ 1 + i \frac{k_B T}{\hbar c} (c(t-t')-s) \right] \bigg) \bigg].
        \end{aligned}
    \end{array}
\end{equation}

In each expression, the first term corresponds to the zero temperature virtual photons correlations of the vacuum, while the second term is associated with the thermal ones, and clearly vanishes at zero temperature. These two correlation functions are the only ones needed for the electric-electric time domain correlations. Indeed, by noting $\Hat{E}_\parallel$ the component of the field parallel to $\mathbf{r}^A - \mathbf{r}^B$ and $\Hat{E}_\perp$ one of the orthogonal components, we can show
\begin{equation}
    \begin{array}{cc}
        \displaystyle \langle \Hat{E}_\parallel^A[t] \Hat{E}_\parallel^B [t'] \rangle = \frac{\hbar c}{8 \pi^2 \epsilon_0} \left( \mathcal{C}_0 - \mathcal{C}_2 \right),
    
        \vspace{0.2cm} \\

        \displaystyle \langle \Hat{E}_\parallel^A[t] \Hat{E}_\perp^B [t'] \rangle = \langle \Hat{E}_\perp^A[t] \Hat{E}_\parallel^B [t'] \rangle = 0,

        \vspace{0.2cm} \\

        \displaystyle \langle \Hat{E}_\perp^A[t] \Hat{E}_\perp^B [t'] \rangle = \frac{\hbar c}{16 \pi^2 \epsilon_0} \left( \mathcal{C}_0 + \mathcal{C}_2 \right),

        \vspace{0.2cm} \\

        \displaystyle \langle \Hat{E}_{\perp}^A[t] \Hat{E}_{\perp'}^B [t'] \rangle = 0 & \perp \neq \perp'
    \end{array}
\end{equation}
where, on the last line, $\perp$ and $\perp'$ stand for two orthogonal components, both orthogonal to the vector $\mathbf{r}^A - \mathbf{r}^B$.

The expressions presented above are very general and can be used for numerical simulations for instance. One can verify them using special cases such as two motionless points. This case corresponds to $s$ being time independent. The literature, focused on formulas in the frequency space, mostly deals with static correlations since the time dependence is easy to handle. However, there is a general interest to work in the frequency domain when treating dynamical cases within linear response theory. Choosing the definition $\Tilde{f}[\omega] = \int_{-\infty}^\infty f[t] e^{i \omega t} \, \mathrm{d}t/2\pi$ and noting that the correlations only depend in the time difference $\Delta t = t-t'$, we have
\begin{equation}
    \begin{array}{c}
        \displaystyle \langle \Tilde{\Hat{E}}_\parallel^A [\omega] \Tilde{\Hat{E}}_\parallel^B [\omega'] \rangle = \int_{-\infty}^\infty \frac{\hbar c}{8 \pi^2 \epsilon_0} (\mathcal{C}_0 [\Delta t] - \mathcal{C}_2 [\Delta t]) e^{i \omega \Delta t} \, \frac{\mathrm{d}\Delta t}{2\pi} \delta[\omega+\omega'],
    
        \vspace{0.2cm} \\

        \displaystyle \langle \Tilde{\Hat{E}}_\perp^A [\omega] \Tilde{\Hat{E}}_\perp^B [\omega'] \rangle = \int_{-\infty}^\infty \frac{\hbar c}{16 \pi^2 \epsilon_0} (\mathcal{C}_0 [\Delta t] + \mathcal{C}_2 [\Delta t]) e^{i \omega \Delta t} \, \frac{\mathrm{d}\Delta t}{2\pi} \delta[\omega+\omega']
    \end{array}
\end{equation}
where we recall the need for the points to remain motionless.

Let us remark that in the general case of two moving points, the real time correlations depend on both $t$ and $t'$ independently, and not only in the time difference $\Delta t$. As a consequence, the argument of the Dirac $\delta$ distribution ceases to be merely $\omega+\omega'$, as in the above equations, which can be understood as an absence of crossover between the modes. Indeed, as we shall see, the modes of the fields can couple to the characteristic frequencies of the points motions. 
Coming back to the static special case, we then need the Fourier transforms of the correlation functions $\mathcal{C}_0$ and $\mathcal{C}_2$, which can be obtained easily from ($\ref{integrals_C}$):
\begin{equation}
    \begin{array}{c}
        \displaystyle \Tilde{\mathcal{C}}_0[\omega] = \frac{2}{c^4} \vert \omega \vert^3 \frac{\sin \left[\frac{\vert \omega \vert s}{c} \right]}{\frac{\vert \omega \vert s}{c}} \bigg( \frac{1}{e^{\hbar \vert \omega \vert / k_B T}-1} + \Theta[\omega] \bigg),
    
        \vspace{0.2cm} \\

        \displaystyle \Tilde{\mathcal{C}}_2 [\omega] = - \frac{2}{c^4} \vert \omega \vert^3 \frac{\bigg( 2 - \left( \frac{\vert \omega \vert s}{c} \right)^2 \bigg) \sin \left[ \frac{\vert \omega \vert s}{c} \right] - 2 \frac{\vert \omega \vert s}{c} \cos \left[ \frac{\vert \omega \vert s}{c} \right]}{\left( \frac{\vert \omega \vert s}{c} \right)^3} \bigg( \frac{1}{e^{\hbar \vert \omega \vert / k_B T}-1} + \Theta[\omega] \bigg) 
    \end{array}
\end{equation}
with $\Theta$ being the Heaviside step function, so that the field correlations in Fourier space can be written as
\begin{equation}
    \begin{array}{c}
        \displaystyle \langle \Tilde{\Hat{E}}_\parallel^A [\omega] \Tilde{\Hat{E}}_\parallel^B [\omega'] \rangle = \frac{\hbar \vert \omega \vert^3}{2 \pi^2 \epsilon_0 c^3} \frac{\sin \left[ \frac{\vert \omega \vert s}{c} \right] - \frac{\vert \omega \vert s}{c} \cos \left[ \frac{\vert \omega \vert s}{c} \right]}{\left( \frac{\vert \omega \vert s}{c} \right)^3} \bigg( \frac{1}{e^{\hbar \vert \omega \vert / k_B T}-1} + \Theta[\omega] \bigg) \delta[\omega+\omega'],
    
        \vspace{0.2cm} \\

        \begin{aligned}
            \langle \Tilde{\Hat{E}}_\perp^A [\omega] \Tilde{\Hat{E}}_\perp^B [\omega'] \rangle = - \frac{\hbar \vert \omega \vert^3}{4 \pi^2 \epsilon_0 c^3} & \frac{\bigg( 1 - \left( \frac{\vert \omega \vert s}{c} \right)^2 \bigg) \sin \left[ \frac{\vert \omega \vert s}{c} \right] - \frac{\vert \omega \vert s}{c} \cos \left[ \frac{\vert \omega \vert s}{c} \right]}{\left( \frac{\vert \omega \vert s}{c} \right)^3} \\
            & \times \bigg( \frac{1}{e^{\hbar \vert \omega \vert / k_B T}-1} + \Theta[\omega] \bigg) \delta[\omega+\omega'].
        \end{aligned} 
    \end{array}
\end{equation}

It is important to note that the previous correlations are not symmetrized, meaning the exchange in the field operators order is not a symmetry. In order to use these correlations in some commutative model, one can symmetrize the expression so that we recover a commutative algebra. For two operators $\Hat{O}$ and $\Hat{Q}$, we simply have to compute \cite{Ford03072017,Milton_2020,ford2013lorentz}:
\begin{equation}\label{symmetrization}
    \langle O Q \rangle = \frac{1}{2} \left( \langle \Hat{O} \Hat{Q} \rangle + \langle \Hat{Q} \Hat{O} \rangle \right).
\end{equation}

In this case, the correlations in Fourier space are given by
\begin{equation}\label{static_corr}
    \begin{array}{c}
        \displaystyle \langle \Tilde{E}_\parallel^A [\omega] \Tilde{E}_\parallel^B [\omega'] \rangle = \frac{\hbar \vert \omega \vert^3}{2 \pi^2 \epsilon_0 c^3} \frac{\sin \left[ \frac{\vert \omega \vert s}{c} \right] - \frac{\vert \omega \vert s}{c} \cos \left[ \frac{\vert \omega \vert s}{c} \right]}{\left( \frac{\vert \omega \vert s}{c} \right)^3} \bigg( \frac{1}{e^{\hbar \vert \omega \vert / k_B T}-1} + \frac{1}{2} \bigg) \delta[\omega+\omega'],
    
        \vspace{0.2cm} \\

        \begin{aligned}
            \langle \Tilde{E}_\perp^A [\omega] \Tilde{E}_\perp^B [\omega'] \rangle = - \frac{\hbar \vert \omega \vert^3}{4 \pi^2 \epsilon_0 c^3} & \frac{\left( 1 - \left( \frac{\vert \omega \vert s}{c} \right)^2 \right) \sin \left[ \frac{\vert \omega \vert s}{c} \right] - \frac{\vert \omega \vert s}{c} \cos \left[ \frac{\vert \omega \vert s}{c} \right]}{\left( \frac{\vert \omega \vert s}{c} \right)^3} \\
            & \times \bigg( \frac{1}{e^{\hbar \vert \omega \vert / k_B T}-1} + \frac{1}{2} \bigg) \delta[\omega+\omega']
        \end{aligned} 
    \end{array}
\end{equation}
which are the expected expressions (see for instance Ref. \cite{Ford03072017}). We have explicitly removed the \textit{hats} from the field operators since the order does not matter anymore, and the correlations can be considered as correlations between stochastic c-number fields \cite{davidson1981stochastic}.

Let us note that, up to now, all the above correlations could have been obtained using usual FDT or a Green tensor formalism \cite{Ford03072017,buhmann2013dispersion}, since they only involve moving points correlations in time domain, where we can just plug the positions of the points into static correlations expressions, or motionless points in Fourier space, where the time dependence exist only in the temporal argument of the fields. In the following, we want to compute such correlations between two points with relative velocities. This is not easily obtained by Fourier transforming the general time correlations presented above, typically because the motion involve a dependence on the two time variables $t$ and $t'$. To perform these calculations, we will go back to the beginning, with the expressions of the electric and magnetic field operators, transformed into Fourier space before computing the correlations. 

\section{Calculation of the electric-electric correlations for two points in opposite and rectilinear uniform motion}\label{Rect_lin_motion}

\begin{figure}
    \centering
    \includegraphics[width=0.5\linewidth]{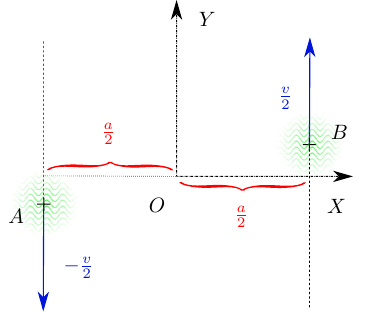}
    \caption{Diagram of two points traveling on rectilinear parallel paths at equal and opposite constant velocities. Field fluctuations at points $A$ and $B$ are symbolized by green wavy lines, a definite "artistic representation" since the field is present anywhere.}
    \label{fig:translation}
\end{figure}

Before entering the exact computation of the field correlations for revolving points, it will be interesting to consider another case, which is closely linked: the opposite and rectilinear uniform motion \cite{Barton_2010,barton2011van,hoye2010casimir1,hoye2011casimir2,hoye2011casimir}. Let us introduce the orthonormal basis unit vectors $\mathbf{I}$, $\mathbf{J}$ and $\mathbf{K}$ associated with the respective axes $X$, $Y$ and $Z$ of the laboratory frame. We can parametrized the points trajectories by
\begin{equation}
    \begin{array}{cc}
        \displaystyle \mathbf{r}^A = - \frac{a}{2} \mathbf{I} - \frac{vt}{2} \mathbf{J}, & \displaystyle \mathbf{r}^B = + \frac{a}{2} \mathbf{I} + \frac{vt}{2} \mathbf{J}
    \end{array}
\end{equation}
where $v$ is the relative velocity of the two points. We set the velocities of the points to be strictly less than the speed of light, which will be useful in the following. A diagram of the system is represented in Fig. \ref{fig:translation}.

The Fourier transform of the electric field seen by point $A$ is then given by 
\begin{equation}
    \Tilde{\Hat{\mathbf{E}}}^A[\omega] = i \sum_{\mathbf{k},\lambda} \sqrt{\frac{\hbar \omega_k}{2 \epsilon_0 V}} \bigg( e^{-i \frac{k_X a}{2}} \Hat{a}_{\mathbf{k},\lambda} \delta \bigg[\omega-\omega_k-\frac{k_Y v}{2} \bigg] \mathbf{e}_{\mathbf{k},\lambda} - e^{+i \frac{k_X a}{2}} \Hat{a}_{\mathbf{k},\lambda}^\dagger \delta \bigg[\omega+\omega_k+\frac{k_Y v}{2} \bigg] \mathbf{e}_{\mathbf{k},\lambda}^* \bigg).
\end{equation}

Using the quantum rules ($\ref{quantum_rules}$), we can again write the correlations, for instance:
\begin{equation}
    \begin{aligned}
        \langle \Tilde{\Hat{E}}_X^A[\omega] \Tilde{\Hat{E}}_X^B[\omega'] \rangle = \sum_{\mathbf{k},\lambda} \frac{\hbar \omega_k}{2 \epsilon_0 V} & \bigg( e^{-i k_X a} \delta \bigg[\omega - \omega_k - \frac{k_Y v}{2} \bigg] \delta \bigg[ \omega' + \omega_k - \frac{k_Y v}{2} \bigg] \left( N_k + 1 \right) \\
        & + e^{i k_X a} \delta \bigg[\omega + \omega_k + \frac{k_Y v}{2} \bigg] \delta \bigg[ \omega' - \omega_k + \frac{k_Y v}{2} \bigg] N_k \bigg) \vert \mathbf{e}_{\mathbf{k},\lambda} \cdot \mathbf{I} \vert^2.
    \end{aligned}
\end{equation}

We can then take the semi-classical limit $\sum_\mathbf{k} \rightarrow \int V/(2\pi)^3 \mathrm{d}\mathbf{k} = \int V/(2 \pi)^3 \mathrm{d}k \sin[\theta] \mathrm{d}\theta \mathrm{d}\phi$, which cancels the volume $V$ in the quadratic averages. This will give us an integral over the wave vectors $\mathbf{k}$, which we choose to represent in spherical coordinates of axis $Y$, the one of the velocity (cf. Fig. \ref{fig:translation}): $\mathbf{k} = k\left( \sin[\theta] \sin[\phi] \mathbf{I} + \cos[\theta] \mathbf{J} + \sin[\theta] \cos[\phi] \mathbf{K} \right)$ with $\theta$, $\phi$ being the polar and azimuthal angles. This leads to
\begin{equation}
    \begin{aligned}
        \langle \Tilde{\Hat{E}}_X^A[\omega] \Tilde{\Hat{E}}_X^B[\omega'] \rangle = & \int k^2 \, \mathrm{d}k \sin[\theta] \, \mathrm{d}\theta \, \mathrm{d}\phi \frac{\hbar \omega_k}{2 \epsilon_0 (2\pi)^3} \\
        & \times \bigg( e^{-i ka \sin[\theta] \sin[\phi]} \left( N_k+1 \right) \\
        & ~~~~~~~~~ \times \delta \bigg[ \omega - kc \left(1+\frac{v}{2c} \cos[\theta] \right) \bigg] \delta \bigg[ \omega' + kc \left( 1 - \frac{v}{2c} \cos[\theta] \right) \bigg] \\
        & ~~~~ + e^{i ka \sin[\theta] \sin[\phi]} N_k \\
        & ~~~~~~~~~ \times \delta \bigg[ \omega + kc \left( 1 + \frac{v}{2c} \cos[\theta] \right) \bigg] \delta \bigg[ \omega' -kc \left( 1 - \frac{v}{2c} \cos[\theta] \right) \bigg] \bigg) \sum_\lambda \vert \mathbf{e}_{\mathbf{k},\lambda} \cdot \mathbf{I} \vert^2.
    \end{aligned}
\end{equation}

We can then simplify the Dirac distributions, which present Doppler shifts coupling the frequencies $\omega$ and $\omega'$ to both the modulus $k$ and the polar angle $\theta$. For instance, we can proceed in two steps:
\begin{equation}
    \begin{aligned}
        & \delta \bigg[ \omega - kc \left(1+\frac{v}{2c} \cos[\theta] \right) \bigg] \delta \bigg[ \omega' + kc \left( 1 - \frac{v}{2c} \cos[\theta] \right) \bigg] \\
        & = \delta \bigg[ \omega - kc \left(1+\frac{v}{2c} \cos[\theta] \right) \bigg] \delta \left[ \omega+\omega' -kv \cos[\theta] \right] \Theta[-\omega'] \\
        & = \delta \bigg[ \frac{\omega-\omega'}{2} - \omega_k \bigg] \delta \left[ \omega+\omega' -kv \cos[\theta] \right] \Theta[\omega] \Theta[-\omega'].
    \end{aligned}
\end{equation}
Let us note that the Heaviside step functions $\Theta$ appear to keep the sign constrains from the original Dirac distributions product. This is actually redundant, as this is naturally appearing back when solving for the system $\omega-\omega'=2\omega_k$, $\omega+\omega'=kv \cos[\theta]$. The important point is to remember that $\omega$ and $\omega'$ have opposite signs, reminding the static case condition $\omega+\omega'=0$. From there, the calculation of the ordered, non-symmetrized correlations could easily be continued with the same steps as the following. However, we choose to symmetrize them in the following, in order to get expressions for semi-classical correlations. Moreover, it will simplify the expressions, since the two Dirac distributions products will match. Namely, we have
\begin{equation}
    \begin{aligned}
        \langle \Tilde{E}_X^A[\omega] \Tilde{E}_X^B[\omega'] \rangle = \int k^2 \, \mathrm{d}k & \sin[\theta] \, \mathrm{d}\theta \, \mathrm{d}\phi \, \frac{\hbar \omega_k}{2 \epsilon_0 (2\pi)^3} e^{-i \text{sgn}[\omega]ka \sin[\theta] \sin[\phi]} \left( N_k+\frac{1}{2} \right) \\
        & \times \delta \bigg[ \frac{\vert \omega-\omega' \vert}{2}-\omega_k \bigg] \delta \left[\omega+\omega'-\text{sgn}[\omega]kv \cos[\theta] \right] \sum_\lambda \vert \mathbf{e}_{\mathbf{k},\lambda} \cdot \mathbf{I} \vert^2
    \end{aligned} 
\end{equation}
where $\text{sgn}$ stands for the sign function.

Using then the polarization summing rule $\sum_{\lambda} (\mathbf{e}_{\mathbf{k},\lambda})_i (\mathbf{e}_{\mathbf{k},\lambda})^*_j = \delta_{ij} - \frac{k_i}{k}\frac{k_j}{k}$, we end up with
\begin{equation}
    \begin{aligned}
        \langle \Tilde{E}_X^A[\omega] \Tilde{E}_X^B[\omega'] \rangle = \frac{\hbar}{2 \epsilon_0 (2\pi)^3} \bigg( \frac{\vert \omega - \omega' \vert}{2c} \bigg)^3 \int & \sin[\theta] \, \mathrm{d}\theta \, \mathrm{d}\phi \, e^{-i\frac{(\omega-\omega')a}{2c} \sin[\theta] \sin[\phi]} \bigg( N \left[ \frac{\vert \omega - \omega' \vert}{2} \right]+\frac{1}{2} \bigg) \\
        & \times \delta \bigg[ \omega+\omega' - \frac{\omega-\omega'}{2} \frac{v}{c} \cos[\theta] \bigg] \left( 1 - \sin^2[\theta] \sin^2[\phi] \right) 
    \end{aligned} 
\end{equation}
where we computed the integral over the modulus $k$ thanks to the Dirac distribution.

We are then left with the two angular integrals. We can begin with the azimuthal one, which gives rise to Bessel functions of the first kind $J_0$ and $J_2$:
\begin{equation}\label{corr_para_XX_before}
    \begin{aligned}
        \langle \Tilde{E}_X^A[\omega] \Tilde{E}_X^B[\omega'] \rangle = ~ & \frac{\hbar}{2 \epsilon_0 (2\pi)^2} \bigg( \frac{\vert \omega - \omega' \vert}{2c} \bigg)^3 \\
        & \times \int_{0}^\pi \sin[\theta] \, \mathrm{d}\theta \, \bigg( J_0 \left[ \frac{\vert \omega-\omega'\vert a}{2c} \sin[\theta] \right] \\
        & ~~~~~~~~~~~~~~~~~~~~~~~~~~ -\frac{1}{2} \bigg( J_0 \left[ \frac{\vert \omega-\omega'\vert a}{2c} \sin[\theta] \right] \\
        & ~~~~~~~~~~~~~~~~~~~~~~~~~~~~~~~~~~ - J_2 \left[ \frac{\vert \omega-\omega'\vert a}{2c} \sin[\theta] \right] \bigg) \sin^2[\theta] \bigg) \\
        & ~~~~~~~~~~~~~~~~~~~~~~~~~~ \times \bigg( N \left[ \frac{\vert \omega - \omega' \vert}{2} \right]+\frac{1}{2} \bigg) \, \delta \bigg[ \omega+\omega' - \frac{\omega-\omega'}{2} \frac{v}{c} \cos[\theta] \bigg].
    \end{aligned}
\end{equation}

We can then easily compute the integral thanks to the Dirac distribution:
\begin{equation}\label{corr_para_XX_after}
    \begin{aligned}
        \langle \Tilde{E}_X^A[\omega] \Tilde{E}_X^B[\omega'] \rangle = ~& \frac{\hbar}{16 \epsilon_0 \pi^2} \bigg( \frac{\vert \omega - \omega' \vert}{2c} \bigg)^3 \\
        & \times \bigg( J_0 {\bigg[ \frac{\vert \omega-\omega'\vert a}{2c} \sqrt{1-\bigg(\frac{\omega+\omega'}{\omega-\omega'} \frac{2c}{v} \bigg)^2} \bigg]} \bigg( 1+\bigg(\frac{\omega+\omega'}{\omega-\omega'} \frac{2c}{v} \bigg)^2 \bigg) \\
        & ~~~~ + J_2 {\bigg[ \frac{\vert \omega-\omega'\vert a}{2c} \sqrt{1-\bigg(\frac{\omega+\omega'}{\omega-\omega'} \frac{2c}{v} \bigg)^2} \bigg]} \bigg( 1-\bigg(\frac{\omega+\omega'}{\omega-\omega'} \frac{2c}{v} \bigg)^2 \bigg) \bigg) \\
        & \times \bigg( N {\left[ \frac{\vert \omega - \omega' \vert}{2} \right]} +\frac{1}{2} \bigg) \Theta \bigg[1-\bigg\vert \frac{\omega+\omega'}{\omega-\omega'}\frac{2c}{v} \bigg\vert \bigg].
    \end{aligned}
\end{equation}

The main difference with the static result stands in the range of frequency yielding a non trivial result. For a fixed frequency $\omega$, the frequency $\omega'$ is highly constrained to $\omega'=-\omega$ in the static case. However, here, the Doppler shift in the Dirac distribution introduces a finite range of possible values for $\omega'$, namely $\omega'/\omega \in [-w,-1/w]$ with $w=(1+\frac{\vert v \vert}{2c})/(1-\frac{\vert v \vert}{2c})$. In particular, one recovers that $\omega$ and $\omega'$ have opposite signs.

While the expression ($\ref{corr_para_XX_after}$) is an exact result, one can be interested in an approximate result, for instance to first order in the velocity $v$. In this case, it is easier to come back to the previous form ($\ref{corr_para_XX_before}$) where the velocity is only appearing in the Dirac distribution. We can then use the following approximation 
\begin{equation}
    \delta[x+vy] \approx \delta[x] + v y \delta'[x]
\end{equation}
where the corrections will be at least of order $v^2$. A way to formalize this identity is to consider a regularized version of the Dirac distribution, such as the function $d$ defined by $d[x]=e^{-x^2/\sigma^2}/\sqrt{\pi}$, where $\sigma$ is a small parameter which we shall have tend to zero.

To first order in $v$, the correlation ($\ref{corr_para_XX_before}$) then becomes
\begin{equation}
    \begin{aligned}
        \langle \Tilde{E}_X^A[\omega] \Tilde{E}_X^B[\omega'] \rangle \approx ~& \frac{\hbar}{16 \epsilon_0 \pi^2} \bigg( \frac{\vert \omega - \omega' \vert}{2c} \bigg)^3 \\
        & \times \int_{-1}^1 \mathrm{d}x \bigg( J_0 {\bigg[ \frac{\vert \omega-\omega'\vert a}{2c} \sqrt{1-x^2} \bigg]} \left( 1+x^2 \right) \\
        & ~~~~~~~~~~~~~~ + J_2 {\bigg[ \frac{\vert \omega-\omega'\vert a}{2c} \sqrt{1-x^2} \bigg]} \left( 1-x^2 \right) \bigg) \bigg( N {\left[ \frac{\vert \omega - \omega' \vert}{2} \right]} +\frac{1}{2} \bigg) \delta[\omega+\omega']
    \end{aligned}
\end{equation}
where the integral proportional to $v$ vanishes by symmetry. Hence, we end up with a similar result as the static one, namely
\begin{equation}
    \langle \Tilde{E}_X^A[\omega] \Tilde{E}_X^B[\omega'] \rangle \approx \frac{\hbar \vert \omega \vert^3}{2 \epsilon_0 \pi^2 c^3} \frac{\sin \left[ \frac{\vert \omega \vert a}{c} \right] - \frac{\vert \omega \vert a}{c} \cos \left[ \frac{\vert \omega \vert a}{c} \right]}{\left( \frac{\vert \omega \vert a}{c} \right)^3} \bigg( N[\vert \omega \vert]+\frac{1}{2} \bigg) \delta[\omega+\omega']
\end{equation}
which corresponds to the expected result ($\ref{static_corr}$).

We can perform the same kind of calculations for the field component of the $Y$ axis to show 
\begin{equation}
    \begin{aligned}
        \langle \Tilde{E}_Y^A[\omega] \Tilde{E}_Y^B[\omega'] \rangle = \frac{\hbar}{8 \epsilon_0 \pi^2} \bigg( \frac{\vert \omega - \omega' \vert}{2c} \bigg)^3 \int_0^\pi & \sin[\theta] \mathrm{d}\theta \, J_0 \bigg[ \frac{\vert \omega-\omega' \vert a}{2c} \sin[\theta] \bigg] \left(1-\cos^2[\theta] \right) \\
        & \times \bigg( N\bigg[ \frac{\vert \omega-\omega' \vert}{2} \bigg] + \frac{1}{2} \bigg) \delta \bigg[ \omega+\omega' - \frac{\omega-\omega'}{2} \frac{v}{c} \cos[\theta] \bigg]
    \end{aligned}
\end{equation}
leading to
\begin{equation}
    \begin{aligned}
        \langle \Tilde{E}_Y^A[\omega] \Tilde{E}_Y^B[\omega'] \rangle = \frac{\hbar}{8 \epsilon_0 \pi^2} \bigg( \frac{\vert \omega - \omega' \vert}{2c} \bigg)^3 & J_0 {\bigg[ \frac{\vert \omega-\omega'\vert a}{2c} \sqrt{1-\bigg(\frac{\omega+\omega'}{\omega-\omega'} \frac{2c}{v} \bigg)^2} \bigg]} \bigg( 1-\bigg(\frac{\omega+\omega'}{\omega-\omega'} \frac{2c}{v} \bigg)^2 \bigg) \\
        & \times \bigg( N {\left[ \frac{\vert \omega - \omega' \vert}{2} \right]} +\frac{1}{2} \bigg) \Theta \bigg[1-\bigg\vert \frac{\omega+\omega'}{\omega-\omega'}\frac{2c}{v} \bigg\vert \bigg]
    \end{aligned}
\end{equation}
or to the approximate expression
\begin{equation}
    \begin{aligned}
        \langle \Tilde{E}_Y^A[\omega] \Tilde{E}_Y^B[\omega'] \rangle & \approx \frac{\hbar}{8 \epsilon_0 \pi^2} \bigg( \frac{\vert \omega-\omega' \vert}{c} \bigg)^3 \int_{-1}^1 \mathrm{d}x \, J_0 \bigg[ \frac{\vert \omega-\omega' \vert a}{2c} \sqrt{1-x^2} \bigg] \left(1-x^2 \right) \\
        & ~~~~~~~~~~~~~~~~~~~~~~~~~~~~~~~~~~~~ \times \bigg( N \bigg[ \frac{\vert \omega-\omega' \vert}{2} \bigg] + \frac{1}{2} \bigg) \delta[\omega+\omega'] \\
        & \approx - \frac{\hbar \vert \omega \vert^3}{4 \epsilon_0 \pi^2 c^3} \frac{\left( 1 - \left( \frac{\vert \omega \vert a}{c} \right)^2 \right) \sin \left[ \frac{\vert \omega \vert a}{c} \right] - \frac{\vert \omega \vert a}{c} \cos \left[ \frac{\vert \omega \vert a}{c} \right]}{\left( \frac{\vert \omega \vert a}{c} \right)^3} \bigg( N[\vert \omega \vert]+\frac{1}{2} \bigg) \delta[\omega+\omega']
    \end{aligned}
\end{equation}
which is again the expected result ($\ref{static_corr}$).

We can also compute the correlations upon the axis out of the plane of motion:
\begin{equation}
    \begin{aligned}
        \langle \Tilde{E}_Z^A[\omega] \Tilde{E}_Z^B[\omega'] \rangle = ~& \frac{\hbar}{16 \epsilon_0 \pi^2} \bigg( \frac{\vert \omega - \omega' \vert}{2c} \bigg)^3 \\
        & \times \int_{0}^\pi \sin[\theta] \, \mathrm{d}\theta \, \bigg( J_0 \left[ \frac{\vert \omega-\omega'\vert a}{2c} \sin[\theta] \right] \left(1+ \cos^2[\theta] \right) \\
        & ~~~~~~~~~~~~~~~~~~~~~~~~~~ - J_2 \left[ \frac{\vert \omega-\omega'\vert a}{2c} \sin[\theta] \right] \left(1- \cos^2[\theta] \right) \bigg) \\
        & \times \bigg( N \left[ \frac{\vert \omega - \omega' \vert}{2} \right]+\frac{1}{2} \bigg) \, \delta \bigg[ \omega+\omega' - \frac{\omega-\omega'}{2} \frac{v}{c} \cos[\theta] \bigg]
    \end{aligned}
\end{equation}
which can be simplified as
\begin{equation}
    \begin{aligned}
        \langle \Tilde{E}_Z^A[\omega] \Tilde{E}_Z^B[\omega'] \rangle = ~& \frac{\hbar}{16 \epsilon_0 \pi^2} \bigg( \frac{\vert \omega - \omega' \vert}{2c} \bigg)^3 \\
        & \times \bigg( J_0 {\bigg[ \frac{\vert \omega-\omega'\vert a}{2c} \sqrt{1-\bigg(\frac{\omega+\omega'}{\omega-\omega'} \frac{2c}{v} \bigg)^2} \bigg]} \bigg( 1+\bigg(\frac{\omega+\omega'}{\omega-\omega'} \frac{2c}{v} \bigg)^2 \bigg) \\
        & ~~~~ - J_2 {\bigg[ \frac{\vert \omega-\omega'\vert a}{2c} \sqrt{1-\bigg(\frac{\omega+\omega'}{\omega-\omega'} \frac{2c}{v} \bigg)^2} \bigg]} \bigg( 1-\bigg(\frac{\omega+\omega'}{\omega-\omega'} \frac{2c}{v} \bigg)^2 \bigg) \bigg) \\
        & \times \bigg( N {\left[ \frac{\vert \omega - \omega' \vert}{2} \right]} +\frac{1}{2} \bigg) \Theta \bigg[1-\bigg\vert \frac{\omega+\omega'}{\omega-\omega'}\frac{2c}{v} \bigg\vert \bigg]
    \end{aligned}
\end{equation}
or give the approximation to first order in $v$:
\begin{equation}
    \begin{aligned}
        \langle \Tilde{E}_Z^A[\omega] \Tilde{E}_Z^B[\omega'] \rangle & \approx \frac{\hbar}{16 \epsilon_0 \pi^2} \bigg( \frac{\vert \omega - \omega' \vert}{2c} \bigg)^3 \\
        & ~~~~ \times \int_{-1}^1 \mathrm{d}x \bigg( J_0 {\bigg[ \frac{\vert \omega-\omega'\vert a}{2c} \sqrt{1-x^2} \bigg]} \left( 1+x^2 \right) \\
        & ~~~~~~~~~~~~~~~~~~~ - J_2 {\bigg[ \frac{\vert \omega-\omega'\vert a}{2c} \sqrt{1-x^2} \bigg]} \left( 1-x^2 \right) \bigg) \bigg( N {\left[ \frac{\vert \omega - \omega' \vert}{2} \right]} +\frac{1}{2} \bigg) \delta[\omega+\omega'] \\
        & \approx - \frac{\hbar \vert \omega \vert^3}{4 \epsilon_0 \pi^2 c^3} \frac{\left( 1 - \left( \frac{\vert \omega \vert a}{c} \right)^2 \right) \sin \left[ \frac{\vert \omega \vert a}{c} \right] - \frac{\vert \omega \vert a}{c} \cos \left[ \frac{\vert \omega \vert a}{c} \right]}{\left( \frac{\vert \omega \vert a}{c} \right)^3} \bigg( N[\vert \omega \vert]+\frac{1}{2} \bigg) \delta[\omega+\omega'].
    \end{aligned}
\end{equation}

From now on, it seems that the correlations to first order in the velocity match with the ones of the static case. However, cross correlations, which are zero in the static case, here are non-vanishing: 
\begin{equation}
    \begin{aligned}
        \langle \Tilde{E}^A_X [\omega] \Tilde{E}^B_Y [\omega'] \rangle = i\frac{\hbar}{8 \epsilon_0 \pi^2} \bigg( \frac{\vert \omega-\omega' \vert }{2c} \bigg)^3 \int_0^\pi & \sin[\theta] \mathrm{d}\theta \, \text{sgn}[\omega] J_1 \bigg[ \frac{\vert \omega-\omega' \vert a}{2c} \sin[\theta] \bigg] \sin[\theta] \cos[\theta] \\
        & \times \bigg( N \left[ \frac{\vert \omega - \omega' \vert}{2} \right]+\frac{1}{2} \bigg) \, \delta \bigg[ \omega+\omega' - \frac{\omega-\omega'}{2} \frac{v}{c} \cos[\theta] \bigg]
    \end{aligned}
\end{equation}
which we can put under the final form
\begin{equation}
    \begin{aligned}
        \langle \Tilde{E}^A_X [\omega] \Tilde{E}^B_Y [\omega'] \rangle = i\frac{\hbar}{8 \epsilon_0 \pi^2} \bigg( \frac{\omega-\omega' }{2c} \bigg)^3 & J_1 \bigg[ \frac{\vert \omega-\omega' \vert a}{2c} \sqrt{1-\bigg(\frac{\omega+\omega'}{\omega-\omega'} \frac{2c}{v} \bigg)^2} \bigg] \\
        & \times \sqrt{1-\bigg(\frac{\omega+\omega'}{\omega-\omega'} \frac{2c}{v} \bigg)^2} \frac{\omega+\omega'}{\omega-\omega'} \frac{2c}{v} \\
        & \times \bigg( N \left[ \frac{\vert \omega - \omega' \vert}{2} \right]+\frac{1}{2} \bigg) \Theta \bigg[1-\bigg\vert \frac{\omega+\omega'}{\omega-\omega'}\frac{2c}{v} \bigg\vert \bigg].
    \end{aligned}
\end{equation}

Hence, in general, the motion induces correlations on the fields seen by two relatively moving points. In this case, the approximation to first order in $v$ is
\begin{equation}\label{corr_para_XY_v1}
    \begin{aligned}
        \langle \Tilde{E}^A_X [\omega] \Tilde{E}^B_Y [\omega'] \rangle & \approx -i\frac{\hbar}{8 \epsilon_0 \pi^2} \bigg( \frac{\vert \omega-\omega' \vert }{2c} \bigg)^4 \int_{-1}^1 J_1 \bigg[ \frac{\vert \omega-\omega' \vert a}{2c} \sqrt{1-x^2} \bigg] x^2 \sqrt{1-x^2} \\
        & ~~~~~~~~~~~~~~~~~~~~~~~~~~~~~~~~~~~~~~~~~~ \times \bigg( N \left[ \frac{\vert \omega - \omega' \vert}{2} \right]+\frac{1}{2} \bigg) v \, \delta' \left[ \omega+\omega' \right] \\
        & \approx -i\frac{\hbar}{4 \epsilon_0 \pi^2} \bigg( \frac{\vert \omega-\omega' \vert }{2c} \bigg)^3 \\
        & ~~~~~~~~ \times \frac{\left(3 - \left( \frac{\vert \omega-\omega' \vert a}{2c} \right)^2 \right) \sin \left[ \frac{\vert \omega-\omega' \vert a}{2c} \right] - 3 \frac{\vert \omega-\omega' \vert a}{2c} \cos \left[ \frac{\vert \omega-\omega' \vert a}{2c} \right]}{\left( \frac{\vert \omega-\omega' \vert a}{2c} \right)^3} \\
        & ~~~~~~~~ \times \bigg( N \left[ \frac{\vert \omega - \omega' \vert}{2} \right]+\frac{1}{2} \bigg) \frac{v}{a} \, \delta' \left[ \omega+\omega' \right]  
    \end{aligned}
\end{equation}
where the zeroth order in the velocity vanished by symmetry, as in the static case. 

Since we now have a strong constraint on the frequency $\omega$ and $\omega'$, one can rewrite the correlations using only one of the two, $\omega$ for instance. To do so, one shall remember the definition of the derivative of the Dirac distribution: for any test function $\varphi$, we have $\int_{-\infty}^\infty \varphi[x] \delta'[x] \, \mathrm{d}x = - \int_{-\infty}^\infty \varphi'[x] \delta[x] \, \mathrm{d}x$. As a consequence, if we take a smooth function $f$, not necessarily compactly supported nor bounded, we have for any test function $\varphi$:
\begin{equation}
     \begin{aligned}
         \int_{-\infty}^\infty \varphi[x] f[x] \delta'[x-y] \, \mathrm{d}x & = - \int_{-\infty}^\infty \left( \varphi'[x] f[x] + \varphi[x] f'[x] \right) \delta[x-y] \, \mathrm{d}x \\
         & = \int_{-\infty}^\infty \varphi[x] f[y] \delta'[x-y] \, \mathrm{d}x -\int_{-\infty}^\infty \varphi[x] f'[y] \delta[x-y] \, \mathrm{d}x
     \end{aligned}
\end{equation}
for any real value of $y$. As a consequence, we have: 
\begin{equation}
    f[x] \delta'[x-y] = f[y] \delta'[x-y]-f'[y] \delta[x-y].
\end{equation}

Applying this result to the last form of the correlation ($\ref{corr_para_XY_v1}$), we obtain
\begin{equation}
    \begin{aligned}
        \langle \Tilde{E}^A_X [\omega] \Tilde{E}^B_Y [\omega'] \rangle \approx -i \frac{\hbar}{4 \epsilon_0 \pi^2 a^3} & \bigg( \frac{v}{a} \bigg( N[\vert \omega \vert]+\frac{1}{2} \bigg) \bigg( \bigg(3 - \left( \frac{\vert \omega \vert a}{c} \right)^2 \bigg) \sin \left[ \frac{\vert \omega \vert a}{c} \right] \\
        & ~~~~~~~~~~~~~~~~~~~~~~~~~~~~~~~~~ - 3 \frac{\vert \omega \vert a}{c} \cos \left[ \frac{\vert \omega-\omega' \vert a}{c} \right] \bigg) \delta'[\omega+\omega'] \\
        & + \frac{v}{2c} \bigg( N[\vert \omega \vert]+\frac{1}{2} \bigg) \frac{\omega a}{c} \bigg( \sin{\left[ \frac{\vert \omega \vert a}{c} \right]} - \frac{\vert \omega \vert a}{c} \cos{\left[ \frac{\vert \omega \vert a}{c} \right]} \bigg) \delta[\omega+\omega'] \\
        & - \frac{v}{2c} \frac{\hbar \vert \omega \vert}{k_B T} N[\vert \omega \vert] \left( N[\vert \omega \vert]+1 \right) \\
        & ~~~~~~~~~~~~~~~~~~~~~~~~ \times \left( \frac{\omega a}{c} \right)^{-1} \bigg( \bigg(3 - \left( \frac{\vert \omega \vert a}{c} \right)^2 \bigg) \sin \left[ \frac{\vert \omega \vert a}{c} \right] \\
        & ~~~~~~~~~~~~~~~~~~~~~~~~~~~~~~~~~~~~~~~~~~~ - 3 \frac{\vert \omega \vert a}{c} \cos \left[ \frac{\vert \omega-\omega' \vert a}{c} \right] \bigg) \delta[\omega+\omega'] \bigg). 
    \end{aligned}
\end{equation}

One can also be interested in the self-correlations in this rectilinear motion case. With the same kind of calculations, we can easily show:
\begin{equation}
    \begin{array}{c}
        \displaystyle \langle \Tilde{E}^A_Y[\omega] \Tilde{E}^A_Y[\omega'] \rangle = \frac{\hbar \vert \omega \vert^3}{8 \epsilon_0 \pi^2 c^3} \int_{-1}^{1} \frac{1-x^2}{\left(1+\frac{v}{2c}x \right)^4} \bigg( N \bigg[\frac{\vert \omega \vert}{1+\frac{v}{2c}x} \bigg] + \frac{1}{2} \bigg) \, \mathrm{d}x \, \delta[\omega+\omega'],
    
        \vspace{0.2cm} \\

        \displaystyle \langle \Tilde{E}^A_X[\omega] \Tilde{E}^A_X[\omega'] \rangle = \displaystyle \langle \Tilde{E}^A_Z[\omega] \Tilde{E}^A_Z[\omega'] \rangle = \frac{\hbar \vert \omega \vert^3}{16 \epsilon_0 \pi^2 c^3} \int_{-1}^{1} \frac{1+x^2}{\left(1+\frac{v}{2c}x \right)^4} \bigg( N \bigg[\frac{\vert \omega \vert}{1+\frac{v}{2c}x} \bigg] + \frac{1}{2} \bigg) \, \mathrm{d}x \, \delta[\omega+\omega']
    \end{array}
\end{equation}
where we have performed the change of variable $x=\cos[\theta]$ for convenience. We can already remark that the expressions are even with respect to the velocity (see the transformation $v \rightarrow -v$ together with the change of variable $x \rightarrow -x$). Hence, there will be no contribution of the first order in the velocity. We can then focus on the exact expressions by computing the above integrals. The non-thermal part is basically a rational function that we can easily integrate. The thermal photons contribution is however less straightforward but can be done through series expansion of the exponential term, which allows us to obtain
\begin{equation}\label{self_corr_parallel}
    \begin{array}{c}
        \begin{aligned}
            \langle \Tilde{E}^A_Y[\omega] \Tilde{E}^A_Y[\omega'] \rangle = \frac{\hbar \vert \omega \vert^3}{4 \epsilon_0 \pi^2 c^3} \bigg( \frac{1}{3} \gamma^4 + \left( \frac{2c}{v} \frac{k_B T}{\hbar \vert \omega \vert} \right)^2 & \bigg( \text{Li}_2 \left[ e^{-\frac{\hbar \vert \omega \vert}{k_B T (1+\frac{v}{2c})}} \right] + \text{Li}_2 \left[ e^{-\frac{\hbar \vert \omega \vert}{k_B T (1-\frac{v}{2c})}} \right] \\
            & - \frac{2c}{v} \frac{k_B T}{\hbar \vert \omega \vert} \gamma^{-2} \bigg( \text{Li}_3 \left[ e^{-\frac{\hbar \vert \omega \vert}{k_B T (1+\frac{v}{2c})}} \right] \\
            & ~~~~~~~~~~~~~~~~~~~~~~~ - \text{Li}_3 \left[ e^{-\frac{\hbar \vert \omega \vert}{k_B T (1-\frac{v}{2c})}} \right] \bigg) \bigg) \bigg) \\
            & ~~~~~~~~~~~~~~~~~~~~~~~~~~~~~~~~~~~~~~~~~~~~~~ \times \delta[\omega+\omega'],
        \end{aligned}
    
        \vspace{0.2cm} \\

        \begin{aligned}
            \langle \Tilde{E}^A_X[\omega] \Tilde{E}^A_X[\omega'] \rangle & = \langle \Tilde{E}^A_Z[\omega] \Tilde{E}^A_Z[\omega'] \rangle \\
            & = \frac{\hbar \vert \omega \vert^3}{8 \epsilon_0 \pi^2 c^3} \bigg( \frac{2}{3} \eta \gamma^4 + \frac{2c}{v} \frac{k_B T}{\hbar \vert \omega \vert} \bigg( \left(1+\frac{v}{2c} \right)^{-2} \text{Li}_1 \left[ e^{-\frac{\hbar \vert \omega \vert}{k_B T (1+\frac{v}{2c})}} \right] \\
            & ~~~~~~~~~~~~~~~~~~~~~~~~~~~~~~~~~~~~~~~~~~~~~~~~ - \left(1-\frac{v}{2c} \right)^{-2} \text{Li}_1 \left[ e^{-\frac{\hbar \vert \omega \vert}{k_B T (1-\frac{v}{2c})}} \right] \\
            & ~~~~~~~~~~~~~~~~~~~~~~~~~~~~~~~~~~~~~~~~~~~~~~~~ - \frac{2c}{v} \frac{k_B T}{\hbar \vert \omega \vert} \gamma^{-2} \bigg( \left(1+\frac{v}{2c} \right)^{-2} \text{Li}_2 \left[ e^{-\frac{\hbar \vert \omega \vert}{k_B T (1+\frac{v}{2c})}} \right] \\
            & ~~~~~~~~~~~~~~~~~~~~~~~~~~~~~~~~~~~~~~~~~~~~~~~~~~~~~~~~~~~~~~~~~~~~~~~ + \left(1-\frac{v}{2c} \right)^{-2} \text{Li}_2 \left[ e^{-\frac{\hbar \vert \omega \vert}{k_B T (1-\frac{v}{2c})}} \right] \bigg) \\
            & ~~~~~~~~~~~~~~~~~~~~~~~~~~~~~~~~~~~~~~~~~~~~~~~~ + \left( \frac{2c}{c} \frac{k_B T}{\hbar \vert \omega \vert} \right)^2 \eta \gamma^{-2} \bigg( \text{Li}_3 \left[ e^{-\frac{\hbar \vert \omega \vert}{k_B T (1+\frac{v}{2c})}} \right] \\
            & ~~~~~~~~~~~~~~~~~~~~~~~~~~~~~~~~~~~~~~~~~~~~~~~~~~~~~~~~~~~~~~~~~~~~~~~~~~~~~~~~~~ + \text{Li}_3 \left[ e^{-\frac{\hbar \vert \omega \vert}{k_B T (1-\frac{v}{2c})}} \right] \bigg) \bigg) \bigg) \\
            & ~~~~~~~~~~~~~~~~~~~~~~~~~~~~~~~~~~~~~~~~~~~~~~~~~~~~~~~~~~~~~~~~~~~~~~~~~~~~~~~~~~~~~~~~~~~~~~~~~~~~~~~~~ \times \delta[\omega+\omega']
        \end{aligned}
    \end{array}
\end{equation}
where we used the relativistic notation for the Lorentz factor $\gamma = 1/\sqrt{1-\left(\frac{v}{2c}\right)^2}$, as well as a factor $\eta$ defined by $\eta=(1+\left(\frac{v}{2c}\right)^2)/(1-\left(\frac{v}{2c}\right)^2)$, recalling that the velocity of the point $A$ verifies $(\frac{v}{2c})^2<1$. We also used the polylogarithm function $\text{Li}_n$ defined by 
\begin{equation}
    \text{Li}_n[x] = \sum_{k=1}^\infty \frac{x^k}{k^n}
\end{equation}
where $n$ is a natural integer and $x$ a real number inside $[0,1)$ for our purpose.

Let us first note that the thermal contribution drastically changed compared to the static case, reinforcing the fact that the Planck distribution is not relativistically invariant (cf. Ref.$\cite{ford2013lorentz}$). On the other hand, the zero point part of the correlations is also slightly modified due to the velocity. Hence, they have no reason to be Lorentz scalars, nor to be simply modified by transforming their arguments. However, they should transform in such a way that observable quantities conserve Lorentz invariance. 

First, let us show that one could have obtained the zero point self-correlations (corresponding to the non-thermal part of Eqs. (\ref{self_corr_parallel})) using simple Lorentz transform. We will add an asterisk in the lower right corner to the quantities defined in the frame where the point is motionless. This frame is moving with velocity $-\frac{v}{2} \mathbf{J}$ compared to the laboratory frame or conversely, the laboratory frame is moving with velocity $+\frac{v}{2} \mathbf{J}$ compared to the rest frame. According to the Lorentz transformation of the electromagnetic field, we have:
\begin{equation}
    \begin{array}{c}
        E^A_Y[t] = (E_*)^{A}_{Y}[t_*],
    
        \vspace{0.2cm} \\

        E^A_X[t] = \gamma \left( (E_*)^{A}_Y[t_*] + \frac{v}{2} (B_*)^{A}_Z[t_*] \right),
    
        \vspace{0.2cm} \\

        E^A_Z[t] = \gamma \left( (E_*)^{A}_Z[t_*] - \frac{v}{2} (B_*)^{A}_X[t_*] \right)
    \end{array}
\end{equation}
where $t_*=\frac{t}{\gamma}$ is the proper time associated with the point $A$ (actually with the frame where $A$ is motionless).

As a consequence, the Fourier transform defined with the laboratory frame time $t$ gives rise to extra gamma factors:
\begin{equation}
    \begin{array}{c}
        \Tilde{E}^A_Y[\omega] = \frac{1}{2\pi} \int_{-\infty}^\infty (E_*)^{A}_Y[t_*] e^{i \omega t} \, \mathrm{d}t = \gamma (\Tilde{E}_*)^{A}_Y [\gamma \omega],
    
        \vspace{0.2cm} \\

        \Tilde{E}^A_X[\omega] = \gamma^2 \left( (\Tilde{E}_*)^{A}_Y[\gamma \omega] + \frac{v}{2} (\Tilde{B}_*)^{A}_Z[\gamma \omega] \right),
    
        \vspace{0.2cm} \\

        \Tilde{E}^A_Z[\omega] = \gamma^2 \left( (\Tilde{E}_*)^{A}_Z[\gamma \omega] - \frac{v}{2} (\Tilde{B}_*)^{A}_X[\gamma \omega] \right).
    \end{array}
\end{equation}

Using then the fact that the static electric-magnetic correlations are vanishing (see section \ref{corr_first_order}), and that the magnetic-magnetic ones are equal to the electric-electric ones up to a factor $c^2$ (see the remark at the end of section \ref{EB}), one can recover the exact zero point correlations of Eqs. (\ref{self_corr_parallel}). Of course, one can also do the same for the thermal contribution on the correlations, but it has to be done before the wave vector integration, since the Planck distribution does not transform trivially (see Ref.\cite{ford2013lorentz}). This contribution however yields frame-dependent outcomes of observable
quantities, due to the existence of the blackbody radiation reference frame. One example of this phenomenon is the Einstein-Hopf friction of a single polarizable object in rectilinear and uniform motion (cf. Refs.\cite{einstein-hopf,Sinha_2022}).

\section{Calculation of the electromagnetic field correlations between revolving points}

\begin{figure}
    \centering
    \includegraphics[width=0.5\linewidth]{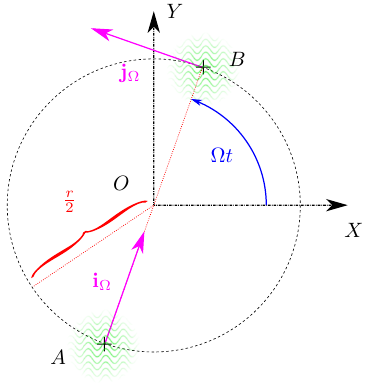}
    \caption{Diagram of two points revolving and diametrically opposed on the same circular trajectory, covered at constant angular velocity $\Omega$. The unit vectors $\mathbf{i}_\Omega$ and $\mathbf{j}_\Omega$ form a direct rotating basis of the motion plane.}
    \label{fig:rotation}
\end{figure}

We now aim to compute correlations between two revolving points. From now on, the two points $A$ and $B$ will have trajectories parametrized by
\begin{equation}\label{rotating_points}
    \begin{array}{cc}
        \displaystyle \mathbf{r}^A = -\frac{r}{2} \mathbf{i}_\Omega, & \displaystyle \mathbf{r}^B = +\frac{r}{2} \mathbf{i}_\Omega
    \end{array}
\end{equation}
where we set $\mathbf{i}_\Omega = \cos[\Omega t] \mathbf{I} + \sin[\Omega t] \mathbf{J}$, with $\Omega$ the constant angular velocity. As set in previous sections, $\mathbf{I}$, $\mathbf{J}$ and $\mathbf{K}$ are the unit vectors from the laboratory frame basis, respectively associated with the axes $X$, $Y$ and $Z$. We consider $\mathbf{k}$ a wave vector. We choose to represent it into spherical coordinates of axis $Z$, since the two points $A$ and $B$ are rotating within the $XY$ plane: $\mathbf{k} = k \left( \sin[\theta] \cos[\phi] \mathbf{I} + \sin[\theta]\sin[\phi] \mathbf{J} + \cos[\theta] \mathbf{K}\right)$ with $\theta$, $\phi$ being the polar and azimuthal angles. The system of two revolving points is sketched in Fig. \ref{fig:rotation}. 

\subsection{Calculation of the electric-electric field correlations for revolving points}\label{EE}

In the following, we use the Jacobi-Anger identity \cite{bateman1953higher}, true for any real values of $x$ and $\phi$: 
\begin{equation}\label{jacobi_anger}
    e^{ix \cos[\phi]} = \sum_{n \in \mathbb{Z}} i^n J_n[x] e^{in\phi}
\end{equation}
where $J_n$ denotes the Bessel function of the first kind and of order $n$ and $\mathbb{Z}$ is the set of all integers, from $-\infty$ to $\infty$. Using Eqs. (\ref{electric_A}), (\ref{rotating_points}) and (\ref{jacobi_anger}), we write the Fourier decomposition of the vacuum electric field seen by point $A$:
\begin{equation}\label{fourier_EA}
    \begin{aligned}
        \Tilde{\hat{\mathbf{E}}}^A[\omega] = i \sum_{\mathbf{k},\lambda} \sum_{n \in \mathbb{Z}} \sqrt{\frac{\hbar \omega_k}{2 \epsilon_0 V}} e^{-in\left(\phi-\frac{\pi}{2}\right)} & \bigg(  J_n\left[-\frac{kr\sin[\theta]}{2}\right] \hat{a}_{\mathbf{k},\lambda} \delta[\omega_n-\omega_k] \mathbf{e}_{\mathbf{k},\lambda} \\
        & -J_n\left[\frac{kr\sin[\theta]}{2}\right] \hat{a}^{\dagger}_{\mathbf{k},\lambda} \delta[\omega_n+\omega_k] \mathbf{e}_{\mathbf{k},\lambda}^* \bigg) 
    \end{aligned}
\end{equation}
where we defined the shifted frequencies $\omega_n = \omega + n \Omega$. Interestingly, the vacuum field in Fourier space depends on the characteristic frequency of the movement of the points through the above integral shifting. 

From Eq. ($\ref{fourier_EA}$), we can actually obtain the electric field at $B$ under the simple transformation $r \rightarrow -r$. This allows us to perform calculations of quadratic vacuum correlations, such as
    \begin{equation}\label{intermediate_elec_elec_correlation}
    \begin{aligned}
        \langle \Tilde{\Hat{E}}_X^A [\omega] \Tilde{\Hat{E}}_X^B [\omega'] \rangle = \sum_{\mathbf{k},\lambda} \sum_{(n,n') \in \mathbb{Z}^2 } \frac{\hbar \omega_k}{2 \epsilon_0 V} e^{-i(n+n')\left( \phi - \frac{\pi}{2} \right)} & \bigg( J_n{\left[ -\frac{k r \sin[\theta]}{2} \right]} J_{n'}{\left[ -\frac{k r \sin[\theta]}{2} \right]} \left(1+N_k \right) \\
        & ~~~~~~~ \times \delta[\omega_n-\omega_k] \delta[\omega'_{n'}+\omega_k]
        \\
        & + J_n{\left[ \frac{k r \sin[\theta]}{2} \right]} J_{n'}{\left[ \frac{k r \sin[\theta]}{2} \right]} N_k \\
        & ~~~~~~~ \times \delta[\omega_n+\omega_k] \delta[\omega'_{n'}-\omega_k] \bigg) \vert \mathbf{e}_{\mathbf{k},\lambda} \cdot \mathbf{I} \vert^2
    \end{aligned}
\end{equation}
using the quantum rules ($\ref{quantum_rules}$).

We can then take again the semi-classical limit $\sum_\mathbf{k} \rightarrow \int V/(2\pi)^3 \mathrm{d}\mathbf{k} = \int V/(2 \pi)^3 \mathrm{d}k \sin[\theta] \mathrm{d}\theta \mathrm{d}\phi$ and use the polarization summing rule $\sum_{\lambda} (\mathbf{e}_{\mathbf{k},\lambda})_i (\mathbf{e}_{\mathbf{k},\lambda})^*_j = \delta_{ij} - \frac{k_i}{k}\frac{k_j}{k}$ together with the azimuthal integral: 
\begin{equation}
    \int_0^{2\pi} e^{-im \left( \phi - \frac{\pi}{2} \right)} e^{il \phi} \, \mathrm{d}\phi = 2\pi i^m \delta_{m,l}
\end{equation}
to perform the integration, where $m$ and $l$ are integers. This last relation will select only a small amount of terms in the $m=n+n'$ sum. The integration on the wave number $k$ will be easily tackled thanks to the Dirac distributions. The non trivial part stands with the integral on $\theta$ involving products of Bessel functions, where the results ($\ref{integral_result_1}$) and ($\ref{integral_result_2}$) of Appendix \ref{lemma} are required. 

To perform the symmetrization over the operators order (\ref{symmetrization}), we require a relation on the Dirac distributions. Let us note that thanks to the fact that $m$ only takes even values, we are allowed to use the following identity, for each mode $n$:
\begin{equation}\label{delta_formula_plus}
    \frac{\delta[\omega-\omega_k]\delta[\omega'+\omega_k]+\delta[\omega+\omega_k]\delta[\omega'-\omega_k]}{2} = \frac{1}{2}\delta[\omega+\omega']\delta[\vert \omega \vert -\omega_k]
\end{equation}
using the formula $\delta[x-a]+\delta[x+a] = 2a \, \delta[x^2-a^2]$ twice.

Applying Eq. (\ref{delta_formula_plus}) in the symmetrization of Eq. (\ref{intermediate_elec_elec_correlation}) yields
\begin{equation}
\begin{aligned}
    \langle \Tilde{E}_X^A[\omega] \Tilde{E}_X^B[\omega'] \rangle = \sum_{n \in \mathbb{Z}} \int & \frac{\hbar \omega_k}{16 \epsilon_0 \pi^2} \left( N_k + \frac{1}{2} \right) k^2 \sin[\theta] \, \mathrm{d}k \, \mathrm{d}\theta \, J_n\left[\frac{kr\sin[\theta]}{2}\right] \\
    & \times \bigg( J_{-n}\left[\frac{kr\sin[\theta]}{2}\right] \left(2-\sin^2[\theta]\right)\delta[\omega_n+\omega'_{-n}]\\
    & ~~~~ + J_{2-n}\left[\frac{kr\sin[\theta]}{2}\right] \frac{\sin^2[\theta]}{2} \delta[\omega_n+\omega'_{2-n}] \\
    & ~~~~ + J_{-2-n}\left[\frac{kr\sin[\theta]}{2}\right] \frac{\sin^2[\theta]}{2} \delta[\omega_n+\omega'_{-2-n}] \bigg) \delta[\vert \omega_n \vert -\omega_k] .
\end{aligned}
\end{equation}

We are then left with the computation of the integrals
\begin{equation}
    \int_0^\pi J_n \left[ \frac{kr \sin[\theta]}{2} \right] J_{m-n} \left[ \frac{kr \sin[\theta]}{2} \right] \left( \left( 2 - \sin^2[\theta] \right) \delta_{m,0} + \frac{\sin^2[\theta]}{2} \left( \delta_{m,2} + \delta_{m,-2} \right) \right) \sin[\theta] \, \mathrm{d}\theta.
\end{equation}

Using now the formulas ($\ref{integral_result_1}$) and ($\ref{integral_result_2}$) from Appendix \ref{lemma}, we can then write
\begin{equation}\label{final_elec_elec_correlation}
    \begin{aligned}
        \textstyle{\langle \Tilde{E}_X^A [\omega] \Tilde{E}_X^B [\omega'] \rangle = \sum_{n} \left( N [\vert \omega_n \vert] + \frac{1}{2} \right) \frac{\hbar \vert \omega_n \vert^3}{16 \epsilon_0 \pi^\frac{3}{2} c^3}} & \textstyle{\bigg( G_n^0 {\left[ \frac{\vert \omega_n \vert r}{c} \right]} \delta[\omega+\omega']} \\
        & \textstyle{+ \, G_n^+ {\left[ \frac{\vert \omega_n \vert r}{c} \right]} \delta[\omega+\omega'+2\Omega]} \\
        & \textstyle{+ \, G_n^- {\left[ \frac{\vert \omega_n \vert r}{c} \right]} \delta[\omega+\omega'-2\Omega] \bigg)} 
    \end{aligned}
\end{equation}
with the notation $N[\omega]=\frac{1}{e^{\hbar \omega/k_B T}-1}$, already introduced in Eq. (\ref{quantum_rules}) and the $G$ correlation functions defined as
\begin{equation}
    \begin{aligned}
        G^0_n \left[x\right] =  2 \, {}_2 F^R_3 \left[\frac{1}{2},1;\frac{3}{2},1-n,1+n;-\left(\frac{x}{2}\right)^2\right] - {}_2 F^R_3 \left[\frac{1}{2},2;\frac{5}{2},1-n,1+n;-\left(\frac{x}{2}\right)^2\right],
    \end{aligned}
\end{equation}

\begin{equation}\label{corr_notations}
    \begin{array}{cc}
        \displaystyle
        G^+_n \left[x\right] = \frac{1}{4}\left(\frac{x}{2}\right)^2 \, {}_2 F^R_3 \left[\frac{3}{2},2;\frac{7}{2},3-n,1+n;-\left(\frac{x}{2}\right)^2\right],

        \vspace{0.2cm}\\
        
        \displaystyle
        G^-_n \left[x\right] = \frac{1}{4}\left(\frac{x}{2}\right)^2 \, {}_2 F^R_3 \left[\frac{3}{2},2;\frac{7}{2},1-n,3+n;-\left(\frac{x}{2}\right)^2\right]
    \end{array}
\end{equation}
which involve regularized hypergeometric functions ${}_p F_q^R$ defined by equation ($\ref{hypergeometric_functions}$) of Appendix \ref{lemma}. 


In expression (\ref{final_elec_elec_correlation}) and in the following, the summation over the index $n$ has to be understood as the sum over all integers, where the ensemble $\mathbb{Z}$ is eluded to lighten the notations.

The expression of the correlation (\ref{final_elec_elec_correlation}) is valid at all orders in $\frac{\Omega r}{c}$. Moreover, we can exchange the points $A$ and $B$ by the simple transformation $r \rightarrow -r$ in the final result. It is also interesting to note that we can easily extract the self-correlation $\langle \Tilde{E}_X^A [\omega] \Tilde{E}_X^A [\omega'] \rangle$ or $\langle \Tilde{E}_X^B [\omega] \Tilde{E}_X^B [\omega'] \rangle$ from the intermediate expression ($\ref{intermediate_elec_elec_correlation}$) by making the transformation $r \rightarrow -r$ in one Bessel factor only, giving rise to a $(-1)^n$ factor within the sum, due to the evenness of $m=n+n'$, coupled with the Bessel functions parity property $J_n[-x]=(-1)^n J_n[x]$:
\begin{equation}\label{ExEx}
    \begin{aligned}
        \textstyle{\langle \Tilde{E}_X^A [\omega] \Tilde{E}_X^A [\omega'] \rangle = \sum_{n} (-1)^n \left( N[\vert \omega_n \vert] + \frac{1}{2} \right) \frac{\hbar \vert \omega_n \vert^3}{16 \epsilon_0 \pi^\frac{3}{2} c^3}} & \textstyle{\bigg( G_n^0 {\left[ \frac{\vert \omega_n \vert r}{c} \right]} \delta[\omega+\omega']} \\
        & \textstyle{+ \, G_n^+ {\left[ \frac{\vert \omega_n \vert r}{c} \right]} \delta[\omega+\omega'+2\Omega]} \\
        & \textstyle{+ \, G_n^- {\left[ \frac{\vert \omega_n \vert r}{c} \right]} \delta[\omega+\omega'-2\Omega] \bigg)}.
    \end{aligned}
\end{equation}

One can first note that Eqs. (\ref{final_elec_elec_correlation}) and (\ref{ExEx}) only differ by the factor $(-1)^n$ within the summation over $\mathbb{Z}$. Interestingly, the velocity and the distance between the two points also appear in the self-correlations. This is due to the fact that the motion is not rectilinear and uniform, so that $\mathbf{r}^A[t]-\mathbf{r}^A[t']$ is not proportional to the time difference $t-t'$. Hence, characteristics of the acceleration, such as the angular velocity or the radius of curvature, remain in the self-correlations expressions. Moreover, we recall that in the above expressions and in the following ones, performing the simple transformation $r \rightarrow -r$ will basically exchange the points $A$ and $B$, so that it is enough to present the correlations $AB$ and $AA$. We can then deal with the other electric-electric fields correlations using an analog method and get for the $Y$ axis:
\begin{equation}\label{EyEy}
\begin{array}{c}
    \begin{aligned}
        \textstyle{\langle \Tilde{E}_Y^A [\omega] \Tilde{E}_Y^B [\omega'] \rangle = \sum_{n} \left( N [\vert \omega_n \vert] + \frac{1}{2} \right) \frac{\hbar \vert \omega_n \vert^3}{16 \epsilon_0 \pi^\frac{3}{2} c^3}} & \textstyle{\bigg( G_n^0 {\left[ \frac{\vert \omega_n \vert r}{c} \right]} \delta[\omega+\omega']} \\
        & \textstyle{- \, G_n^+ {\left[ \frac{\vert \omega_n \vert r}{c} \right]} \delta[\omega+\omega'+2\Omega]} \\
        & \textstyle{- \, G_n^- {\left[ \frac{\vert \omega_n \vert r}{c} \right]} \delta[\omega+\omega'-2\Omega] \bigg) },
    \end{aligned}
    
    \vspace{0.2cm} \\
    
    \begin{aligned}
        \textstyle{\langle \Tilde{E}_Y^A [\omega] \Tilde{E}_Y^A [\omega'] \rangle = \sum_{n} (-1)^n \left( N[\vert \omega_n \vert] + \frac{1}{2} \right) \frac{\hbar \vert \omega_n \vert^3}{16 \epsilon_0 \pi^\frac{3}{2} c^3}} & \textstyle{\bigg( G_n^0 {\left[ \frac{\vert \omega_n \vert r}{c} \right]} \delta[\omega+\omega']} \\
        & \textstyle{- \, G_n^+ {\left[ \frac{\vert \omega_n \vert r}{c} \right]} \delta[\omega+\omega'+2\Omega]} \\
        & \textstyle{- \, G_n^- {\left[ \frac{\vert \omega_n \vert r}{c} \right]} \delta[\omega+\omega'-2\Omega] \bigg) }
    \end{aligned}
\end{array}
\end{equation}
and the $Z$ axis, which is also affected by the in-plane motion:
\begin{equation}\label{EzEz}
    \begin{array}{c}
        \langle \Tilde{E}_Z^A [\omega] \Tilde{E}_Z^B [\omega'] \rangle = \sum_{n} \left( N [\vert \omega_n \vert] + \frac{1}{2} \right) \frac{\hbar \vert \omega_n \vert^3}{16 \epsilon_0 \pi^\frac{3}{2} c^3} G_n^Z {\left[ \frac{\vert \omega_n \vert r}{c} \right]} \delta[\omega+\omega'],

        \vspace{0.2cm} \\

        \langle \Tilde{E}_Z^A [\omega] \Tilde{E}_Z^A [\omega'] \rangle = \sum_{n} (-1)^n \left( N[\vert \omega_n \vert]+ \frac{1}{2} \right) \frac{\hbar \vert \omega_n \vert^3}{16 \epsilon_0 \pi^\frac{3}{2} c^3} G_n^Z {\left[ \frac{\vert \omega_n \vert r}{c} \right]} \delta[\omega+\omega'] 
    \end{array}
\end{equation}
where we defined the $G$ correlation function along $Z$:
\begin{equation}
    G^Z_n\left[x\right] = 2 ~ {}_2 F^R_3 \left[\frac{1}{2},2;\frac{5}{2},1-n,1+n;-\left(\frac{x}{2}\right)^2\right].
\end{equation}

Because of the rotation, the axes $X$ and $Y$ are coupled in the general case. Hence, it is also useful to compute some cross-correlations:
\begin{equation}\label{ExEy}
    \begin{array}{c}
        \begin{aligned}
            \textstyle{\langle \Tilde{E}_X^A [\omega] \Tilde{E}_Y^B [\omega'] \rangle} & \textstyle{ ~ = -i \sum_{n} \left( N[\vert \omega_n \vert] + \frac{1}{2} \right) \frac{\hbar \vert \omega_n \vert^3}{16 \epsilon_0 \pi^\frac{3}{2} c^3} \bigg( G_n^+ {\left[ \frac{\vert \omega_n \vert r}{c} \right]} \delta[\omega+\omega'+2\Omega]} \\
            & ~~~~~~~~~~~~~~~~~~~~~~~~~~~~~~~~~~~~~~~~~~~~~~~~~~ \textstyle{- \, G_n^- {\left[ \frac{\vert \omega_n \vert r}{c} \right]} \delta[\omega+\omega'-2\Omega] \bigg)} \\
            & ~ \textstyle{= \langle \Tilde{E}_Y^A [\omega ] \Tilde{E}_X^B [\omega'] \rangle}, 
        \end{aligned}
    
        \vspace{0.2cm} \\

        \begin{aligned}
            \textstyle{\langle \Tilde{E}_X^A [\omega] \Tilde{E}_Y^A [\omega'] \rangle} & \textstyle{ ~ = -i \sum_{n} (-1)^n \left( N[\vert \omega_n \vert] + \frac{1}{2} \right) \frac{\hbar \vert \omega_n \vert^3}{16 \epsilon_0 \pi^\frac{3}{2} c^3} \bigg( G_n^+ {\left[ \frac{\vert \omega_n \vert r}{c} \right]} \delta[\omega+\omega'+2\Omega]} \\
            & ~~~~~~~~~~~~~~~~~~~~~~~~~~~~~~~~~~~~~~~~~~~~~~~~~~~~~~~~~~~~ \textstyle{- \, G_n^- {\left[ \frac{\vert \omega_n \vert r}{c} \right]} \delta[\omega+\omega'-2\Omega] \bigg)} \\
            & ~ \textstyle{= \langle \Tilde{E}_Y^A [\omega ] \Tilde{E}_X^A [\omega'] \rangle},
        \end{aligned}
    \end{array}
\end{equation}

\begin{equation}
    \langle \Tilde{E}_{\substack{X \\ Z}}^A [\omega] \Tilde{E}_{\substack{Z \\ X}}^B [\omega'] \rangle = \langle \Tilde{E}_{\substack{Y \\ Z}}^A [\omega] \Tilde{E}_{\substack{Z \\ Y}}^B [\omega'] \rangle = \langle \Tilde{E}_{\substack{X \\ Z}}^A [\omega] \Tilde{E}_{\substack{Z \\ X}}^A [\omega'] \rangle = \langle \Tilde{E}_{\substack{Y \\ Z}}^A [\omega] \Tilde{E}_{\substack{Z \\ Y}}^A [\omega'] \rangle = 0.
\end{equation}

One can note that some field projections are still uncorrelated, as in the static case, for any value of the angular velocity $\Omega$ and the distance $r$. In Eqs. (\ref{EyEy}), (\ref{EzEz}) and (\ref{ExEy}) as already noted for the $X$ component, the self correlation only differs from the two-point correlation by the factor $(-1)^n$ within the summation over $\mathbb{Z}$. We can also note that the $G$ correlation functions presented above are associated with the integrals $\mathcal{I}_{l,m,n}$ of the Appendix $\ref{lemma}$ where $m$ is always even. Hence, these correlation functions are even with respect to their argument, which is useful to know in connection with the transformation $r \rightarrow -r$.

\subsection{Calculation of the electric-magnetic correlations for revolving points}\label{EB}

To go further, it is possible with this method to compute other fields correlations than the electric ones. In the following, we will focus on the ones between the electric and magnetic fields.

Using the same procedure as for the electric-electric correlations, we can obtain the electric-magnetic ones with the sum rule over polarization states: $\sum_\lambda (\mathbf{e}_{\mathbf{k},\lambda})_i (\mathbf{k} \times \mathbf{e}_{\mathbf{k},\lambda}^*)_j = \varepsilon_{ijl} k_l$ with $\varepsilon_{ijl}$ being the usual Levi-Civita symbol. However, it is important to note a major change in the symmetrization process; for the electric-magnetic correlations, the values of the integer $m = n+n'$ are odd, so that, instead of Eq. (\ref{delta_formula_plus}), we need the property:
\begin{equation}
    \frac{\delta[\omega-\omega_k]\delta[\omega'+\omega_k]-\delta[\omega+\omega_k]\delta[\omega'-\omega_k]}{2} = \frac{1}{2} \frac{\omega}{\omega_k} \delta[\omega+\omega']\delta[\vert \omega \vert -\omega_k]
\end{equation}
leading to
\begin{equation}\label{ExBz}
    \begin{array}{c}
        \begin{aligned}
        \textstyle{\langle \Tilde{E}^A_X [ \omega ] \Tilde{B}^B_Z [ \omega' ] \rangle} & \textstyle{ ~ = \sum_{n} \left( N[\vert \omega_n \vert] + \frac{1}{2} \right) \frac{\hbar \omega_n \vert \omega_n \vert^2}{16 \epsilon_0 \pi^\frac{3}{2} c^4} \bigg( H^+_n {\left[ \frac{\vert \omega_n \vert r}{c} \right]} \delta [\omega+\omega'+\Omega]} \\
        & ~~~~~~~~~~~~~~~~~~~~~~~~~~~~~~~~~~~~~~~~~~~~~~ \textstyle{- \, H^-_n {\left[ \frac{\vert \omega_n \vert r}{c} \right]} \delta [\omega+\omega'-\Omega] \bigg)} \\
        & \textstyle{ ~ = -\langle \Tilde{E}^A_Z [ \omega ] \Tilde{B}^B_X [ \omega' ] \rangle},
        \end{aligned}
    
        \vspace{0.2cm} \\

        \begin{aligned}
            \textstyle{\langle \Tilde{E}_X^A [\omega] \Tilde{B}_Z^A [\omega'] \rangle} & \textstyle{ ~ = - \sum_{n} (-1)^n \left( N[\vert \omega_n \vert] + \frac{1}{2} \right) \frac{\hbar \omega_n \vert \omega_n \vert^2}{16 \epsilon_0 \pi^\frac{3}{2} c^4} \bigg( H_n^+ {\left[ \frac{\vert \omega_n \vert r}{c} \right]} \delta[\omega+\omega'+\Omega]} \\
            & ~~~~~~~~~~~~~~~~~~~~~~~~~~~~~~~~~~~~~~~~~~~~~~~~~~~~~~~~~~~ \textstyle{- \, H_n^- {\left[ \frac{\vert \omega_n \vert r}{c} \right]} \delta[\omega+\omega'-\Omega] \bigg)} \\
            & \textstyle{ ~ = -\langle \Tilde{E}_Z^A [\omega] \Tilde{B}_X^A [\omega'] \rangle }.
        \end{aligned}
    \end{array}
\end{equation}

We can then compute with an analogous method:
\begin{equation}\label{EyBz}
    \begin{array}{c}
        \begin{aligned}
            \textstyle{\langle \Tilde{E}^A_Y [ \omega ] \Tilde{B}^B_Z [ \omega' ] \rangle} & \textstyle{ ~ = -i\sum_{n} \left( N[\vert \omega_n \vert] + \frac{1}{2} \right) \frac{\hbar \omega_n \vert \omega_n \vert^2}{16 \epsilon_0 \pi^\frac{3}{2} c^4} \bigg( H^+_n {\left[ \frac{\vert \omega_n \vert r}{c} \right]} \delta [\omega+\omega'+\Omega]} \\
            & ~~~~~~~~~~~~~~~~~~~~~~~~~~~~~~~~~~~~~~~~~~~~~~~~~~ \textstyle{+ \, H^-_n {\left[ \frac{\vert \omega_n \vert r}{c} \right]} \delta [\omega+\omega'-\Omega] \bigg)} \\
            & \textstyle{ ~ = -\langle \Tilde{E}^A_Z [ \omega ] \Tilde{B}^B_Y [ \omega' ] \rangle },
        \end{aligned}
        
        \vspace{0.2cm} \\

        \begin{aligned}
            \textstyle{\langle \Tilde{E}_Y^A [\omega] \Tilde{B}_Z^A [\omega'] \rangle} & \textstyle{~ = i \sum_{n} (-1)^n \left( N[\vert \omega_n \vert] + \frac{1}{2} \right) \frac{\hbar \omega_n \vert \omega_n \vert^2}{16 \epsilon_0 \pi^\frac{3}{2} c^4} \bigg( H_n^+ {\left[ \frac{\vert \omega_n \vert r}{c} \right]} \delta[\omega+\omega'+\Omega]} \\
            & ~~~~~~~~~~~~~~~~~~~~~~~~~~~~~~~~~~~~~~~~~~~~~~~~~~~~~~~~~ \textstyle{+ \, H_n^- {\left[ \frac{\vert \omega_n \vert r}{c} \right]} \delta[\omega+\omega'-\Omega] \bigg)} \\
            & \textstyle{ ~= - \langle \Tilde{E}_Z^A [\omega] \Tilde{B}_Y^A [\omega'] \rangle} 
        \end{aligned}
    \end{array}
\end{equation}
and show the vanishing of the correlations:
\begin{equation}
    \begin{array}{c}
        \begin{aligned}
            & \langle \Tilde{E}^A_X [\omega] \Tilde{B}^B_X [\omega'] \rangle = \langle \Tilde{E}^A_Y [\omega] \Tilde{B}^B_Y [\omega'] \rangle = \langle \Tilde{E}^A_Z [\omega] \Tilde{B}^B_Z [\omega'] \rangle \\
            & = \langle \Tilde{E}^A_X [\omega] \Tilde{B}^A_X [\omega'] \rangle = \langle \Tilde{E}^A_Y [\omega] \Tilde{B}^A_Y [\omega'] \rangle = \langle \Tilde{E}^A_Z [\omega] \Tilde{B}^A_Z [\omega'] \rangle = 0,
        \end{aligned}
    
        \vspace{0.2cm} \\

        \langle \Tilde{E}^A_X [ \omega ] \Tilde{B}^B_Y [ \omega' ] \rangle = \langle \Tilde{E}^A_Y [ \omega ] \Tilde{B}^B_X [ \omega' ] \rangle =  \langle \Tilde{E}^A_X [ \omega ] \Tilde{B}^A_Y [ \omega' ] \rangle = \langle \Tilde{E}^A_Y [ \omega ] \Tilde{B}^A_X [ \omega' ] \rangle = 0
    \end{array}
\end{equation}
where we defined the electric-magnetic field correlation $H$ functions
\begin{equation}\label{definition_Hn}
\begin{array}{cc}
    \displaystyle
    H^+_n \left[ x \right] = \frac{1}{2} \frac{x}{2} ~ {}_2 F^R_3 \left[1,\frac{3}{2};\frac{5}{2},2-n,1+n;-\left(\frac{x}{2}\right)^2 \right],

    \vspace{0.2cm} \\
        
    \displaystyle
    H^-_n \left[ x \right] = \frac{1}{2} \frac{x}{2} ~ {}_2 F^R_3 \left[1,\frac{3}{2};\frac{5}{2},1-n,2+n;-\left(\frac{x}{2}\right)^2 \right].
\end{array}
\end{equation}

One could also be interested in computing the magnetic-magnetic correlations: this is simply obtained from the electric-electric ones (cf. section \ref{EE}), remarking the relation of polarization vectors: $\sum_\lambda (\mathbf{k} \times \mathbf{e}_{\mathbf{k},\lambda})_i (\mathbf{k} \times \mathbf{e}_{\mathbf{k},\lambda})_j^* = k^2 \sum_\lambda (\mathbf{e}_{\mathbf{k},\lambda})_i (\mathbf{e}_{\mathbf{k},\lambda})_j^*$. Moreover, these $H$ correlations functions are derived from integrals $\mathcal{I}_{l,m,n}$ with odd values of $m$ and are thus odd functions of their argument. In equations (\ref{ExBz}) and (\ref{EyBz}) we explicitly wrote $\omega_n \vert \omega_n \vert^2$ instead of $\omega_n^3$ in order to make the unevenness of the correlations even clearer.

We also propose in Appendices \ref{EdE} and \ref{EdB} the correlations between the electromagnetic field and its own spatial derivatives, which can also be useful, for instance when the fields acts on an electric dipole.

\subsection{Rewriting of the correlations with the circular fields}

Because of the rotation, it will actually be convenient to work with the circularly transformed fields \cite{PRL_Herve}
\begin{equation}
    \begin{array}{cc}
        \displaystyle \Tilde{E}^A_\pm = \frac{\Tilde{E}^A_X \pm i \Tilde{E}^A_Y}{2}, & \displaystyle \Tilde{E}^B_\pm = \frac{\Tilde{E}^B_X \pm i \Tilde{E}^B_Y}{2},

        \vspace{0.2cm} \\

        \displaystyle \Tilde{B}^A_\pm = \frac{\Tilde{B}^A_X \pm i \Tilde{B}^A_Y}{2}, & \displaystyle \Tilde{B}^B_\pm = \frac{\Tilde{B}^B_X \pm i \Tilde{B}^B_Y}{2}
    \end{array}
\end{equation}
where the circular projections $\pm$ should be read line by line.

Together with the $Z$ projections, these fields $E_\pm$ and $B_\pm$ will be referred to as circular fields. From the expressions of the previous sections \ref{EE}, \ref{EB} as well as Appendices \ref{EdE} and \ref{EdB}, one can show for the circular electric-electric two-point and self-correlations:
\begin{equation}
    \begin{array}{c}
        \langle \Tilde{E}_\pm^A [\omega] \Tilde{E}_\mp^B [\omega'] \rangle = \frac{1}{2} \frac{\hbar}{16 \epsilon_0 \pi^\frac{3}{2} r^3} \sum_n \left( N[\vert \omega_n \vert]+\frac{1}{2} \right) \left( \frac{\vert \omega_n \vert r}{c} \right)^3 G_n^0 \left[ \frac{\vert \omega_n \vert r}{c} \right] \delta[\omega+\omega'],
    
        \vspace{0.2cm} \\

        \langle \Tilde{E}_\pm^A [\omega] \Tilde{E}_\mp^A [\omega'] \rangle = \frac{1}{2} \frac{\hbar}{16 \epsilon_0 \pi^\frac{3}{2} r^3} \sum_n (-1)^n \left( N[\vert \omega_n \vert]+\frac{1}{2} \right) \left( \frac{\vert \omega_n \vert r}{c} \right)^3 G_n^0 \left[ \frac{\vert \omega_n \vert r}{c} \right] \delta[\omega+\omega'],
    \end{array}
\end{equation}

\begin{equation}
    \begin{array}{c}
        \langle \Tilde{E}_\pm^A [\omega] \Tilde{E}_\pm^B [\omega'] \rangle = \frac{\hbar}{16 \epsilon_0 \pi^\frac{3}{2} r^3} \sum_n \left( N[\vert \omega_n \vert]+\frac{1}{2} \right) \left( \frac{\vert \omega_n \vert r}{c} \right)^3 G_n^\pm \left[ \frac{\vert \omega_n \vert r}{c} \right] \delta[\omega+\omega'\pm 2 \Omega],
    
        \vspace{0.2cm} \\

        \langle \Tilde{E}_\pm^A [\omega] \Tilde{E}_\pm^A [\omega'] \rangle = \frac{\hbar}{16 \epsilon_0 \pi^\frac{3}{2} r^3} \sum_n (-1)^n \left( N[\vert \omega_n \vert]+\frac{1}{2} \right) \left( \frac{\vert \omega_n \vert r}{c} \right)^3 G_n^\pm \left[ \frac{\vert \omega_n \vert r}{c} \right] \delta[\omega+\omega'\pm 2 \Omega],
    \end{array}
\end{equation}

\begin{equation}
    \langle \Tilde{E}_\pm^A [\omega] \Tilde{E}_Z^B [\omega'] \rangle = \langle \Tilde{E}_Z^A [\omega] \Tilde{E}_\pm^B [\omega'] \rangle = \langle \Tilde{E}_\pm^A [\omega] \Tilde{E}_Z^A [\omega'] \rangle = \langle \Tilde{E}_Z^A [\omega] \Tilde{E}_\pm^A [\omega'] \rangle = 0,
\end{equation}
and for the circular electric-magnetic ones:
\begin{equation}
    \begin{array}{c}
        \begin{aligned}
            \langle \Tilde{E}_\pm^A [\omega] \Tilde{B}_Z^B [\omega'] \rangle & = \textstyle{\pm \frac{\hbar}{16 \epsilon_0 \pi^\frac{3}{2} c r^3} \sum_n \left( N[\vert \omega_n \vert]+\frac{1}{2} \right) \left( \frac{\omega_n r}{c} \right)^3 H_n^\pm \left[ \frac{\vert \omega_n \vert r}{c} \right] \delta[\omega+\omega' \pm \Omega]} \\
            & = -\langle \Tilde{E}_Z^A [\omega] \Tilde{B}_\pm^B [\omega'] \rangle,
        \end{aligned}
    
        \vspace{0.2cm} \\

        \begin{aligned}
            \langle \Tilde{E}_\pm^A [\omega] \Tilde{B}_Z^A [\omega'] \rangle & = \textstyle{\mp \frac{\hbar}{16 \epsilon_0 \pi^\frac{3}{2} c r^3} \sum_n (-1)^n \left( N[\vert \omega_n \vert]+\frac{1}{2} \right) \left( \frac{\omega_n r}{c} \right)^3 H_n^\pm \left[ \frac{\vert \omega_n \vert r}{c} \right] \delta[\omega+\omega' \pm \Omega]} \\
            & = -\langle \Tilde{E}_Z^A [\omega] \Tilde{B}_\pm^A [\omega'] \rangle,
        \end{aligned}
    \end{array}
\end{equation}

\begin{equation}
    \langle \Tilde{E}_\pm^A [\omega] \Tilde{B}^B_\pm [\omega'] \rangle = \langle \Tilde{E}_\pm^A [\omega] \Tilde{B}^B_\mp [\omega'] \rangle = \langle \Tilde{E}_\pm^A [\omega] \Tilde{B}^A_\pm [\omega'] \rangle = \langle \Tilde{E}_\pm^A [\omega] \Tilde{B}^A_\mp [\omega'] \rangle = 0.
\end{equation}

An important feature of the correlations of the circular fields is that the frequency shift is equal to the sum of the $+$ and $-$ symbols. For instance $\langle \Tilde{E}_+^A [\omega] \Tilde{E}_-^B [\omega'] \rangle$ is associated with the Dirac distribution $\delta[\omega+\omega'+\Omega-\Omega]$ while $\langle \Tilde{E}_+^A [\omega] \Tilde{E}_+^B [\omega'] \rangle$ is proportional to $\delta[\omega+\omega'+\Omega+\Omega]$ (cf. table \ref{tab:corr_func}).

One can easily write correlations into the laboratory basis using combinations of the above circularly transformed ones. For instance, we have: $\langle \Tilde{E}_X^A [\omega] \Tilde{E}_X^B [\omega'] \rangle = \langle \Tilde{E}^A_+ [\omega] \Tilde{E}^B_+ [\omega'] \rangle + \langle \Tilde{E}^A_+ [\omega] \Tilde{E}^B_- [\omega'] \rangle + \langle \Tilde{E}^A_- [\omega] \Tilde{E}^B_+ [\omega'] \rangle + \langle \Tilde{E}^A_- [\omega] \Tilde{E}^B_- [\omega'] \rangle$, so that the two-point XX correlation indeed contains frequency shifts of $0$ and $\pm 2 \Omega$.

As evoked previously, the expressions for the electromagnetic field correlations with its spatial derivatives, using circular notations, are presented in Appendix \ref{circ_notations_EdEB}. A brief summary of the non-zero field correlations regarding the associated correlation functions and their properties can be found in table \ref{tab:corr_summary}.

\begin{table}[ht]
    \centering
    \begin{tabular}{|c|c|c|c|}
        \hline
        Field correlation & Correlation function & Symmetry under $\omega \leftrightarrow \omega'$ & Transformation under $\Omega \rightarrow -\Omega$ 
        \\ \hline
        $\langle \Tilde{E}_\pm [\omega] \Tilde{E}_\mp [\omega'] \rangle$ & $G_n^0$ & even & $\langle \Tilde{E}_\mp [\omega] \Tilde{E}_\pm [\omega'] \rangle$
        \\ \hline
        $\langle \Tilde{E}_\pm [\omega] \Tilde{E}_\pm [\omega'] \rangle$ & $G_n^\pm$ & even & $\langle \Tilde{E}_\mp [\omega] \Tilde{E}_\mp [\omega'] \rangle$
        \\ \hline
        $\langle \Tilde{E}_Z [\omega] \Tilde{E}_Z [\omega'] \rangle$ & $G_n^Z$ & even & $\langle \Tilde{E}_Z [\omega] \Tilde{E}_Z [\omega'] \rangle$ 
        \\ \hline 
        $\langle \Tilde{E}_\pm [\omega] \Tilde{B}_Z [\omega'] \rangle$ & $H_n^\pm$ & odd & $-\langle \Tilde{E}_\mp [\omega] \Tilde{B}_Z [\omega'] \rangle$
        \\ \hline
        $\langle \Tilde{E}_\pm [\omega] \partial_\pm \Tilde{E}_\pm [\omega'] \rangle$ & $P_n^{3\pm}$ & even & $\langle \Tilde{E}_\mp [\omega] \partial_\mp \Tilde{E}_\mp [\omega'] \rangle$
        \\ \hline
        $\langle \Tilde{E}_\pm [\omega] \partial_\pm \Tilde{E}_\mp [\omega'] \rangle$ & $P_n^{\div \pm}$ & even & $\langle \Tilde{E}_\mp [\omega] \partial_\mp \Tilde{E}_\pm [\omega'] \rangle$
        \\ \hline
        $\langle \Tilde{E}_\pm [\omega] \partial_\mp \Tilde{E}_\pm [\omega'] \rangle$ & $P_n^{\times \pm}$ & even & $\langle \Tilde{E}_\mp [\omega] \partial_\pm \Tilde{E}_\mp [\omega'] \rangle$
        \\ \hline
        $\langle \Tilde{E}_\pm [\omega] \partial_Z \Tilde{E}_Z [\omega'] \rangle$ & $P_n^{Z\pm}$ & even & $\langle \Tilde{E}_\mp [\omega] \partial_Z \Tilde{E}_Z [\omega'] \rangle$
        \\ \hline
        $\langle \Tilde{E}_Z[\omega] \partial_\pm \Tilde{E}_Z [\omega'] \rangle$ & $P_n^{\times \pm}$ & even & $\langle \Tilde{E}_Z[\omega] \partial_\mp \Tilde{E}_Z [\omega'] \rangle$
        \\ \hline
        $\langle \Tilde{B}_\pm [\omega] \partial_\pm \Tilde{E}_Z [\omega'] \rangle$ & $G_n^{\pm}$ & odd & $-\langle \Tilde{B}_\mp [\omega] \partial_\mp \Tilde{E}_Z [\omega'] \rangle$
        \\ \hline
        $\langle \Tilde{B}_\pm [\omega] \partial_\mp \Tilde{E}_Z [\omega'] \rangle$ & $G_n^Z$ & odd & $-\langle \Tilde{B}_\mp [\omega] \partial_\pm \Tilde{E}_Z [\omega'] \rangle$
        \\ \hline
        $\langle \Tilde{B}_\pm [\omega] \partial_Z \Tilde{E}_\mp [\omega']$ & $G_n^Z$ & odd & $-\langle \Tilde{B}_\mp [\omega] \partial_Z \Tilde{E}_\pm [\omega']$
        \\ \hline
    \end{tabular}
    \caption{Summary of the characteristics of the non-zero field correlations expressed using circular notations. Each field correlation is presented with the associated correlation function and with properties under exchange of frequency arguments and switch of the rotation direction. These properties being valid for both the cross and self-correlations, we did not labeled any point $A$ or $B$ on the fields, but it should still be understood as evaluated at one of them. One can in particular remark the interesting property of the circular notation: under the transformation $\Omega \rightarrow -\Omega$, the $+$ projections changes to the $-$ ones, and vice versa.}
    \label{tab:corr_summary}
\end{table}

\section{Approximations of the correlations to first order in the velocity}\label{corr_first_order}

The expressions given above are exact field correlations. However, one may be interested in approximate correlations, e.g. up to first order in velocity, which are valid in a non-relativistic treatment. For this we need some formulas involving summations of the correlation functions, which are presented in Appendix \ref{formulas_corr}.

Using again the circular notations, we can apply the results of the Appendix \ref{formulas_corr} to derive the expressions of the correlations which are non trivially vanishing to first order in $\Omega r/c$. The correlations $AB$ and $AA$ are then given by
\begin{equation}
    \begin{array}{c}
        \langle \Tilde{E}_\pm^A [\omega] \Tilde{E}_\mp^B [\omega'] \rangle \approx \frac{\hbar}{16 \epsilon_0 \pi^2 r^3} \left( N[\vert \omega \vert]+\frac{1}{2} \right) \bigg( \bigg( 1 + \left( \frac{\vert \omega \vert r}{c} \right)^2 \bigg) \sin \left[ \frac{\vert \omega \vert r}{c} \right] - \frac{\vert \omega \vert r}{c} \cos \left[ \frac{\vert \omega \vert r}{c} \right] \bigg) \delta[\omega+\omega'],
    
        \vspace{0.2cm} \\

        \langle \Tilde{E}_\pm^A [\omega] \Tilde{E}_\mp^A [\omega'] \rangle \approx \frac{\hbar \vert \omega \vert^3}{12 \epsilon_0 \pi^2 c^3} \left( N[\vert \omega \vert]+\frac{1}{2} \right) \delta[\omega+\omega'],
    \end{array}
\end{equation}

\begin{equation}
    \begin{array}{c}
        \begin{aligned}
            & \textstyle{\langle \Tilde{E}^A_\pm [\omega] \Tilde{E}^B_\pm [\omega'] \rangle \approx \frac{\hbar}{16 \epsilon_0 \pi^2 r^3} \bigg( \left( N[\vert \omega \vert]+\frac{1}{2} \right) \bigg( \bigg( 3 - \left( \frac{\vert \omega \vert r}{c} \right)^2 \bigg) \sin \left[ \frac{\vert \omega \vert r}{c} \right] - 3 \frac{\vert \omega \vert r}{c} \cos \left[ \frac{\vert \omega \vert r}{c} \right] \bigg)} \\
            & ~~~~~~~~~~~~~~~~~~~~~~~~~~~~~~~~~~~~~~~~~ \textstyle{\pm \left( N[\vert \omega \vert]+\frac{1}{2} \right) \frac{\Omega r}{c} \frac{\omega r}{c} \bigg( \sin \left[ \frac{\vert \omega \vert r}{c} \right] - \frac{\vert \omega \vert r}{c} \cos \left[ \frac{\vert \omega \vert r}{c} \right] \bigg)} \\
            & ~~~~~~~~~~~~~~~~~~~~~~~~~~~~~~~~~~~~~~~~~ \textstyle{\mp \, \frac{\hbar \vert \omega \vert}{k_B T} N[\vert \omega \vert] \left( N[\vert \omega \vert]+1 \right) \frac{\Omega r}{c} \left( \frac{\omega r}{c} \right)^{-1} \bigg( \bigg( 3 - \left( \frac{\vert \omega \vert r}{c} \right)^2 \bigg) \sin \left[ \frac{\vert \omega \vert r}{c} \right]} \\
            & ~~~~~~~~~~~~~~~~~~~~~~~~~~~~~~~~~~~~~~~~~~~~~~~~~~~~~~~~~~~~~~~~~~~~~~~~~~~~~~~~~~~~~~~~~~~~~~ - \textstyle{3 \frac{\vert \omega \vert r}{c} \cos \left[ \frac{\vert \omega \vert r}{c} \right] \bigg) \bigg) \delta[\omega+\omega' \pm 2\Omega]},
        \end{aligned}
    
        \vspace{0.2cm} \\

        \langle \Tilde{E}^A_\pm [\omega] \Tilde{E}^A_\pm [\omega'] \rangle \approx 0,
    \end{array}
\end{equation}

\begin{equation}
    \begin{array}{c}
        \langle \Tilde{E}_Z^A [\omega] \Tilde{E}_Z^B [\omega'] \rangle \approx - \frac{\hbar}{4 \epsilon_0 \pi^2 r^3} \left( N[\vert \omega \vert]+\frac{1}{2} \right) \bigg( \bigg( 1 - \left( \frac{\vert \omega \vert r}{c} \right)^2 \bigg) \sin \left[ \frac{\vert \omega \vert r}{c} \right] - \frac{\vert \omega \vert r}{c} \cos \left[ \frac{\vert \omega \vert r}{c} \right] \bigg) \delta[\omega+\omega'],
    
        \vspace{0.2cm} \\

        \langle \Tilde{E}_Z^A [\omega] \Tilde{E}_Z^A [\omega'] \rangle \approx \frac{\hbar \vert \omega \vert^3}{6 \epsilon_0 \pi^2 c^3} \left( N[\vert \omega \vert]+\frac{1}{2} \right) \delta[\omega+\omega']
    \end{array}
\end{equation}
where we did not explicitly mentioned the $ O \left( \left( \Omega r/c \right)^2 \right)$ to lighten the notations. We however keep the approximate sign $\approx$ to avoid any ambiguity. One can remark that the approximate correlations on the $Z$ axis are unaffected by the points motion, contrary to the exact expressions (\ref{EzEz}). However, even to zeroth order, some other correlations involve frequency shifts of $\pm 2 \Omega$ within the Dirac distributions. 

Such expressions can also be obtained for the electric-magnetic correlations, using again the formulas of Appendix \ref{formulas_corr}:
\begin{equation}
    \begin{array}{c}
        \begin{aligned}
            \langle \Tilde{E}_\pm^A [\omega] \Tilde{B}_Z^B [\omega'] \rangle & =  \langle \Tilde{E}_Z^A [\omega] \Tilde{B}_\pm^B [\omega'] \rangle \\
            & \textstyle{~\approx \pm \frac{\hbar}{16 \epsilon_0 \pi^2 c r^3} \bigg( \left( N[\vert \omega \vert]+\frac{1}{2} \right) 2 \frac{\omega r}{c} \bigg( \sin \left[ \frac{\vert \omega \vert r}{c} \right] - \frac{\vert \omega \vert r}{c} \cos \left[ \frac{\vert \omega \vert r}{c} \right] \bigg)} \\
            & ~~~~~~~~~~~~~~~~~~~~~ \textstyle{\pm \left( N[\vert \omega \vert]+\frac{1}{2} \right) \frac{\Omega r}{c} \bigg( \bigg( 1 + \left( \frac{\vert \omega \vert r}{c} \right)^2 \bigg) \sin \left[ \frac{\vert \omega \vert r}{c} \right] - \frac{\vert \omega \vert r}{c} \cos \left[ \frac{\vert \omega \vert r}{c} \right] \bigg)} \\
            & ~~~~~~~~~~~~~~~~~~~~~ \textstyle{\mp \, \frac{\hbar \vert \omega \vert}{k_B T} N[\vert \omega \vert] \left( N[\vert \omega \vert]+1 \right) \frac{\Omega r}{c} \bigg( \sin \left[ \frac{\vert \omega \vert r}{c} \right]} \\
            & ~~~~~~~~~~~~~~~~~~~~~~~~~~~~~~~~~~~~~~~~~~~~~~~~~~~~~~~~~~~~~~~~~ - \textstyle{\frac{\vert \omega \vert r}{c} \cos \left[ \frac{\vert \omega \vert r}{c} \right] \bigg) \bigg) \delta[\omega+\omega' \pm \Omega] }, 
        \end{aligned}
    
        \vspace{0.2cm} \\

        \begin{aligned}
            \textstyle{\langle \Tilde{E}_\pm^A [\omega] \Tilde{B}_Z^A [\omega'] \rangle} & = - \textstyle{\langle \Tilde{E}_Z^A [\omega] \Tilde{B}_\pm^A [\omega'] \rangle} \\
            & \approx \textstyle{\frac{\hbar \vert \omega \vert^3}{48 \epsilon_0 \pi^2 c^4} \bigg( \left( N[\vert \omega \vert]+\frac{1}{2} \right) 4 \frac{\Omega r}{c} - \frac{\hbar \vert \omega \vert}{k_B T} N[\vert \omega \vert] \left( N[\vert \omega \vert]+1 \right) \frac{\Omega r}{c} \bigg) \delta[\omega+\omega' \pm \Omega] },  
        \end{aligned} 
    \end{array}
\end{equation}

Finally, the reader can find in Appendix \ref{approx_EdEB} the approximate expressions of the electromagnetic field correlations with its spatial derivatives, up to first order terms in $\Omega r/c$.

\section{Discussion and conclusion}

We first expressed two-point correlations in the time domain, where the movement can be easily introduced. We then entered in Fourier space with the unbounded case where two points are moving along two parallel trajectories in opposite direction. This case is important for the study of atomic collisions with non-zero impact parameter, see for instance Refs.\cite{Barton_2010,barton2011van,hoye2010casimir1,hoye2011casimir,hoye2011casimir2}. The latter references were using only first quantization formalism, where the vacuum field was the exactly vanishing classical electromagnetic field. The results we are here presenting can hence be used to derive a radiation correction to their friction, arising from the dynamical correlations presented in our paper.

Secondly, we focused on the case of two rotating points. The dynamical correlations of such a system show up in two types of terms involving the rotation speed $\Omega$: some corrections, of order $\Omega r/c$, where the ratio of a point velocity to the speed of light naturally occurs, appear along frequency shifts $m \Omega$, with $m$ an integer, in the Dirac distributions, and not compared with any other characteristic frequency. The latter cannot be easily neglected and have some physical meaning related to time delay representation in Fourier space: it might be necessary to keep the frequency shifts, even if one is interested in terms of the zeroth order in the ratio of the velocity to the speed of light. For instance, the $XX$ electric-electric correlations are, in time domain, equal to the $YY$ ones, up to the delay to travel a quarter of the circle:
\begin{equation}
    \begin{array}{cc}
        \delta[\omega+\omega'] e^{i(\omega+\omega')\frac{\pi}{2 \Omega}} = \delta[\omega+\omega'], & \delta[\omega+\omega' \pm 2 \Omega] e^{i(\omega+\omega')\frac{\pi}{2 \Omega}} = -\delta[\omega+\omega' \pm 2 \Omega]
    \end{array}
\end{equation}
lead to
\begin{equation}
    \langle E_Y^A [t] E_Y^B [t] \rangle = \bigg\langle E_X^A \left[t-\frac{\pi}{2 \Omega} \right] E_X^B \left[t-\frac{\pi}{2 \Omega} \right] \bigg\rangle
\end{equation}
with $\pi/2\Omega$ being the delay to cover a quarter-circle within the rotation motion. 

The approximate correlations to first order in $\Omega$, in the rotating case, can also be proven to match with the correlations of two points moving uniformly and rectilinearly with velocity $v$ along the $Y$ axis, case discussed in section $\ref{Rect_lin_motion}$. It however requires to consider $\Omega$ as a small frequency scale compared to any other in the problem, since it appears directly in the shifts, which was not the case in section $\ref{Rect_lin_motion}$ where the velocity $v$ was naturally compared to $c$ due to usual Doppler effect. Under such assumption on the angular velocity $\Omega$, we can argue that the present correlations expressions to first order in the velocity can be extrapolated to more complicated trajectories.

The method presented here has the advantage of being material independent and allows us to derive properties related only to the electromagnetic field. It can be applied to two points involved in many types of motion, as long as the Fourier transform has a usable closed form, and can also be used to compute quartic and even higher order field products. 

As a summary, in this paper, we derived several expressions of the vacuum field correlations using only the usual quantum field operators, with or without a symmetrization process, in frequency domain, where the point motion has an important contribution. The symmetrized results can be used in semi-classical models and match, in the static case, with correlations obtained from the FDT (cf. Ref.\cite{Ford03072017}). For instance, we are currently using most of the results of the rotating case presented above for the consistent calculation of the attraction and friction forces between a pair of revolving atomic oscillators \cite{Vaz1,Vaz2}. Finally, one can further note that these field quadratic averages all have a $\hbar$ prefactor, denoting their quantum nature, without considering matter quantization.

\begin{acknowledgments}
First, we would like to thank Stefan Buhmann and the SPHYNX group for very interesting discussions. We also thank the Ecole Normale Supérieure Paris-Saclay for M.V.'s PhD founding.
\end{acknowledgments}

\section*{Authors declaration}
\subsection*{Conflict of interest}
The authors have no conflicts to disclose.

\subsection*{Authors contributions}
\textbf{Michael Vaz}: conceptualization (equal); formal analysis (lead); writing/original draft preparation (lead); writing, review and editing (equal)

\textbf{Hervé Bercegol}: conceptualization (equal); supervision (lead); writing, review and editing (equal)

\section*{Data availability}

Data sharing is not applicable to this article as no new data were created or analyzed in this study.

\appendix

\section{Digamma function and Fourier transform properties}\label{digamma}


We are interested in $\psi$, the logarithmic derivative of Euler's Gamma Function \cite{andrews1999special}, also called digamma function. The Gauss integral representation of the digamma function gives, for any real value of $x$:
\begin{equation}
    \psi[1+ix] = \int_0^\infty \bigg( \frac{e^{-k}}{k} - \frac{e^{-k}}{1-e^{-k}} e^{-ikx} \bigg) \, \mathrm{d}k.
\end{equation}

Hence, we have the interesting property:
\begin{equation}
    \psi[1+ix]-\psi[1-ix] = \int_0^\infty \frac{1}{e^k-1} \big( e^{ikx} - e^{-ikx} \big) \, \mathrm{d}k
\end{equation}
which is very close to a Fourier decomposition. By performing the change of variable $k \rightarrow -k$ in one of the terms, one can show
\begin{equation}
    \Im[\psi[1+ix]] = \frac{i}{2} \int_{-\infty}^\infty \frac{\text{sgn}[k]}{e^{\vert k \vert}-1} e^{-ikx} \, \mathrm{d}k
\end{equation}
where $\Im$ stands for the imaginary part. This result is directly the Fourier expansion of the left hand side.

With the same idea, we can set, for any real numbers $x$ and $y$:
\begin{equation}
    \begin{aligned}
        \psi[1+i(x+y)]-\psi[1+i(x-y)] & = \int_0^\infty \frac{1}{e^k-1} \big( e^{-ik(x-y)}-e^{-ik(x+y)} \big) \, \mathrm{d}k  \\
        & = 2i \int_0^\infty \frac{1}{e^k-1} \sin[ky] e^{-ikx} \, \mathrm{d}k
    \end{aligned}
\end{equation}
which is a useful property for the correlations calculations.

By differentiating once with respect to $y$, we can prove
\begin{equation}
    \psi'[1+i(x+y)]+\psi'[1+i(x-y)] = 2 \int_0^\infty \frac{k}{e^k-1} \cos[ky] e^{-ikx} \, \mathrm{d}k.
\end{equation}

Then, by differentiating again with respect to $y$, we have
\begin{equation}
    \psi''[1+i(x+y)]-\psi''[1+i(x-y)] = 2i \int_0^\infty \frac{k^2}{e^k-1} \sin[ky] e^{-ikx} \, \mathrm{d}k.
\end{equation}

\section{Integration of the product of two Bessel functions}\label{lemma}
In the following, we want to compute integrals of the form
\begin{equation}
    \begin{array}{ccc}
        \displaystyle \int_0^\pi \cos[\theta] \sin^l[\theta] J_n [\kappa \sin[\theta]] J_{m-n} [\kappa \sin[\theta]] \, \mathrm{d}\theta & \text{or} & \displaystyle \int_0^\pi \sin^l[\theta] J_n [\kappa \sin[\theta]] J_{m-n} [\kappa \sin[\theta]] \, \mathrm{d}\theta
    \end{array}
\end{equation}
where $J_n$ is the Bessel functions of the first kind and of order $n$, $\kappa$ being a real positive number, $l$, $m$ and $n$ being integers and $l \ge 0$.

The first integral can be shown to be zero by symmetry between $[0,\frac{\pi}{2}]$ and $[\frac{\pi}{2},\pi]$ of the sinuses and cosines functions. With the same symmetries, we can rewrite the second integral using only the interval $[0,\frac{\pi}{2}]$, so that we are left with the computation of integrals of the form
\begin{equation}
    \mathcal{I}_{l,m,n}[\kappa] = \int_0^\frac{\pi}{2} \sin[\theta]^l J_n[\kappa \sin[\theta]] J_{m-n}[\kappa \sin[\theta]] \, \mathrm{d}\theta
\end{equation}
where again $\kappa$ is a real positive number, $l$, $m$ and $n$ being integers and $l \ge 0$. 

Let us consider first the case $m \ge 0$. We can then use the formula from Ref.\cite{luke1969special} page 216 to write
\begin{equation}
    \begin{aligned}
        J_n [\kappa \sin[\theta]] J_{m-n} [\kappa \sin[\theta]] = & \left( \frac{\kappa \sin[\theta]}{2} \right)^m \Gamma[1+m] \\
        & \times {}_2 F_3^R \left[ \frac{1+m}{2},\frac{2+m}{2};1+n,1+m-n,1+m; - \left( \kappa \sin[\theta] \right)^2 \right]
    \end{aligned}
\end{equation}
involving the regularized hypergeometric functions
\begin{equation}\label{hypergeometric_functions}
    {}_p F^R_q [\{a_i\}_{1\leq i \leq p};\{b_j\}_{1 \leq j \leq q};x] = \displaystyle \sum_{k=0}^\infty \frac{\prod_{i=1}^p (a_i)_k}{\prod_{j=1}^q \Gamma[b_j +k]} \frac{x^k}{k!}
\end{equation}
where $\{a_i\}_{1\leq i \leq p}$ and $\{b_j\}_{1 \leq j \leq q}$ are finite sets of rationals of respective sizes $p$ and $q$ verifying $p \le q$ (to have infinite convergence radius), $x$ is a real parameter and $(a)_k$ represents the Pochhammer symbol, defined as $(a)_k = \Gamma[a+k]/\Gamma[a]$ with $\Gamma$ being the Euler Gamma function. One can note that the order of the elements inside a set is irrelevant.

Hence we can write
\begin{equation}
    \begin{aligned}
        \mathcal{I}_{l,m,n}[\kappa] = & \left( \frac{\kappa}{2} \right)^m \Gamma[1+m] \\
        & \times \sum_{k=0}^\infty \frac{\left( \frac{1+m}{2} \right)_k \left( \frac{2+m}{2} \right)_k}{\Gamma[1+n+k] \Gamma[1+m-n+k] \Gamma[1+m+k]} \frac{(-\kappa^2)^k}{k!} \int_0^\frac{\pi}{2} \sin^{l+m+2k} [\theta] \, \mathrm{d}\theta 
    \end{aligned}
\end{equation}
which gives, for any $m \ge 0$,
\begin{equation}\label{integral_result_1}
    \begin{aligned}
        \mathcal{I}_{l,m,n}[\kappa] = \, & \frac{\sqrt{\pi}}{2} \left( \frac{\kappa}{2} \right)^m m! \, \Gamma\left[ \frac{1+l+m}{2}\right] \\
        & \times {}_3 F_4^R \left[ \frac{1+m}{2},\frac{2+m}{2},\frac{1+l+m}{2};1+n,1+m-n,1+m,\frac{2+l+m}{2}; -\kappa^2 \right]
    \end{aligned}
\end{equation}
where we used the famous Wallis formula to get to the last line (cf. Ref.\cite{guo2015wallis} for instance).

To obtain the value of the integral for $m<0$, we can simply remark:
\begin{equation}
    J_n[\kappa \sin[\theta]] J_{m-n} [\kappa \sin[\theta]] = (-1)^m J_{-n}[\kappa \sin[\theta]] J_{-m-(-n)} [\kappa \sin[\theta]]
\end{equation}
from the property of the Bessel functions $J_{-n}[x] = (-1)^n J_n[x]$.

As a consequence, for $m>0$, we also have
\begin{equation}\label{integral_result_2}
    \begin{aligned}
        \mathcal{I}_{l,-m,n}[\kappa] = \, & \frac{\sqrt{\pi}}{2} \left( -\frac{\kappa}{2} \right)^m m! \, \Gamma\left[ \frac{1+l+m}{2}\right] \\
        & \times {}_3 F_4^R \left[ \frac{1+m}{2},\frac{2+m}{2},\frac{1+l+m}{2};1-n,1+m+n,1+m,\frac{2+l+m}{2}; -\kappa^2 \right].
    \end{aligned}
\end{equation}

The two integral results ($\ref{integral_result_1}$) and ($\ref{integral_result_2}$) are useful for the vacuum field correlations. In fact, a requirement for $\mathcal{I}_{l,m,n}$ to be well defined is the condition $l+m \ge 0$, which is always naturally fulfilled in the correlations expressions. Moreover, it is useful to note the parity property of these results: for even values of $m$, the function $\mathcal{I}_{l,m,n}$ is an even function of the variable $\kappa$ and on the opposite, whenever $m$ is odd, it is an odd function of the variable $\kappa$.

\section{Calculation of the correlations between the electric field and its spatial derivatives for revolving points}\label{EdE}

For some models, it is also useful to derive the correlations between the electric field and its spatial derivatives, which are not trivially obtained from the electric-electric ones since no individual position remains in the final expressions of the latter, only the relative distance. In particular, one should come back to the original field expression Eq. (\ref{quantum_EB}) and then apply spatial derivatives:
\begin{equation}\label{quantum_partial_EB}
    \begin{array}{c}
        \begin{aligned}
        \partial_j \hat{\mathbf{E}}[\Hat{\mathbf{x}},t] = -\sum_{\mathbf{k},\lambda} \sqrt{\frac{\hbar \omega_k}{2 \epsilon_0 V}} k_j \left( \hat{a}_{\mathbf{k},\lambda} e^{i(\mathbf{k}\cdot \Hat{\mathbf{x}} - \omega_k t)} \mathbf{e}_{\mathbf{k},\lambda} + \hat{a}^{\dagger}_{\mathbf{k},\lambda} e^{-i(\mathbf{k}\cdot \Hat{\mathbf{x}} - \omega_k t)} \mathbf{e}_{\mathbf{k},\lambda}^* \right),
        \end{aligned}
    
        \vspace{0.2cm} \\

        \begin{aligned}
        \partial_j \hat{\mathbf{B}}[\Hat{\mathbf{x}},t] = - \sum_{\mathbf{k},\lambda} \sqrt{\frac{\hbar}{2 \epsilon_0 \omega_k V}} k_j \left( \hat{a}_{\mathbf{k},\lambda} e^{i(\mathbf{k}\cdot \Hat{\mathbf{x}} - \omega_k t)} (\mathbf{k} \times \mathbf{e}_{\mathbf{k},\lambda}) + \hat{a}^{\dagger}_{\mathbf{k},\lambda} e^{-i(\mathbf{k}\cdot \Hat{\mathbf{x}} - \omega_k t)} (\mathbf{k} \times \mathbf{e}_{\mathbf{k},\lambda}^*) \right)
        \end{aligned}
    \end{array}
\end{equation}
with $j \in \{X,Y,Z \}$. Afterwards, the spatial derivatives can be evaluated at points $A$ or $B$ precise location and the steps presented in the main body can be applied.

In the following, we will present the cross-correlations together with the self-correlations for components along specific axes.

\paragraph*{Correlations of the electric field and its spatial derivative along the $X$ axis:}

\begin{equation}\label{ExdxEx}
    \begin{array}{c}
         \begin{aligned}
         \textstyle{\langle \Tilde{E}^A_X [\omega] \partial_X \Tilde{E}_X^B [\omega'] \rangle = - \sum_{n} \left( N[\vert \omega_n \vert] + \frac{1}{2} \right)} \textstyle{\frac{\hbar \vert \omega_n \vert^4}{16 \epsilon_0 \pi^\frac{3}{2} c^4}} & \textstyle{\bigg( P_n^{\parallel +} {\left[ \frac{\vert \omega_n \vert r}{c} \right]} \delta [\omega+\omega'+\Omega]} \\
         & \textstyle{+ \, P_n^{\parallel -} {\left[ \frac{\vert \omega_n \vert r}{c} \right]} \delta [\omega+\omega'-\Omega] } \\
         & \textstyle{+ \, P_n^{3 +} {\left[ \frac{\vert \omega_n \vert r}{c} \right]} \delta [\omega+\omega'+3\Omega]} \\
         & \textstyle{+ \, P_n^{3-} {\left[ \frac{\vert \omega_n \vert r}{c} \right]} \delta [\omega+\omega'-3\Omega] \bigg) },
         \end{aligned}
    
         \vspace{0.2cm} \\

         \begin{aligned}
         \textstyle{\langle \Tilde{E}_X^A [\omega] \partial_X \Tilde{E}_X^A [\omega'] \rangle = \sum_{n} (-1)^n \left( N[\vert \omega_n \vert] + \frac{1}{2} \right) \frac{\hbar \vert \omega_n \vert^4}{16 \epsilon_0 \pi^\frac{3}{2} c^4}} & \textstyle{ \bigg( P_n^{\parallel +} {\left[ \frac{\vert \omega_n \vert r}{c} \right]} \delta[\omega+\omega'+\Omega]} \\
         & \textstyle{+ \, P_n^{\parallel -} {\left[ \frac{\vert \omega_n \vert r}{c} \right]} \delta[\omega+\omega'-\Omega]} \\
         & \textstyle{+ \, P_n^{3+} {\left[ \frac{\vert \omega_n \vert r}{c} \right] \delta [\omega+\omega'+3\Omega]}} \\
         & \textstyle{+ \, P_n^{3-} {\left[ \frac{\vert \omega_n \vert r}{c} \right]} \delta[\omega+\omega'-3\Omega] \bigg) }.
         \end{aligned}
    \end{array}
\end{equation}

\paragraph*{Correlations of the electric field and its spatial derivative along the $Y$ axis:}

\begin{equation}\label{EydyEy}
    \begin{array}{c}
        \begin{aligned}
        \textstyle{\langle \Tilde{E}_Y^A [\omega] \partial_Y \Tilde{E}_Y^B [\omega'] \rangle = i \sum_{n} \left( N[\vert \omega_n \vert] + \frac{1}{2} \right) \frac{\hbar \vert \omega_n \vert^4}{16 \epsilon_0 \pi^\frac{3}{2} c^4}} & \textstyle{ \bigg( P_n^{\parallel +} {\left[ \frac{\vert \omega_n \vert r}{c} \right]} \delta [\omega+\omega'+\Omega]} \\
        & \textstyle{- \, P_n^{\parallel -} {\left[ \frac{\vert \omega_n \vert r}{c} \right]} \delta [\omega+\omega'-\Omega] } \\
        & \textstyle{ -\, P_n^{3+} {\left[ \frac{\vert \omega_n \vert r}{c} \right]} \delta [\omega+\omega'+3\Omega]} \\
        & \textstyle{+ \, P_n^{3-} {\left[ \frac{\vert \omega_n \vert r}{c} \right]} \delta[\omega+\omega'-3\Omega] \bigg) },
        \end{aligned}
    
        \vspace{0.2cm} \\

        \begin{aligned}
        \textstyle{\langle \Tilde{E}_Y^A [\omega] \partial_Y \Tilde{E}_Y^A [\omega'] \rangle = -i \sum_{n} (-1)^n \left( N[\vert \omega_n \vert] + \frac{1}{2} \right)  \frac{\hbar \vert \omega_n \vert^4}{16 \epsilon_0 \pi^\frac{3}{2} c^4}} & \textstyle{ \bigg( P_n^{\parallel +} {\left[ \frac{\vert \omega_n \vert r}{c} \right]} \delta[\omega+\omega'+\Omega]} \\
        & \textstyle{- \, P_n^{\parallel -} {\left[ \frac{\vert \omega_n \vert r}{c} \right]} \delta[\omega+\omega'-\Omega]} \\
        & \textstyle{ - \, P_n^{3+} {\left[ \frac{\vert \omega_n \vert r}{c} \right]} \delta[\omega+\omega'+3\Omega]} \\
        & \textstyle{+ \, P_n^{3-} {\left[ \frac{\vert \omega_n \vert r}{c} \right]} \delta[\omega+\omega'-3\Omega] \bigg) }.
        \end{aligned}
    \end{array}
\end{equation}

\paragraph*{Cross-axes $XY$ correlations of the electric field with its spatial derivative along $X$:}

\begin{equation}\label{ExdxEy}
    \begin{array}{c}
        \begin{alignedat}{2}
            \textstyle{\langle \Tilde{E}_X^A [\omega] \partial_X \Tilde{E}_Y^B [\omega'] \rangle} & \textstyle{~= -i \sum_{n} \left( N[\vert \omega_n \vert] + \frac{1}{2} \right) \frac{\hbar \vert \omega_n \vert^4}{16 \epsilon_0 \pi^\frac{3}{2} c^4}} && \textstyle{\bigg( P_n^{\times +} {\left[ \frac{\vert \omega_n \vert r}{c} \right]} \delta [\omega+\omega'+\Omega]} \\
            & && \textstyle{- \, P_n^{\times -} {\left[ \frac{\vert \omega_n \vert r}{c} \right]} \delta[\omega+\omega'-\Omega]} \\
            & && \textstyle{ - P_n^{3+} {\left[ \frac{\vert \omega_n \vert r}{c} \right]} \delta[\omega+\omega'+3\Omega]} \\
            & && \textstyle{+ \, P_n^{3-} {\left[ \frac{\vert \omega_n \vert r}{c} \right]} \delta[\omega+\omega'-3\Omega] \bigg)} \\
            & ~\textstyle{= \langle \Tilde{E}^A_Y [\omega] \partial_X \Tilde{E}_X^B [\omega'] \rangle}, 
        \end{alignedat}
    
        \vspace{0.2cm} \\

        \begin{alignedat}{2}
            \textstyle{\langle \Tilde{E}_X^A [\omega] \partial_X \Tilde{E}_Y^A [\omega'] \rangle} & \textstyle{ ~= i \sum_{n} (-1)^n \left( N[\vert \omega_n \vert] + \frac{1}{2} \right) \frac{\hbar \vert \omega_n \vert^4}{16 \epsilon_0 \pi^\frac{3}{2} c^4}} && \textstyle{ \bigg( P_n^{\times +} {\left[ \frac{\vert \omega_n \vert r}{c} \right]} \delta[\omega+\omega'+\Omega]} \\ 
            & && \textstyle{- \, P_n^{\times -} {\left[ \frac{\vert \omega_n \vert r}{c} \right]} \delta[\omega+\omega'-\Omega]} \\
            & && \textstyle{- \, P_n^{3+} {\left[ \frac{\vert \omega_n \vert r}{c} \right]} \delta[\omega+\omega'+3\Omega]} \\
            & && \textstyle{+ \, P_n^{3-} {\left[ \frac{\vert \omega_n \vert r}{c} \right]} \delta[\omega+\omega'-3\Omega] \bigg)} \\
            & = \langle \Tilde{E}_Y^A [\omega] \partial_X \Tilde{E}_X^A [\omega'] \rangle.
        \end{alignedat}
    \end{array}
\end{equation}

\paragraph*{Cross-axes $XY$ correlations of the electric field with its spatial derivative along $Y$:}

\begin{equation}\label{EydyEx}
    \begin{array}{c}
        \begin{alignedat}{2}
            \textstyle{\langle \Tilde{E}_Y^A [\omega] \partial_Y \Tilde{E}_X^B [\omega'] \rangle} & \textstyle{ ~= \sum_{n} \left( N[\vert \omega_n \vert]+ \frac{1}{2} \right) \frac{\hbar \vert \omega_n \vert^4}{16 \epsilon_0 \pi^\frac{3}{2} c^4}} && \textstyle{\bigg( P_n^{\times +} {\left[ \frac{\vert \omega_n \vert r}{c} \right]} \delta[\omega+\omega'+\Omega]} \\
            & && \textstyle{+ \, P_n^{\times -} {\left[ \frac{\vert \omega_n \vert r}{c} \right]} \delta[\omega+\omega'-\Omega]} \\
            & && \textstyle{+ \, P_n^{3+} {\left[ \frac{\vert \omega_n \vert r}{c} \right]} \delta[\omega+\omega'+3\Omega]} \\
            & && \textstyle{+ \, P_n^{3-} {\left[ \frac{\vert \omega_n \vert r}{c} \right]} \delta[\omega+\omega'-3\Omega] \bigg)} \\
            & = \langle \Tilde{E}_X^A [\omega] \partial_Y \Tilde{E}_Y^B [\omega'] \rangle,
        \end{alignedat}
    
        \vspace{0.2cm} \\

        \begin{alignedat}{2}
            \textstyle{\langle \Tilde{E}_Y^A [\omega] \partial_Y \Tilde{E}_X^A [\omega'] \rangle} & \textstyle{~ = -\sum_{n} (-1)^n \left( N[\vert \omega_n \vert]+ \frac{1}{2} \right) \frac{\hbar \vert \omega_n \vert^4}{16 \epsilon_0 \pi^\frac{3}{2} c^4}} && \textstyle{ \bigg( P_n^{\times +} {\left[ \frac{\vert \omega_n \vert r}{c} \right]} \delta[\omega+\omega'+\Omega]} \\
            & && \textstyle{+ \, P_n^{\times -} {\left[ \frac{\vert \omega_n \vert r}{c} \right]} \delta[\omega+\omega'-\Omega]} \\
            & && \textstyle{+ \, P_n^{3+} {\left[ \frac{\vert \omega_n \vert r}{c} \right]} \delta[\omega+\omega'+3\Omega]} \\
            & && \textstyle{+ \, P_n^{3-} {\left[ \frac{\vert \omega_n \vert r}{c} \right]} \delta[\omega+\omega'-3\Omega] \bigg)} \\
            & = \langle \Tilde{E}_X^A [\omega] \partial_Y \Tilde{E}_Y^A [\omega'] \rangle.
        \end{alignedat}
    \end{array}
\end{equation}

\paragraph*{Cross axes $XZ$ correlations of the electric field with its spatial derivative along $Z$:}

\begin{equation}\label{EzdzEx}
    \begin{array}{c}
        \begin{alignedat}{2}
        \textstyle{\langle \Tilde{E}_Z^A [\omega] \partial_Z \Tilde{E}_X^B [\omega'] \rangle} & \textstyle{~ = \sum_{n} \left( N[\vert \omega_n \vert]+\frac{1}{2} \right) \frac{\hbar \vert \omega_n \vert^4}{16 \epsilon_0 \pi^\frac{3}{2} c^4}} && \textstyle{\bigg( P_n^{Z+} {\left[ \frac{\vert \omega_n \vert r}{c} \right]} \delta[\omega+\omega'+\Omega]} \\
        & && \textstyle{+ \, P_n^{Z-} {\left[ \frac{\vert \omega_n \vert r}{c} \right]} \delta[\omega+\omega'-\Omega] \bigg)} \\
        & = \textstyle{\langle \Tilde{E}_X^A [\omega] \partial_Z \Tilde{E}_Z^B [\omega'] \rangle},
        \end{alignedat}
    
        \vspace{0.2cm} \\

        \begin{alignedat}{2}
        \textstyle{\langle \Tilde{E}_Z^A [\omega] \partial_Z \Tilde{E}_X^A [\omega'] \rangle} & \textstyle{ ~ = -\sum_{n} (-1)^n \left( N[\vert \omega_n \vert]+\frac{1}{2} \right) \frac{\hbar \vert \omega_n \vert^4}{16 \epsilon_0 \pi^\frac{3}{2} c^4}} && \textstyle{\bigg( P_n^{Z+} {\left[ \frac{\vert \omega_n \vert r}{c} \right]} \delta[\omega+\omega'+\Omega]} \\
        & && \textstyle{+ \, P_n^{Z-} {\left[ \frac{\vert \omega_n \vert r}{c}\right]} \delta[\omega+\omega'-\Omega] \bigg) } \\
        & = \textstyle{\langle \Tilde{E}_X^A [\omega] \partial_Z \Tilde{E}_Z^A [\omega'] \rangle}.
        \end{alignedat}
    \end{array}
\end{equation}

\paragraph*{Cross axes $YZ$ correlations of the electric field with its spatial derivative along $Z$:}

\begin{equation}\label{EzdzEy}
    \begin{array}{c}
        \begin{alignedat}{2}
            \textstyle{\langle \Tilde{E}_Z^A [\omega] \partial_Z \Tilde{E}_Y^B [\omega'] \rangle} & \textstyle{ ~ = -i\sum_{n} \left( N[\vert \omega_n \vert] + \frac{1}{2} \right) \frac{\hbar \vert \omega_n \vert^4}{16 \epsilon_0 \pi^\frac{3}{2} c^4}} && \textstyle{ \bigg( P_n^{Z+} {\left[ \frac{\vert \omega_n \vert r}{c} \right]} \delta[\omega+\omega'+\Omega]} \\
            & && \textstyle{- \, P_n^{Z-} {\left[ \frac{\vert \omega_n \vert r}{c} \right]} \delta[\omega+\omega'-\Omega] \bigg)} \\
            & = \textstyle{\langle \Tilde{E}_Y^A [\omega] \partial_Z \Tilde{E}_Z^B [\omega'] \rangle},
        \end{alignedat}
    
        \vspace{0.2cm} \\

        \begin{alignedat}{2}
            \textstyle{\langle \Tilde{E}_Z^A [\omega] \partial_Z \Tilde{E}_Y^A [\omega']\rangle} & \textstyle{ ~ = i \sum_{n} (-1)^n \left( N[\vert \omega_n \vert]+\frac{1}{2} \right) \frac{\hbar \vert \omega_n \vert^4}{16 \epsilon_0 \pi^\frac{3}{2} c^4}} && \textstyle{ \bigg( P_n^{Z+} {\left[ \frac{\vert \omega_n \vert r}{c} \right]} \delta[\omega+\omega'+\Omega]} \\
            & && \textstyle{- \, P_n^{Z-} {\left[ \frac{\vert \omega_n \vert r}{c} \right]} \delta[\omega+\omega'-\Omega] \bigg) } \\
            & = \textstyle{\langle \Tilde{E}_Y^A [\omega] \partial_Z \Tilde{E}_Z^A [\omega']\rangle}.
        \end{alignedat}
    \end{array}
\end{equation}

We can also derive the vanishing correlations:
\begin{equation}
    \begin{array}{c}
        \displaystyle \langle \Tilde{E}_Z^A [\omega] \partial_Z \Tilde{E}_Z^B [\omega'] \rangle = \langle \Tilde{E}_Z^A [\omega] \partial_Z \Tilde{E}_Z^A [\omega'] \rangle = 0,
    
        \vspace{0.2cm} \\

        \displaystyle \langle \Tilde{E}_{\substack{X \\ Y}}^A [\omega] \partial_{\substack{X \\ Y}} \Tilde{E}_Z^B [\omega'] \rangle = \langle \Tilde{E}_Z^A [\omega] \partial_{\substack{X \\ Y}} \Tilde{E}_{\substack{X \\ Y}}^B [\omega'] \rangle = \langle \Tilde{E}_{\substack{X \\ Y}}^A [\omega] \partial_{\substack{X \\ Y}} \Tilde{E}_Z^A [\omega'] \rangle = \langle \Tilde{E}^A_Z [\omega] \partial_{\substack{X \\ Y}} \Tilde{E}^A_{\substack{X \\ Y}} [\omega'] \rangle = 0
    \end{array}
\end{equation}
where the sub-stack notations should be read line by line.

In the above correlations, we used six new correlation functions, particular to the correlations between the electric field and its spatial derivatives, noted with $P$ and defined by
\begin{equation}\label{Ppara}
    \begin{array}{c}
        \begin{aligned}
            P_n^{\parallel +}[x] = \frac{1}{2}\frac{x}{2} \bigg( {}_2 F_3^R \left[ 1,\frac{3}{2};\frac{5}{2},2-n,1+n;- \left(\frac{x}{2} \right)^2 \right] & - \frac{3}{4} \, {}_2 F_3^R \left[1,\frac{3}{2};\frac{7}{2},2-n,1+n;-\left(\frac{x}{2} \right)^2 \right] \\
            & - \frac{3}{4} \, {}_2 F_3^R \left[ \frac{3}{2},2;\frac{7}{2},2-n,1+n;-\left(\frac{x}{2} \right)^2 \right] \bigg),
        \end{aligned}
    
        \vspace{0.2cm} \\

        \begin{aligned}
            P_n^{\parallel -}[x] = \frac{1}{2}\frac{x}{2} \bigg( {}_2 F_3^R \left[ 1,\frac{3}{2};\frac{5}{2},1-n,2+n;- \left(\frac{x}{2} \right)^2 \right] & - \frac{3}{4} \, {}_2 F_3^R \left[1,\frac{3}{2};\frac{7}{2},1-n,2+n;-\left(\frac{x}{2} \right)^2 \right] \\
            & - \frac{3}{4} \, {}_2 F_3^R \left[ \frac{3}{2},2;\frac{7}{2},1-n,2+n;-\left(\frac{x}{2} \right)^2 \right] \bigg),
        \end{aligned}
    \end{array}
\end{equation}

\begin{equation}\label{Ptimes}
    \begin{array}{c}
        \begin{aligned}
            P_n^{\times +} [x] = \frac{1}{8}\frac{x}{2} \bigg( {}_2 F_3^R \left[ 1,\frac{3}{2};\frac{7}{2},2-n,1+n;-\left(\frac{x}{2}\right)^2 \right] + {}_2 F_3^R \left[ \frac{3}{2},2;\frac{7}{2},2-n,1+n;-\left( \frac{x}{2} \right)^2 \right] \bigg),
        \end{aligned}
    
        \vspace{0.2cm} \\

        \begin{aligned}
            P_n^{\times -} [x] = \frac{1}{8}\frac{x}{2} \bigg( {}_2 F_3^R \left[ 1,\frac{3}{2};\frac{7}{2},1-n,2+n;-\left(\frac{x}{2}\right)^2 \right] + {}_2 F_3^R \left[ \frac{3}{2},2;\frac{7}{2},1-n,2+n;-\left( \frac{x}{2} \right)^2 \right] \bigg),
        \end{aligned}
    \end{array}
\end{equation}

\begin{equation}
    \begin{array}{cc}
        \displaystyle P_n^{3+} [x] = \frac{3}{16} \left( \frac{x}{2} \right)^3 \, {}_2 F_3^R \left[ 2,\frac{5}{2};\frac{9}{2},4-n,1+n;-\left( \frac{x}{2}\right)^2 \right],

        \vspace{0.2cm} \\

        \displaystyle P_n^{3-} [x] = \frac{3}{16} \left( \frac{x}{2} \right)^3 \, {}_2 F_3^R \left[ 2,\frac{5}{2};\frac{9}{2},1-n,4+n;-\left( \frac{x}{2}\right)^2 \right],
    \end{array}
\end{equation}

\begin{equation}
    \begin{array}{cc}
        \displaystyle P_n^{Z+} [x] = \frac{1}{4}\frac{x}{2} \, {}_2 F_3^R \left[ 1,\frac{3}{2};\frac{7}{2},2-n,1+n;-\left( \frac{x}{2} \right)^2 \right],
    
        \vspace{0.2cm} \\

        \displaystyle P_n^{Z-} [x] = \frac{1}{4}\frac{x}{2} \, {}_2 F_3^R \left[ 1,\frac{3}{2};\frac{7}{2},1-n,2+n;-\left( \frac{x}{2} \right)^2 \right]
    \end{array}
\end{equation}
where we used again the hypergeometric functions notations (\ref{hypergeometric_functions}) defined in Appendix \ref{lemma}.

These correlations are clearly different from the ones used for the electric-electric quadratic averages, reinforcing the idea that it is not possible to obtain directly one from the other. Again, because of the rotation-induced coupling of the axes of the $XY$ plane, it is also interesting to compute correlations where the spatial derivative projection is different from the field ones.

\paragraph*{Correlations of the $X$ axis projections of the electric field with its spatial derivative along $Y$:}

\begin{equation}\label{ExdyEx}
    \begin{array}{c}
        \begin{aligned}
        \textstyle{\langle \Tilde{E}_X^A [\omega] \partial_Y \Tilde{E}_X^B [\omega'] \rangle = -i \sum_{n} \left( N[\vert \omega_n \vert] + \frac{1}{2} \right) \frac{\hbar \vert \omega_n \vert^4}{16 \epsilon_0 \pi^\frac{3}{2} c^4}} & \textstyle{ \bigg( P_n^{\nparallel +} {\left[ \frac{\vert \omega_n \vert r}{c} \right]} \delta[\omega+\omega'+\Omega]} \\
        & \textstyle{- \, P_n^{\nparallel -} {\left[ \frac{\vert \omega_n \vert r}{c} \right]} \delta[\omega+\omega'-\Omega]} \\
        & \textstyle{- \, P_n^{3+} {\left[ \frac{\vert \omega_n \vert r}{c} \right]} \delta[\omega+\omega'+3\Omega]} \\
        & \textstyle{+ \, P_n^{3-} {\left[ \frac{\vert \omega_n \vert r}{c} \right]} \delta[\omega+\omega'-3\Omega] \bigg) },
        \end{aligned}
        
        \vspace{0.2cm} \\

        \begin{aligned}
        \textstyle{\langle \Tilde{E}_X^A [\omega] \partial_Y \Tilde{E}_X^A [\omega'] \rangle = i \sum_{n} (-1)^n \left( N[\vert \omega_n \vert] + \frac{1}{2} \right) \frac{\hbar \vert \omega_n \vert^4}{16 \epsilon_0 \pi^\frac{3}{2} c^4}} & \textstyle{ \bigg( P_n^{\nparallel +} {\left[ \frac{\vert \omega_n \vert r}{c} \right]} \delta[\omega+\omega'+\Omega]} \\
        & \textstyle{- \, P_n^{\nparallel -} {\left[ \frac{\vert \omega_n \vert r}{c} \right]} \delta[\omega+\omega'-\Omega]} \\
        & \textstyle{- \, P_n^{3+} {\left[ \frac{\vert \omega_n \vert r}{c} \right]} \delta[\omega+\omega'+3\Omega]} \\
        & \textstyle{+ \, P_n^{3-} {\left[ \frac{\vert \omega_n \vert r}{c} \right]} \delta[\omega+\omega'-3\Omega] \bigg)}.
        \end{aligned}
    \end{array}
\end{equation}

\paragraph*{Correlations of the $Y$ axis projections of the electric field with its spatial derivative along $X$:}

\begin{equation}\label{EydxEy}
    \begin{array}{c}
        \begin{aligned}
            \textstyle{\langle \Tilde{E}_Y^A [\omega] \partial_X \Tilde{E}_Y^B [\omega'] \rangle = \sum_{n} \left( N[\vert \omega_n \vert] + \frac{1}{2} \right) \frac{\hbar \vert \omega_n \vert^4}{16 \epsilon_0 \pi^\frac{3}{2} c^4}} & \textstyle{ \bigg( P_n^{\nparallel +} {\left[ \frac{\vert \omega_n \vert r}{c} \right]} \delta[\omega+\omega'+\Omega]} \\
            & \textstyle{+ \, P_n^{\nparallel -} {\left[ \frac{\vert \omega_n \vert r}{c} \right]} \delta[\omega+\omega'-\Omega]} \\
            & \textstyle{+ \, P_n^{3+} {\left[ \frac{\vert \omega_n \vert r}{c} \right]} \delta[\omega+\omega'+3\Omega]} \\
            & \textstyle{+ \, P_n^{3-} {\left[ \frac{\vert \omega_n \vert r}{c} \right]} \delta[\omega+\omega'-3\Omega] \bigg)}, 
        \end{aligned}
    
        \vspace{0.2cm} \\

        \begin{aligned}
            \textstyle{\langle \Tilde{E}_Y^A [\omega] \partial_X \Tilde{E}_Y^A [\omega'] \rangle = -\sum_{n} (-1)^n \left( N[\vert \omega_n \vert] + \frac{1}{2} \right) \frac{\hbar \vert \omega_n \vert^4}{16 \epsilon_0 \pi^\frac{3}{2} c^4}} & \textstyle{ \bigg( P_n^{\nparallel +} {\left[ \frac{\vert \omega_n \vert r}{c} \right]} \delta[\omega+\omega'+\Omega]} \\
            & \textstyle{+ \, P_n^{\nparallel -} {\left[ \frac{\vert \omega_n \vert r}{c} \right]} \delta[\omega+\omega'-\Omega]} \\
            & \textstyle{+ \, P_n^{3+} {\left[ \frac{\vert \omega_n \vert r}{c} \right]} \delta[\omega+\omega'+3\Omega]} \\
            & \textstyle{+ \, P_n^{3-} {\left[ \frac{\vert \omega_n \vert r}{c} \right]} \delta[\omega+\omega'-3\Omega] \bigg) }.
        \end{aligned}
    \end{array}
\end{equation}

\paragraph*{Correlations of the $Z$ axis projections of the electric field with its spatial derivative along $X$:}

\begin{equation}\label{EzdxEz}
    \begin{array}{c}
        \begin{aligned}
            \textstyle{\langle \Tilde{E}_Z^A [\omega] \partial_X \Tilde{E}_Z^B [\omega'] \rangle = - \sum_{n} \left( N[\vert \omega_n \vert] + \frac{1}{2} \right) \frac{\hbar \vert \omega_n \vert^4}{16 \epsilon_0 \pi^\frac{3}{2} c^4}} & \textstyle{\bigg( 4 P_n^{\times+} {\left[ \frac{\vert \omega_n \vert r}{c} \right]} \delta[\omega+\omega'+\Omega]} \\
            & \textstyle{+ \, 4 P_n^{\times -} {\left[ \frac{\vert \omega_n \vert r}{c} \right]} \delta[\omega+\omega'-\Omega] \bigg)},
        \end{aligned}
    
        \vspace{0.2cm} \\

        \begin{aligned}
            \textstyle{\langle \Tilde{E}_Z^A [\omega] \partial_X \Tilde{E}_Z^A [\omega'] \rangle = \sum_{n} (-1)^n \left( N[\vert \omega_n \vert] + \frac{1}{2} \right) \frac{\hbar \vert \omega_n^4 \vert}{16 \epsilon_0 \pi^\frac{3}{2} c^4}} & \textstyle{\bigg( 4 P_n^{\times+} {\left[ \frac{\vert \omega_n \vert r}{c} \right]} \delta[\omega+\omega'+\Omega]} \\
            & \textstyle{+ \, 4 P_n^{\times -} {\left[ \frac{\vert \omega_n \vert r}{c} \right]} \delta[\omega+\omega'-\Omega] \bigg)}.
        \end{aligned}
    \end{array}
\end{equation}

\paragraph*{Correlations of the $Z$ axis projections of the electric field with its spatial derivative along $Y$:}

\begin{equation}\label{EzdyEz}
    \begin{array}{c}
        \begin{aligned}
            \textstyle{\langle \Tilde{E}_Z^A [\omega] \partial_Y \Tilde{E}_Z^B [\omega'] \rangle = i\sum_{n} \left( N[\vert \omega_n \vert] + \frac{1}{2} \right) \frac{\hbar \vert \omega_n \vert^4}{16 \epsilon_0 \pi^\frac{3}{2} c^4}} & \textstyle{ \bigg( 4 P_n^{\times+} {\left[ \frac{\vert \omega_n \vert r}{c} \right]} \delta[\omega+\omega'+\Omega]} \\
            & \textstyle{- \, 4 P_n^{\times -} {\left[ \frac{\vert \omega_n \vert r}{c} \right]} \delta[\omega+\omega'-\Omega] \bigg)},
        \end{aligned}
    
        \vspace{0.2cm} \\

        \begin{aligned}
            \textstyle{\langle \Tilde{E}_Z^A [\omega] \partial_Y \Tilde{E}_Z^A [\omega'] \rangle = -i\sum_{n} (-1)^n \left( N[\vert \omega_n \vert] + \frac{1}{2} \right) \frac{\hbar \vert \omega_n \vert^4}{16 \epsilon_0 \pi^\frac{3}{2} c^4}} & \textstyle{ \bigg( 4 P_n^{\times+} {\left[ \frac{\vert \omega_n \vert r}{c} \right]} \delta[\omega+\omega'+\Omega]} \\
            & \textstyle{- \, 4 P_n^{\times -} {\left[ \frac{\vert \omega_n \vert r}{c} \right]} \delta[\omega+\omega'-\Omega] \bigg)}.
        \end{aligned}
    \end{array}
\end{equation}

As previously, a bunch of correlations are vanishing:
\begin{equation}
    \begin{array}{c}
        \langle \Tilde{E}^A_{\substack{X \\ Y}} [\omega] \partial_{\substack{Y \\ X}} \Tilde{E}^B_Z [\omega'] \rangle =  \langle \Tilde{E}_Z^A [\omega] \partial_{\substack{X \\ Y}} \Tilde{E}_{\substack{Y \\ X}}^B [\omega'] \rangle = \langle \Tilde{E}^A_{\substack{X \\ Y}} [\omega] \partial_{\substack{Y \\ X}} \Tilde{E}^A_Z [\omega'] \rangle = \langle \Tilde{E}_Z^A [\omega] \partial_{\substack{X \\ Y}} \Tilde{E}_{\substack{Y \\ X}}^A [\omega'] \rangle = 0,
    
        \vspace{0.2cm} \\

        \langle \Tilde{E}_{\substack{X \\ Y}}^A [\omega] \partial_Z \Tilde{E}_{\substack{X \\ Y}}^B [\omega'] \rangle = \langle \Tilde{E}_{\substack{X \\ Y}}^A [\omega] \partial_Z \Tilde{E}_{\substack{Y \\ X}}^B [\omega'] \rangle = \langle \Tilde{E}_{\substack{X \\ Y}}^A [\omega] \partial_Z \Tilde{E}_{\substack{X \\ Y}}^A [\omega'] \rangle = \langle \Tilde{E}_{\substack{X \\ Y}}^A [\omega] \partial_Z \Tilde{E}_{\substack{Y \\ X}}^A [\omega'] \rangle = 0. 
    \end{array}
\end{equation}

The above correlations involved the pair of correlation functions defined by
\begin{equation}\label{Pnpara}
    \begin{array}{c}
        \begin{aligned}
            P_n^{\nparallel +}[x] = - \frac{1}{2} \frac{x}{2} \bigg( {}_2 F_3^R \left[ 1,\frac{3}{2};\frac{5}{2},2-n,1+n;-\left(\frac{x}{2}\right)^2 \right] & - \frac{1}{4} \, {}_2 F_3^R \left[ 1,\frac{3}{2};\frac{7}{2},2-n,1+n;-\left(\frac{x}{2}\right)^2 \right] \\
            & - \frac{1}{4} \, {}_2 F_3^R \left[ \frac{3}{2},2;\frac{7}{2},2-n,1+n;-\left(\frac{x}{2}\right)^2 \right] \bigg),
        \end{aligned}
    
        \vspace{0.2cm} \\

        \begin{aligned}
            P_n^{\nparallel -}[x] = - \frac{1}{2}\frac{x}{2} \bigg( {}_2 F_3^R \left[ 1,\frac{3}{2};\frac{5}{2},1-n,2+n;-\left(\frac{x}{2}\right)^2 \right] & - \frac{1}{4} \, {}_2 F_3^R \left[ 1,\frac{3}{2};\frac{7}{2},1-n,2+n;-\left(\frac{x}{2}\right)^2 \right] \\
            & - \frac{1}{4} \, {}_2 F_3^R \left[ \frac{3}{2},2;\frac{7}{2},1-n,2+n;-\left(\frac{x}{2}\right)^2 \right] \bigg).
        \end{aligned}
    \end{array}
\end{equation} 

Just as the correlation functions between the electric and magnetic fields, these $P$ correlations functions are linear combinations of the $\mathcal{I}_{l,m,n}$ integrals where $m$ takes odd values. As a consequence, the $P$ correlation functions are odd with respect to their argument. In particular, the transformation $r \rightarrow -r$ will give rise to a minus sign in the correlations expressions.

\section{Calculation of the correlations between the magnetic field and the spatial derivatives of the electric field for revolving points}\label{EdB}

We are left with the correlations between the electric field and the spatial derivatives of the magnetic field or vice versa. In this appendix, we will specifically present the expressions of the correlations between the magnetic field and the spatial derivatives of the electric field. The ones between the electric field with the spatial derivatives of the magnetic field are closely related, as it is discussed at the end of the appendix.

\paragraph*{Cross-axes $XZ$ (resp. $YZ$) correlations of the magnetic field and the spatial derivative along $X$ (resp. $Y$) of the electric field:}

\begin{equation}\label{BzdxEx}
    \begin{array}{c}
        \begin{aligned}
        \langle \Tilde{B}_Z^A [\omega] \partial_X \Tilde{E}_X^B [\omega'] \rangle & \textstyle{~ = - \sum_{n} \left( N[\vert \omega_n \vert]+\frac{1}{2} \right) \frac{\hbar \omega_n \vert \omega_n \vert^3}{16 \epsilon_0 \pi^\frac{3}{2} c^5} \bigg( G_n^+ {\left[ \frac{\vert \omega_n \vert r}{c} \right]} \delta[\omega+\omega'+2\Omega]} \\
        & ~~~~~~~~~~~~~~~~~~~~~~~~~~~~~~~~~~~~~~~~~~~~~~~~~~ \textstyle{- \, G_n^- {\left[ \frac{\vert \omega_n \vert r}{c} \right]} \delta[\omega+\omega'-2\Omega] \bigg)} \\
        & = - \langle \Tilde{B}_X^A [\omega] \partial_X \Tilde{E}_Z^B [\omega'] \rangle = -\langle \Tilde{B}_Z^A [\omega] \partial_Y \Tilde{E}_Y^B [\omega'] \rangle = \langle \Tilde{B}_Y^A [\omega] \partial_Y \Tilde{E}_Z^B [\omega'] \rangle, 
        \end{aligned}
    
        \vspace{0.2cm} \\

        \begin{aligned}
        \langle \Tilde{B}_Z^A [\omega] \partial_X \Tilde{E}_X^A [\omega'] \rangle & \textstyle{ ~ = - \sum_{n} (-1)^n \left( N[\vert \omega_n \vert]+\frac{1}{2} \right) \frac{\hbar \omega_n \vert \omega_n \vert^3}{16 \epsilon_0 \pi^\frac{3}{2} c^5} \bigg( G_n^+ {\left[ \frac{\vert \omega_n \vert r}{c} \right]} \delta[\omega+\omega'+2\Omega]} \\
        & ~~~~~~~~~~~~~~~~~~~~~~~~~~~~~~~~~~~~~~~~~~~~~~~~~~~~~~~~~~~~ \textstyle{- \, G_n^- {\left[ \frac{\vert \omega_n \vert r}{c} \right]} \delta[\omega+\omega'-2\Omega] \bigg)} \\
        & = -\langle \Tilde{B}_X^A [\omega] \partial_X \Tilde{E}_Z^A [\omega'] \rangle = -\langle \Tilde{B}_Z^A [\omega] \partial_Y \Tilde{E}_Y^A [\omega'] \rangle = -\langle \Tilde{B}_Y^A [\omega] \partial_Y \Tilde{E}_Z^A [\omega'] \rangle. 
    \end{aligned}
    \end{array}
\end{equation}

Interestingly, the correlations between the electric field and the spatial derivative of the magnetic field (or vice versa) involve the same correlation functions as the electric-electric ones. However, the parity properties of the overall correlations are not the same due to the prefactor $\omega_n$. Most of the remaining correlations can also be expressed using the electric-electric correlation functions.

\paragraph*{Cross-axes $YZ$ correlations of the magnetic field with the spatial derivative along $X$ of the electric field:}

\begin{equation}\label{BzdxEy}
    \begin{array}{c}
        \begin{alignedat}{2}
        \langle \Tilde{B}_Z^A [\omega] \partial_X \Tilde{E}_Y^B [\omega'] \rangle & \textstyle{ ~ = -i \sum_{n} \left( N[\vert \omega_n \vert] + \frac{1}{2} \right) \frac{\hbar \omega_n \vert \omega_n \vert^3}{16 \epsilon_0 \pi^\frac{3}{2} c^5}} && \textstyle{\bigg( \frac{1}{2} G_n^Z {\left[ \frac{\vert \omega_n \vert r}{c} \right]} \delta[\omega+\omega']} \\
        & && \textstyle{- \, G_n^+ {\left[ \frac{\vert \omega_n \vert r}{c} \right]} \delta[\omega+\omega'+2\Omega]} \\
        & && \textstyle{- \, G_n^- {\left[ \frac{\vert \omega_n \vert r}{c} \right]} \delta[\omega+\omega'-2\Omega] \bigg)} \\
        & = - \langle \Tilde{B}_Y^A [\omega] \partial_X \Tilde{E}_Z^B [\omega'] \rangle, 
        \end{alignedat}
    
        \vspace{0.2cm} \\

        \begin{alignedat}{2}
            \langle \Tilde{B}_Z^A [\omega] \partial_X \Tilde{E}_Y^A [\omega'] \rangle & \textstyle{= -i \sum_{n} (-1)^n \left( N[\vert \omega_n \vert] + \frac{1}{2} \right) \frac{\hbar \omega_n \vert \omega_n \vert^3}{16 \epsilon_0 \pi^\frac{3}{2} c^5}} && \textstyle{ \bigg( \frac{1}{2} G_n^Z {\left[ \frac{\vert \omega_n \vert r}{c} \right]} \delta[\omega+\omega']} \\
            & && \textstyle{- \, G_n^+ {\left[ \frac{\vert \omega_n \vert r}{c} \right]} \delta[\omega+\omega'+2\Omega]} \\
            & && \textstyle{- \, G_n^- {\left[ \frac{\vert \omega_n \vert r}{c} \right]} \delta[\omega+\omega'-2\Omega] \bigg)} \\
            & = -\langle \Tilde{B}_Y^A [\omega] \partial_X \Tilde{E}_Z^A [\omega'] \rangle. 
        \end{alignedat}
    \end{array}
\end{equation}

\paragraph*{Cross-axes $XZ$ correlations of the magnetic field with the spatial derivative along $Y$ of the electric field:}

\begin{equation}\label{BzdyEx}
    \begin{array}{c}
        \begin{alignedat}{2}
            \langle \Tilde{B}_Z^A [\omega] \partial_Y \Tilde{E}_X^B [\omega'] \rangle & \textstyle{ ~ = i \sum_{n \in \mathbb{Z}} \left( N[\vert \omega_n \vert] + \frac{1}{2} \right) \frac{\hbar \omega_n \vert \omega_n \vert^3}{16 \epsilon_0 \pi^\frac{3}{2} c^5}} && \textstyle{ \bigg( \frac{1}{2} G_n^Z {\left[ \frac{\vert \omega_n \vert r}{c} \right]} \delta[\omega+\omega']} \\
            & && \textstyle{ + \, G_n^+ {\left[ \frac{\vert \omega_n \vert r}{c} \right]} \delta[\omega+\omega'+2\Omega]} \\
            & && \textstyle{+ \, G_n^- {\left[ \frac{\vert \omega_n \vert r}{c} \right]} \delta[\omega+\omega'-2\Omega] \bigg)} \\
            & = -\langle \Tilde{B}_X^A [\omega] \partial_Y \Tilde{E}_Z^B [\omega'] \rangle, 
        \end{alignedat}
        
        \vspace{0.2cm} \\
    
        \begin{alignedat}{2}
            \langle \Tilde{B}_Z^A [\omega] \partial_Y \Tilde{E}_X^A [\omega'] \rangle & \textstyle{ ~ = i \sum_{n \in \mathbb{Z}} (-1)^n \left( N[\vert \omega_n \vert] + \frac{1}{2} \right) \frac{\hbar \omega_n \vert \omega_n \vert^3}{16 \epsilon_0 \pi^\frac{3}{2} c^5}} && \textstyle{ \bigg( \frac{1}{2} G_n^Z {\left[ \frac{\vert \omega_n \vert r}{c} \right]} \delta[\omega+\omega']} \\
            & && \textstyle{+ ~ G_n^+ {\left[ \frac{\vert \omega_n \vert r}{c} \right]} \delta[\omega+\omega'+2\Omega]} \\
            & && \textstyle{+ \, G_n^- {\left[ \frac{\vert \omega_n \vert r}{c} \right]} \delta[\omega+\omega'-2\Omega] \bigg)} \\
            & = -\langle \Tilde{B}_X^A [\omega] \partial_Y \Tilde{E}_Z^A [\omega'] \rangle. 
        \end{alignedat}
    \end{array}
\end{equation}

\paragraph*{Cross-axes $XY$ correlations of the magnetic field with the spatial derivative along $Z$ of the electric field:}

\begin{equation}\label{BxdzEy}
    \begin{array}{c}
        \begin{aligned}
        \langle \Tilde{B}_X^A [\omega] \partial_Z \Tilde{E}_Y^B [\omega'] \rangle & \textstyle{ ~ = i \sum_{n \in \mathbb{Z}} \left( N[\vert \omega_n \vert] + \frac{1}{2} \right) \frac{\hbar \omega_n \vert \omega_n \vert^3}{16 \epsilon_0 \pi^\frac{3}{2} c^5} Q_n^Z \left[ \frac{\vert \omega_n \vert r}{c} \right] \delta[\omega+\omega']} \\
        & = - \langle \Tilde{B}_Y^A [\omega] \partial_Z \Tilde{E}_X^B [\omega'] \rangle, 
        \end{aligned}
    
        \vspace{0.2cm} \\

        \begin{aligned}
        \langle \Tilde{B}_X^A [\omega] \partial_Z \Tilde{E}_Y^A [\omega'] \rangle & \textstyle{ ~ = i \sum_{n \in \mathbb{Z}} (-1)^n \left( N[\vert \omega_n \vert] + \frac{1}{2} \right) \frac{\hbar \omega_n \vert \omega_n \vert^3}{16 \epsilon_0 \pi^\frac{3}{2} c^5} Q_n^Z \left[ \frac{\vert \omega_n \vert r}{c} \right] \delta[\omega+\omega']} \\
        & = - \langle \Tilde{B}_Y^A [\omega] \partial_Z \Tilde{E}_X^A [\omega'] \rangle 
        \end{aligned}
    \end{array}
\end{equation}
where we used the new $Q$ correlation function defined by
\begin{equation}
    Q_n^Z [x] = {}_2 F_3^R \left[\frac{1}{2},1;\frac{5}{2},1-n,1+n;-\left( \frac{x}{2} \right)^2 \right].
\end{equation}

Most of the correlations between the magnetic field and the spatial derivatives of the electric field are actually zero:
\begin{equation}
    \begin{array}{c}
        \begin{aligned}
            & \langle \Tilde{B}_{\substack{X \\ Y \\ Z}}^A [\omega] \partial_{\substack{X \\ Y \\ Z}} \Tilde{E}_{\substack{X \\ Y \\ Z}}^B [\omega'] \rangle = \langle \Tilde{B}_{\substack{X \\ Y \\ Z}}^A [\omega] \partial_{\substack{Y \\ Z \\ X}} \Tilde{E}_{\substack{X \\ Y \\ Z}}^B [\omega'] \rangle = \langle \Tilde{B}^A_{\substack{X \\ Y \\ Z}} [\omega] \partial_{\substack{Z \\ X \\ Y}} \Tilde{E}_{\substack{X \\ Y \\ Z}}^B [\omega'] \rangle \\
            & = \langle \Tilde{B}_{\substack{X \\ Y \\ Z}}^A [\omega] \partial_{\substack{X \\ Y \\ Z}} \Tilde{E}_{\substack{X \\ Y \\ Z}}^A [\omega'] \rangle = \langle \Tilde{B}_{\substack{X \\ Y \\ Z}}^A [\omega] \partial_{\substack{Y \\ Z \\ X}} \Tilde{E}_{\substack{X \\ Y \\ Z}}^A [\omega'] \rangle = \langle \Tilde{B}^A_{\substack{X \\ Y \\ Z}} [\omega] \partial_{\substack{Z \\ X \\ Y}} \Tilde{E}_{\substack{X \\ Y \\ Z}}^A [\omega'] \rangle = 0,
        \end{aligned}

        \vspace{0.2cm} \\

        \displaystyle \langle \Tilde{B}^A_{\substack{X \\ Y}} [\omega] \partial_{\substack{X \\ Y}} \Tilde{E}_{\substack{Y \\ X}}^B [\omega'] \rangle = \langle \Tilde{B}^A_{\substack{X \\ Y}} [\omega] \partial_{\substack{Y \\ X}} \Tilde{E}_{\substack{Y \\ X}}^B [\omega'] \rangle = \langle \Tilde{B}^A_{\substack{X \\ Y}} [\omega] \partial_{\substack{X \\ Y}} \Tilde{E}_{\substack{Y \\ X}}^A [\omega'] \rangle = \langle \Tilde{B}^A_{\substack{X \\ Y}} [\omega] \partial_{\substack{Y \\ X}} \Tilde{E}_{\substack{Y \\ X}}^A [\omega'] \rangle = 0,

        \vspace{0.2cm} \\

        \displaystyle \langle \Tilde{B}_{\substack{X \\ Z}}^A [\omega] \partial_{\substack{Z \\ Z}} \Tilde{E}_{\substack{Z \\ X}}^B [\omega'] \rangle = \langle \Tilde{B}_{\substack{Y \\ Z}}^A [\omega] \partial_{\substack{Z \\ Z}} \Tilde{E}_{\substack{Z \\ Y}}^B [\omega'] \rangle = \langle \Tilde{B}_{\substack{X \\ Z}}^A [\omega] \partial_{\substack{Z \\ Z}} \Tilde{E}_{\substack{Z \\ X}}^A [\omega'] \rangle = \langle \Tilde{B}_{\substack{Y \\ Z}}^A [\omega] \partial_{\substack{Z \\ Z}} \Tilde{E}_{\substack{Z \\ Y}}^A [\omega'] \rangle = 0

    \end{array}
\end{equation}

If one is now interested in the correlations of the electric field with the spatial derivatives of the magnetic field, it can be easily obtained from the results of this section. Indeed, thanks to the relation $\sum_\lambda (\mathbf{e}_{\mathbf{k},\lambda})_i (\mathbf{k} \times \mathbf{e}_{\mathbf{k},\lambda})_j = - \sum_\lambda (\mathbf{k} \times \mathbf{e}_{\mathbf{k},\lambda})_i (\mathbf{e}_{\mathbf{k},\lambda})_j$, the exchange of the roles of $E$ and $B$ is reduced to an extra minus sign.
Furthermore, if one is looking for the correlations between the magnetic field and its own spatial derivatives, it can easily be obtained from the correlations between the electric field and its spatial derivatives (cf. Appendix \ref{EdE}). Moreover, as already discussed, the $G$ correlation functions are even with respect to their argument and this property is also verified by the above $Q$ correlation function.

We present in Appendix \ref{prop_corr_func} some properties and symmetries of the correlations functions introduced in the previous sections. These properties are of great use when studying the symmetries of the field correlations, such as the exchange of the argument frequencies $\omega$ and $\omega'$. From their expressions (\ref{Ppara}), (\ref{Ptimes}) and (\ref{Pnpara}), we can remark the important relations linking the $P$ correlation functions:
\begin{equation}\label{relation-P}
    P_n^{\parallel \pm} [x] + P_n^{\nparallel \pm} [x] + 2 P_n^{\times \pm} [x] = 0.
\end{equation}

Actually, thanks to these identities, only the differences $P_n^{\parallel \pm} - P_n^{\nparallel \pm}$ remains. To take this into account, we set new correlation functions only to simplify our expressions:
\begin{equation}
    P_n^{\div \pm} = \frac{P_n^{\parallel \pm} - P_n^{\nparallel \pm}}{4}.
\end{equation}

In terms of the hypergeometric functions, this yields
\begin{equation}
    \begin{array}{c}
        \begin{aligned}
            P_n^{\div +} = \frac{1}{4}\frac{x}{2} \bigg( {}_2 F_3^R \left[ 1,\frac{3}{2};\frac{5}{2},2-n,1+n;-\left(\frac{x}{2}\right)^2 \right] & - \frac{1}{2} \, {}_2 F_3^R \left[ 1,\frac{3}{2};\frac{7}{2},2-n,1+n;-\left( \frac{x}{2} \right)^2 \right] \\
            & - \frac{1}{2} \, {}_2 F_3^R \left[ \frac{3}{2},2;\frac{7}{2},2-n,1+n;-\left( \frac{x}{2}\right)^2 \right] \bigg),
        \end{aligned}
    
        \vspace{0.2cm} \\

        \begin{aligned}
            P_n^{\div -} = \frac{1}{4}\frac{x}{2} \bigg( {}_2 F_3^R \left[ 1,\frac{3}{2};\frac{5}{2},1-n,2+n;-\left(\frac{x}{2} \right)^2 \right] & - \frac{1}{2} \, {}_2 F_3^R \left[ 1,\frac{3}{2};\frac{7}{2},1-n,2+n;-\left( \frac{x}{2} \right)^2 \right] \\
            & -\frac{1}{2} \, {}_2 F_3^R \left[ \frac{3}{2},2;\frac{7}{2},1-n,2+n;-\left( \frac{x}{2} \right)^2 \right] \bigg)
        \end{aligned}
    \end{array}
\end{equation}
which allows us to remark that these new correlations functions also verify the same properties as the $P_n^{\times \pm}$ (see Appendix \ref{prop_corr_func}).
Hence, instead of using $P_n^\parallel$, $P_n^\nparallel$ and $P_n^\times$, we can use only $P_n^\times$ and $P_n^\div$. We recap the useful correlations functions and their expressions in Appendix \ref{prop_corr_func}.

\section{Index symmetries of the correlation functions}\label{prop_corr_func}

In this section, we will present some interesting properties of the correlations function $G$, $H$, $P$ and $Q$ introduced in the sections \ref{EE}, \ref{EB} and Appendices \ref{EdE} and \ref{EdB}.

Alongside the already mentioned parity properties of the correlation functions with respect to their argument, the latter have some exchange properties with respect to their integer index:
\begin{equation}\label{G_n+-_sym}
    \begin{array}{cc}
        \displaystyle G_n^\Box [x] = G_{-n}^\Box [x] & \Box \in \{0,Z\}, 
        \vspace{0.2cm}\\
        \multicolumn{2}{c}{\displaystyle G_n^+ [x] = G_{-n}^- [x],} 
        \vspace{0.2cm}\\
        \multicolumn{2}{c}{ Q_n^Z [x] = Q_{-n}^Z [x],}
        
        \vspace{0.2cm} 
        \\
        
        G_n^\pm [x] = G_{\pm 2 - n}^\pm [x], & G_{n+2}^+ [x] = G_n^- [x],
    \end{array}
\end{equation}

\begin{equation}\label{H_n+-_sym}
    \begin{array}{cc}
        \multicolumn{2}{c}{\displaystyle H^+_n [x] = H^-_{-n} [x],}
        
        \vspace{0.2cm} 
        \\
        
        \displaystyle H^\pm_n [x] = H^\pm_{\pm 1 -n} [x], & \displaystyle H_{n+1}^+ [x] = H^-_n [x],
    \end{array}
\end{equation}

\begin{equation}
    \begin{array}{cccc}
        \multicolumn{2}{c}{P_n^{\Box +}[x] = P_{-n}^{\Box -} [x]} & \multicolumn{2}{c}{\Box \in \{ \parallel, \nparallel, \times, \div, Z,3 \},}

        \vspace{0.2cm} 
        \\

        P_n^{\Box \pm} [x] = P_{\pm 1 -n}^{\Box \pm} [x], & P_{n+1}^{\Box +} [x] = P_n^{\Box -} [x] & \multicolumn{2}{c}{\Box \in \{ \parallel, \nparallel, \times, \div, Z \},} 

        \vspace{0.2cm} 
        \\

        \multicolumn{2}{c}{P_n^{3\pm} [x] = P_{\pm 3-n}^{3 \pm} [x],} & \multicolumn{2}{c}{P_{n+3}^{3 +}[x] = P_n^{3-}[x].}
    \end{array}
\end{equation}

We recap the useful correlation functions together with their expressions inside table \ref{tab:corr_func}.

\begin{table}[ht]
    \centering
    \begin{tabular}{|c|c|c|}
    \hline
    Correlation function & Expression & Associated frequency shift 
    \\
    \hline
    \multirow{2}{*}{$G_n^0$} & $2 \, {}_2 F^R_3 \left[\frac{1}{2},1;\frac{3}{2},1-n,1+n;-\left(\frac{x}{2}\right)^2\right]$ & \multirow{2}{*}{$0$}
    \\
    & $- ~ {}_2 F^R_3 \left[\frac{1}{2},2;\frac{5}{2},1-n,1+n;-\left(\frac{x}{2}\right)^2\right]$ & 
    \\
    \hline
    $G_n^{\pm}$ & $\frac{1}{4}\left(\frac{x}{2}\right)^2 \, {}_2 F^R_3 \left[\frac{3}{2},2;\frac{7}{2},3 \mp n,1 \pm n;-\left(\frac{x}{2}\right)^2\right]$ & $\pm 2 \Omega$
    \\
    \hline
    $G_n^Z$ & $2 ~ {}_2 F^R_3 \left[\frac{1}{2},2;\frac{5}{2},1-n,1+n;-\left(\frac{x}{2}\right)^2\right]$ & $0$
    \\
    \hline
    $Q_n^Z$ & ${}_2 F_3^R \left[\frac{1}{2},1;\frac{5}{2},1-n,1+n;-\left( \frac{x}{2} \right)^2 \right]$ & $0$
    \\
    \hline
    $H_n^\pm$ & $\frac{1}{2} \frac{x}{2} ~ {}_2 F^R_3 \left[1,\frac{3}{2};\frac{5}{2},2\mp n,1\pm n;-\left(\frac{x}{2}\right)^2 \right]$ & $\pm \Omega$ 
    \\
    \hline
    \multirow{2}{*}{$P_n^{\times \pm}$} & $\frac{1}{8}\frac{x}{2} \bigg( {}_2 F_3^R \left[ 1,\frac{3}{2};\frac{7}{2},2-n,1+n;-\left(\frac{x}{2}\right)^2 \right]$ & \multirow{2}{*}{$\pm \Omega$}
    \\
    & $~~~~~~~~~~ + ~ {}_2 F_3^R \left[ \frac{3}{2},2;\frac{7}{2},2-n,1+n;-\left( \frac{x}{2} \right)^2 \right] \bigg)$ & 
    \\
    \hline
    \multirow{3}{*}{$P_n^{\div \pm}$} & $\frac{1}{4}\frac{x}{2} \bigg( {}_2 F_3^R \left[ 1,\frac{3}{2};\frac{5}{2},2\mp n,1\pm n;-\left(\frac{x}{2}\right)^2 \right]$ & \multirow{3}{*}{$\pm \Omega$}
    \\
    & $~~~~~~~~~ - \frac{1}{2} ~ {}_2 F_3^R \left[ 1,\frac{3}{2};\frac{7}{2},2\mp n,1\pm n;-\left( \frac{x}{2} \right)^2 \right]$ & 
    \\
    & $~~~~~~~~~~~~~ - \frac{1}{2} ~ {}_2 F_3^R \left[ \frac{3}{2},2;\frac{7}{2},2\mp n,1\pm n;-\left( \frac{x}{2}\right)^2 \right] \bigg)$ &
    \\
    \hline
    $P_n^{3\pm}$ & $\frac{3}{16} \left( \frac{x}{2} \right)^3 \, {}_2 F_3^R \left[ 2,\frac{5}{2};\frac{9}{2},4\mp n,1\pm n;-\left( \frac{x}{2}\right)^2 \right]$ & $\pm 3 \Omega$
    \\
    \hline
    $P_n^{Z\pm}$ & $\frac{1}{4}\frac{x}{2} \, {}_2 F_3^R \left[ 1,\frac{3}{2};\frac{7}{2},2 \mp n,1\pm n;-\left( \frac{x}{2} \right)^2 \right]$ & $\pm \Omega$
    \\
    \hline
    \end{tabular}
    \caption{Table of the correlation functions used in the expressions of the vacuum field correlations between two revolving points. We present, for each of them, their expression alongside the frequency shifts arising in the associated Dirac delta distributions. The hypergeometric functions ${}_p F_q^R$ are defined in Eq. (\ref{hypergeometric_functions}).}
    \label{tab:corr_func}
\end{table}

\section{Rewriting of the correlations involving spatial derivatives using circular notations}\label{circ_notations_EdEB}

For the spatial derivatives, it is also useful to set the circular derivatives
\begin{equation}
    \partial_\pm = \frac{\partial_X \pm i \partial_Y}{2}
\end{equation}
allowing us to write the following correlations in the circular notation: 
\begin{equation}
    \begin{array}{c}
        \langle \Tilde{E}_\pm^A [\omega] \partial_\pm \Tilde{E}_\pm^B [\omega'] \rangle = - \frac{\hbar}{16 \epsilon_0 \pi^\frac{3}{2} r^4} \sum_n \left( N[\vert \omega_n \vert]+\frac{1}{2} \right) \left( \frac{\omega_n r}{c} \right)^4 P_n^{3\pm} \left[ \frac{\vert \omega_n \vert r}{c} \right] \delta[\omega+\omega' \pm 3 \Omega],
    
        \vspace{0.2cm} \\

        \langle \Tilde{E}_\pm^A [\omega] \partial_\pm \Tilde{E}_\pm^A [\omega'] \rangle = \frac{\hbar}{16 \epsilon_0 \pi^\frac{3}{2} r^4} \sum_n (-1)^n \left( N[\vert \omega_n \vert]+\frac{1}{2} \right) \left( \frac{\omega_n r}{c} \right)^4 P_n^{3\pm} \left[ \frac{\vert \omega_n \vert r}{c} \right] \delta[\omega+\omega' \pm 3 \Omega],
    \end{array}
\end{equation}

\begin{equation}
    \begin{array}{c}
        \begin{aligned}
            \langle \Tilde{E}^A_\pm [\omega] \partial_\pm \Tilde{E}^B_\mp [\omega'] \rangle & = \textstyle{ -\frac{\hbar}{16 \epsilon_0 \pi^\frac{3}{2} r^4} \sum_n \left( N[\vert \omega_n \vert]+\frac{1}{2} \right) \left( \frac{\omega_n r}{c} \right)^4 P_n^{\div \pm} \left[ \frac{\vert \omega_n \vert r}{c} \right] \delta[\omega+\omega' \pm \Omega]} \\
            & = \langle \Tilde{E}^A_\mp [\omega] \partial_\pm \Tilde{E}^B_\pm [\omega'] \rangle,
        \end{aligned}
    
        \vspace{0.2cm} \\

        \begin{aligned}
            \langle \Tilde{E}^A_\pm [\omega] \partial_\pm \Tilde{E}^A_\mp [\omega'] \rangle & = \textstyle{\frac{\hbar}{16 \epsilon_0 \pi^\frac{3}{2} r^4} \sum_n (-1)^n \left( N[\vert \omega_n \vert]+\frac{1}{2} \right) \left( \frac{\omega_n r}{c} \right)^4 P_n^{\div \pm} \left[ \frac{\vert \omega_n \vert r}{c} \right] \delta[\omega+\omega' \pm \Omega]} \\
            & = \langle \Tilde{E}^A_\mp [\omega] \partial_\pm \Tilde{E}^A_\pm [\omega'] \rangle,
        \end{aligned}
    \end{array}
\end{equation}

\begin{equation}
    \begin{array}{c}
        \langle \Tilde{E}_\pm^A [\omega] \partial_\mp \Tilde{E}^B_\pm [\omega'] \rangle = \frac{\hbar}{16 \epsilon_0 \pi^\frac{3}{2} r^4} \sum_n \left( N[\vert \omega_n \vert]+\frac{1}{2} \right) \left( \frac{\omega_n r}{c} \right)^4 P_n^{\times \pm} \left[ \frac{\vert \omega_n \vert r}{c} \right] \delta[\omega+\omega' \pm \Omega],
    
        \vspace{0.2cm} \\

        \langle \Tilde{E}_\pm^A [\omega] \partial_\mp \Tilde{E}^A_\pm [\omega'] \rangle = -\frac{\hbar}{16 \epsilon_0 \pi^\frac{3}{2} r^4} \sum_n (-1)^n \left( N[\vert \omega_n \vert]+\frac{1}{2} \right) \left( \frac{\omega_n r}{c} \right)^4 P_n^{\times \pm} \left[ \frac{\vert \omega_n \vert r}{c} \right] \delta[\omega+\omega' \pm \Omega],
    \end{array}
\end{equation}

\begin{equation}
    \begin{array}{c}
        \begin{aligned}
            \langle \Tilde{E}_\pm^A [\omega] \partial_Z \Tilde{E}_Z^B [\omega'] \rangle & = \textstyle{\frac{\hbar}{16 \epsilon_0 \pi^\frac{3}{2} r^4} \sum_n \left( N[\vert \omega_n \vert]+\frac{1}{2} \right) \left( \frac{\omega_n r}{c} \right)^4 P_n^{Z\pm} \left[ \frac{\vert \omega_n \vert r}{c} \right] \delta[\omega+\omega' \pm \Omega]} \\
            & = \langle \Tilde{E}_Z^A [\omega] \partial_Z \Tilde{E}_\pm^B [\omega'] \rangle,
        \end{aligned}
    
        \vspace{0.2cm} \\

        \begin{aligned}
            \langle \Tilde{E}_\pm^A [\omega] \partial_Z \Tilde{E}_Z^A [\omega'] \rangle & = \textstyle{-\frac{\hbar}{16 \epsilon_0 \pi^\frac{3}{2} r^4} \sum_n (-1)^n \left( N[\vert \omega_n \vert]+\frac{1}{2} \right) \left( \frac{\omega_n r}{c} \right)^4 P_n^{Z\pm} \left[ \frac{\vert \omega_n \vert r}{c} \right] \delta[\omega+\omega' \pm \Omega]} \\
            & = \langle \Tilde{E}_Z^A [\omega] \partial_Z \Tilde{E}_\pm^A [\omega'] \rangle,
        \end{aligned}
    \end{array}
\end{equation}

\begin{equation}
    \begin{array}{c}
        \langle \Tilde{E}_Z^A [\omega] \partial_\pm \Tilde{E}_Z^B [\omega'] \rangle = - 4 \frac{\hbar}{16 \epsilon_0 \pi^\frac{3}{2} r^4} \sum_n \left( N[\vert \omega_n \vert]+\frac{1}{2} \right) \left( \frac{\omega_n r}{c} \right)^4 P_n^{\times \pm} \left[ \frac{\vert \omega_n \vert r}{c} \right] \delta[\omega+\omega' \pm \Omega],
    
        \vspace{0.2cm} \\

        \langle \Tilde{E}_Z^A [\omega] \partial_\pm \Tilde{E}_Z^A [\omega'] \rangle = 4 \frac{\hbar}{16 \epsilon_0 \pi^\frac{3}{2} r^4} \sum_n (-1)^n \left( N[\vert \omega_n \vert]+\frac{1}{2} \right) \left( \frac{\omega_n r}{c} \right)^4 P_n^{\times \pm} \left[ \frac{\vert \omega_n \vert r}{c} \right] \delta[\omega+\omega' \pm \Omega],
    \end{array}
\end{equation}

\begin{equation}
    \begin{array}{c}
        \langle \Tilde{E}^A_\pm [\omega] \partial_\pm \Tilde{E}_Z^B [\omega'] \rangle = \langle \Tilde{E}^A_Z [\omega] \partial_\pm \Tilde{E}_\pm^B [\omega'] \rangle = \langle \Tilde{E}^A_\pm [\omega] \partial_\pm \Tilde{E}_Z^A [\omega'] \rangle = \langle \Tilde{E}^A_Z [\omega] \partial_\pm \Tilde{E}_\pm^A [\omega'] \rangle = 0,
         
        \vspace{0.2cm} \\

        \langle \Tilde{E}^A_\pm [\omega] \partial_\mp \Tilde{E}_Z^B [\omega'] \rangle = \langle \Tilde{E}^A_Z [\omega] \partial_\mp \Tilde{E}_\pm^B [\omega'] \rangle = \langle \Tilde{E}^A_\pm [\omega] \partial_\mp \Tilde{E}_Z^A [\omega'] \rangle = \langle \Tilde{E}^A_Z [\omega] \partial_\mp \Tilde{E}_\pm^A [\omega'] \rangle = 0,

        \vspace{0.2cm} \\

        \langle \Tilde{E}^A_\pm [\omega] \partial_Z \Tilde{E}^B_\pm [\omega'] \rangle = \langle \Tilde{E}^A_\pm [\omega] \partial_Z \Tilde{E}^B_\mp [\omega'] \rangle = \langle \Tilde{E}^A_\pm [\omega] \partial_Z \Tilde{E}^A_\pm [\omega'] \rangle = \langle \Tilde{E}^A_\pm [\omega] \partial_Z \Tilde{E}^A_\mp [\omega'] \rangle = 0
    \end{array}
\end{equation}
and finally
\begin{equation}
    \begin{array}{c}
        \begin{aligned}
            \textstyle{\langle \Tilde{B}_\pm^A [\omega] \partial_\pm \Tilde{E}_Z^B [\omega'] \rangle} & \textstyle{~= \pm \frac{\hbar}{16 \epsilon_0 \pi^\frac{3}{2} c r^4} \sum_n \left( N[\vert \omega_n \vert]+\frac{1}{2} \right) \frac{\omega_n r}{c} \left( \frac{\vert \omega_n \vert r}{c} \right)^3 G_n^\pm \left[ \frac{\vert \omega_n \vert r}{c} \right] \delta[\omega+\omega' \pm 2\Omega]} \\
            & = -\langle \Tilde{B}_Z^A [\omega] \partial_\pm \Tilde{E}_\pm^B [\omega'] \rangle, 
        \end{aligned}
    
        \vspace{0.2cm} \\

        \begin{aligned}
            \textstyle{\langle \Tilde{B}_\pm^A [\omega] \partial_\pm \Tilde{E}_Z^A [\omega'] \rangle} & \textstyle{~= \pm \frac{\hbar}{16 \epsilon_0 \pi^\frac{3}{2} c r^4} \sum_n (-1)^n \left( N[\vert \omega_n \vert]+\frac{1}{2} \right) \frac{\omega_n r}{c} \left( \frac{\vert \omega_n \vert r}{c} \right)^3 G_n^\pm \left[ \frac{\vert \omega_n \vert r}{c} \right] \delta[\omega+\omega' \pm 2\Omega]} \\
            & = -\langle \Tilde{B}_Z^A [\omega] \partial_\pm \Tilde{E}_\pm^A [\omega'] \rangle, 
        \end{aligned}
    \end{array}
\end{equation}

\begin{equation}
    \begin{array}{c}
        \begin{aligned}
            \textstyle{\langle \Tilde{B}_\pm^A [\omega] \partial_\mp \Tilde{E}_Z^B [\omega'] \rangle} & = \textstyle{ \mp \frac{1}{4} \frac{\hbar}{16 \epsilon_0 \pi^\frac{3}{2} c r^4} \sum_n \left( N[\vert \omega_n \vert]+\frac{1}{2} \right) \frac{\omega_n r}{c} \left( \frac{\vert \omega_n \vert r}{c} \right)^3 G_n^Z \left[ \frac{\vert \omega_n \vert r}{c} \right] \delta[\omega+\omega']} \\ 
            & = -\langle \Tilde{B}_Z^A [\omega] \partial_\mp \Tilde{E}_\pm^B [\omega'] \rangle, 
        \end{aligned}
    
        \vspace{0.2cm} \\

        \begin{aligned}
            \textstyle{\langle \Tilde{B}_\pm^A [\omega] \partial_\mp \Tilde{E}_Z^A [\omega'] \rangle} &= \textstyle{ \mp \frac{1}{4} \frac{\hbar}{16 \epsilon_0 \pi^\frac{3}{2} c r^4} \sum_n (-1)^n \left( N[\vert \omega_n \vert]+\frac{1}{2} \right) \frac{\omega_n r}{c} \left( \frac{\vert \omega_n \vert r}{c} \right)^3 G_n^Z \left[ \frac{\vert \omega_n \vert r}{c} \right] \delta[\omega+\omega']} \\
            & = -\langle \Tilde{B}_Z^A [\omega] \partial_\mp \Tilde{E}_\pm^A [\omega'] \rangle, 
        \end{aligned}
    \end{array}
\end{equation}

\begin{equation}
    \begin{array}{c}
        \langle \Tilde{B}^A_\pm [\omega] \partial_Z \Tilde{E}^B_\mp [\omega'] \rangle = \pm \frac{1}{2} \frac{\hbar}{16 \epsilon_0 \pi^\frac{3}{2} c r^4} \sum_n \left( N[\vert \omega_n \vert]+\frac{1}{2} \right) \frac{\omega_n r}{c} \left( \frac{\vert \omega_n \vert r}{c} \right)^3 Q_n^Z \left[ \frac{\vert \omega_n \vert r}{c} \right] \delta[\omega+\omega'], 
    
        \vspace{0.2cm} \\

        \langle \Tilde{B}^A_\pm [\omega] \partial_Z \Tilde{E}^A_\mp [\omega'] \rangle = \pm \frac{1}{2} \frac{\hbar}{16 \epsilon_0 \pi^\frac{3}{2} c r^4} \sum_n (-1)^n \left( N[\vert \omega_n \vert]+\frac{1}{2} \right) \frac{\omega_n r}{c} \left( \frac{\vert \omega_n \vert r}{c} \right)^3 Q_n^Z \left[ \frac{\vert \omega_n \vert r}{c} \right] \delta[\omega+\omega'], 
    \end{array}
\end{equation}

\begin{equation}
    \begin{array}{c}
        \begin{aligned}
            & \langle \Tilde{B}^A_\pm [\omega] \partial_\pm \Tilde{E}^B_\pm [\omega'] \rangle = \langle \Tilde{B}^A_\pm [\omega] \partial_\mp \Tilde{E}_\mp^B [\omega'] \rangle = \langle \Tilde{B}_\pm^A [\omega] \partial_\mp \Tilde{E}_\pm^B [\omega'] \rangle = \langle \Tilde{B}^A_\pm [\omega] \partial_\pm \Tilde{E}_\mp^B [\omega'] \rangle \\
            & = \langle \Tilde{B}^A_\pm [\omega] \partial_\pm \Tilde{E}^A_\pm [\omega'] \rangle = \langle \Tilde{B}^A_\pm [\omega] \partial_\mp \Tilde{E}_\mp^A [\omega'] \rangle = \langle \Tilde{B}_\pm^A [\omega] \partial_\mp \Tilde{E}_\pm^A [\omega'] \rangle = \langle \Tilde{B}^A_\pm [\omega] \partial_\pm \Tilde{E}_\mp^A [\omega'] \rangle = 0,
        \end{aligned}
    
        \vspace{0.2cm} \\

        \begin{aligned}
            & \langle \Tilde{B}^A_\pm [\omega] \partial_Z \Tilde{E}_Z^B [\omega'] \rangle = \langle \Tilde{B}^A_Z [\omega] \partial_\pm \Tilde{E}_Z^B [\omega'] \rangle = \langle \Tilde{B}^A_Z [\omega] \partial_Z \Tilde{E}_\pm^B [\omega'] \rangle \\
            & = \langle \Tilde{B}^A_\pm [\omega] \partial_Z \Tilde{E}_Z^A [\omega'] \rangle = \langle \Tilde{B}^A_Z [\omega] \partial_\pm \Tilde{E}_Z^A [\omega'] \rangle = \langle \Tilde{B}^A_Z [\omega] \partial_Z \Tilde{E}_\pm^A [\omega'] \rangle = 0,
        \end{aligned}

        \vspace{0.2cm} \\

        \langle \Tilde{B}_\pm^A [\omega] \partial_Z \Tilde{E}_\pm^B [\omega'] \rangle = \langle \Tilde{B}_\pm^A [\omega] \partial_Z \Tilde{E}_\pm^A [\omega'] \rangle = 0. 




    \end{array}
\end{equation}

\section{Summation formulas of the correlation functions to first order in the angular velocity}\label{formulas_corr}

We are now going to derive some summation properties of the correlation functions. First of all, recalling that $1/\Gamma[-m]=0$ for $m \in \mathbb{N}$, the set of natural numbers including zero, and using some binomial formulas, one can show:
\begin{equation}
\begin{array}{ccc}
        \displaystyle
        \sum_{n \in \mathbb{Z}} \frac{1}{\Gamma [1-n+k] \Gamma [1+n+k]} = \frac{\sqrt{\pi}}{\Gamma[1+k] \Gamma\left[\frac{1}{2}+k\right]}, & \displaystyle
        \sum_{n \in \mathbb{Z}} \frac{(-1)^n}{\Gamma [1-n+k] \Gamma [1+n+k]} = \delta_{k,0} & k \in \mathbb{N},
    
        \vspace{0.2cm} \\
        
        \displaystyle
        \sum_{n \in \mathbb{Z}} \frac{1}{\Gamma [2-n+k] \Gamma [1+n+k]} = \frac{\sqrt{\pi}}{\Gamma \left[\frac{3}{2}+k\right] \Gamma[1+k]}, & \displaystyle
        \sum_{n \in \mathbb{Z}} \frac{(-1)^n}{\Gamma [2-n+k] \Gamma [1+n+k]} = 0 & k \in \mathbb{N},
        
        \vspace{0.2cm} \\
        
        \displaystyle
        \sum_{n \in \mathbb{Z}} \frac{1}{\Gamma [3-n+k] \Gamma [1+n+k]} = \frac{\sqrt{\pi}}{\Gamma[2+k] \Gamma\left[\frac{3}{2}+k\right]}, & \displaystyle
        \sum_{n \in \mathbb{Z}} \frac{(-1)^n}{\Gamma [3-n+k] \Gamma [1+n+k]} = 0 & k \in \mathbb{N},

        \vspace{0.2cm} \\

        \displaystyle 
        \sum_{n \in \mathbb{Z}} \frac{1}{\Gamma[4-n+k] \Gamma[1+n+k]} = \frac{\sqrt{\pi}}{\Gamma \left[\frac{5}{2}+k \right] \Gamma[2+k]}, & \displaystyle \sum_{n \in \mathbb{Z}} \frac{(-1)^n}{\Gamma[4-n+k] \Gamma[1+n+k]} = 0 & k \in \mathbb{N},
    \end{array}
\end{equation}
which are useful for the computation of the correlations to zeroth order in $\Omega r/c$.

To obtain the first order, one need the following ones:
\begin{equation}
    \begin{array}{ccc}
        \multicolumn{2}{c}{\displaystyle \sum_{n \in \mathbb{Z}} \frac{n}{\Gamma [1-n+k] \Gamma [1+n+k]} = \sum_{n \in \mathbb{Z}} \frac{(-1)^n n}{\Gamma [1-n+k] \Gamma [1+n+k]} = 0} & k \in \mathbb{N},
        
        \vspace{0.2cm} \\
    
        \displaystyle \sum_{n \in \mathbb{Z}} \frac{n}{\Gamma [2-n+k] \Gamma[1+n+k]} = \frac{1}{2} \frac{\sqrt{\pi}}{\Gamma \left[ \frac{3}{2} +k \right] \Gamma[1+k]}, & \displaystyle \sum_{n \in \mathbb{Z}} \frac{(-1)^n n}{\Gamma [2-n+k] \Gamma[1+n+k]} = - \delta_{k,0} & k \in \mathbb{N},
        
        \vspace{0.2cm} \\
        
        \displaystyle \sum_{n \in \mathbb{Z}} \frac{n}{\Gamma[3-n+k] \Gamma [1+n+k]} = \frac{\sqrt{\pi}}{\Gamma[2+k] \Gamma \left[ \frac{3}{2} +k \right]}, & \displaystyle \sum_{n \in \mathbb{Z}} \frac{(-1)^n n}{\Gamma[3-n+k] \Gamma [1+n+k]} = 0 & k \in \mathbb{N},

        \vspace{0.2cm} \\

        \displaystyle \sum_{n \in \mathbb{Z}} \frac{n}{\Gamma[4-n+k] \Gamma [1+n+k]} = \frac{3}{2} \frac{\sqrt{\pi}}{\Gamma\left[ \frac{5}{2}+k \right] \Gamma[2+k]}, & \displaystyle \sum_{n \in \mathbb{Z}} \frac{(-1)^n n}{\Gamma[4-n+k] \Gamma[1+n+k]} = 0 & k \in \mathbb{N}.
    \end{array}
\end{equation}

With these relations, one can then derive the following series expansions, for any real numbers $x$ and $y$:
\begin{equation}
    \begin{array}{c}
        \begin{aligned}
            \sum_{n \in \mathbb{Z}} \left( \frac{1}{e^{z \vert x+ny \vert }-1} +\frac{1}{2} \right) \frac{\vert x+ny \vert^{2k+3}}{\Gamma [1-n+k] \Gamma[1+n+k]} = \frac{\sqrt{\pi}}{\Gamma [1+k] \Gamma \left[ \frac{1}{2} +k \right]} \left( \frac{1}{e^{z \vert x \vert}-1} +\frac{1}{2} \right) & \vert x \vert^{2k+3} \\
            & + O \left( y^2 \right),
        \end{aligned}
    
        \vspace{0.2cm} \\

        \begin{aligned}
            \sum_{n \in \mathbb{Z}} \left( \frac{1}{e^{z \vert x+ny \vert}-1}+\frac{1}{2} \right) \frac{\left(x+ny \right) \vert x+ny \vert^{2k+3}}{\Gamma[1-n+k] \Gamma[1+n+k]} = \frac{\sqrt{\pi}}{\Gamma[1+k] \Gamma\left[\frac{1}{2}+k \right]} \left( \frac{1}{e^{z \vert x \vert}-1} + \frac{1}{2} \right) & x \vert x \vert^{2k+3} \\
            & + O \left( y^2 \right),
        \end{aligned}
    \end{array}
\end{equation}

\begin{equation}
    \begin{array}{c}
        \begin{aligned}
            & \sum_{n \in \mathbb{Z}} \left( \frac{1}{e^{z \vert x+ny \vert}-1} +\frac{1}{2} \right) \frac{\left(x+ny \right) \vert x +ny \vert^{2k+3}}{\Gamma[2-n+k] \Gamma[1+n+k]} \\
            & = \frac{\sqrt{\pi}}{\Gamma \left[\frac{3}{2}+k \right] \Gamma[1+k]} \bigg( \left( \frac{1}{e^{z \vert x \vert}-1} + \frac{1}{2} \right) x \vert x \vert^{2k+3} \\
            & ~~~~~~~~~~~~~~~~~~~~~~~~~~~~~~~~~ + \frac{1}{2} \left( \frac{1}{e^{z \vert x \vert}-1} + \frac{1}{2} \right) \left(2k+4 \right) \vert x \vert^{2k+3} y - \frac{1}{2} \frac{e^{z \vert x \vert}}{\left( e^{z \vert x \vert}-1 \right)^2} z x^{2k+4} y \bigg) + O\left( y^2 \right),
        \end{aligned}
    
        \vspace{0.2cm} \\

        \begin{aligned}
            & \sum_{n \in \mathbb{Z}} \left( \frac{1}{e^{z \vert x+ny \vert}-1} +\frac{1}{2} \right) \frac{\vert x+ny \vert^{2k+5}}{\Gamma[2-n+k] \Gamma[1+n+k]} \\
            & = \frac{\sqrt{\pi}}{\Gamma \left[\frac{3}{2}+k \right] \Gamma[1+k]} \bigg( \left( \frac{1}{e^{z \vert x \vert} -1} + \frac{1}{2} \right) \vert x \vert^{2k+5} \\
            & ~~~~~~~~~~~~~~~~~~~~~~~~~~~~~~~~~ + \frac{1}{2} \left( \frac{1}{e^{z \vert x \vert}-1} +\frac{1}{2} \right) \left(2k+5 \right) x \vert x \vert^{2k+3} y - \frac{1}{2} \frac{e^{z \vert x \vert}}{\left( e^{z \vert x \vert}-1 \right)^2} z x^{2k+5} y \bigg) + O \left( y^2 \right),
        \end{aligned}
    \end{array}
\end{equation}

\begin{equation}
    \begin{array}{c}
        \begin{aligned}
            & \sum_{n \in \mathbb{Z}} \left( \frac{1}{e^{z \vert x+ny \vert}-1} +\frac{1}{2} \right) \frac{\vert x +ny \vert^{2k+5}}{\Gamma[3-n+k] \Gamma[1+n+k]} \\
            & = \frac{\sqrt{\pi}}{\Gamma[2+k] \Gamma \left[\frac{3}{2}+k \right]} \bigg( \left( \frac{1}{e^{z \vert x \vert}-1} +\frac{1}{2} \right) \vert x \vert^{2k+5} \\
            & ~~~~~~~~~~~~~~~~~~~~~~~~~~~~~~~~~ + \left( \frac{1}{e^{z \vert x \vert}-1} +\frac{1}{2} \right) \left(2k+5 \right) x \vert x \vert^{2k+3} y - \frac{e^{z \vert x \vert}}{\left( e^{z \vert x \vert}-1 \right)^2} z x^{2k+5} y \bigg) + O \left( y^2 \right),
        \end{aligned}
    
        \vspace{0.2cm} \\

        \begin{aligned}
            & \sum_{n \in \mathbb{Z}} \left( \frac{1}{e^{z \vert x+ny \vert}-1} +\frac{1}{2} \right) \frac{(x+ny) \vert x +ny \vert^{2k+5}}{\Gamma[3-n+k] \Gamma[1+n+k]} \\
            & = \frac{\sqrt{\pi}}{\Gamma[2+k] \Gamma \left[\frac{3}{2}+k \right]} \bigg( \left( \frac{1}{e^{z \vert x \vert}-1} +\frac{1}{2} \right) x \vert x \vert^{2k+5} \\
            & ~~~~~~~~~~~~~~~~~~~~~~~~~~~~~~~~~ + \left( \frac{1}{e^{z \vert x \vert}-1} +\frac{1}{2} \right) \left(2k+6 \right) \vert x \vert^{2k+5} y - \frac{e^{z \vert x \vert}}{\left( e^{z \vert x \vert}-1 \right)^2} z x^{2k+6} y \bigg) + O \left( y^2 \right),
        \end{aligned}
    \end{array}
\end{equation}

\begin{equation}
    \begin{aligned}
        & \sum_{n \in \mathbb{Z}} \left( \frac{1}{e^{z \vert x+ny \vert}-1} +\frac{1}{2} \right) \frac{\vert x+ny \vert^{2k+7}}{\Gamma[4-n+k] \Gamma[1+n+k]} \\
        & = \frac{\sqrt{\pi}}{\Gamma \left[\frac{5}{2}+k \right] \Gamma [2+k]} \bigg( \left( \frac{1}{e^{z \vert x \vert}-1} +\frac{1}{2} \right) \vert x \vert^{2k+7} \\
        & ~~~~~~~~~~~~~~~~~~~~~~~~~~~~~~~~~ + \frac{3}{2} \left( \frac{1}{e^{z \vert x \vert}-1} +\frac{1}{2} \right) (2k+7) x \vert x \vert^{2k+5} y - \frac{3}{2} \frac{e^{z \vert x \vert}}{\left( e^{z \vert x \vert}-1 \right)^2} z x^{2k+7} y \bigg) + O \left( y^2 \right)
    \end{aligned}
\end{equation}
together with
\begin{equation}
    \begin{array}{c}
        \displaystyle \sum_{n \in \mathbb{Z}} (-1)^n \left( \frac{1}{e^{z \vert x+ny \vert}-1} +\frac{1}{2} \right) \frac{\vert x+ny \vert^{2k+3}}{\Gamma[1-n+k] \Gamma[1+n+k]} = \delta_{k,0} \left( \frac{1}{e^{z \vert x \vert}-1} + \frac{1}{2} \right) \vert x \vert^3 + O \left( y^2 \right),
    
        \vspace{0.2cm} \\

        \displaystyle \sum_{n \in \mathbb{Z}} (-1)^n \left( \frac{1}{e^{z\vert x+ny \vert}-1}+\frac{1}{2} \right) \frac{(x+ny) \vert x+ny \vert^3}{\Gamma [1-n+k] \Gamma[1+n+k]} = \delta_{k,0} \left( \frac{1}{e^{z \vert x \vert}-1}+\frac{1}{2} \right) x \vert x \vert^3 + O \left( y^2 \right),
    \end{array}
\end{equation}

\begin{equation}
    \begin{array}{c}
        \begin{aligned}
            \sum_{n \in \mathbb{Z}} (-1)^n \left( \frac{1}{e^{z \vert x+ny \vert}-1} +\frac{1}{2} \right) \frac{(x+ny) \vert x+ny \vert^{2k+3}}{\Gamma[2-n+k] \Gamma[1+n+k]} = - \delta_{k,0} & \bigg( \left( \frac{1}{e^{z \vert x \vert}-1} + \frac{1}{2} \right) 4 \vert x \vert^3 y \\
            & - \frac{e^{z \vert x \vert}}{\left( e^{z \vert x \vert}-1 \right)^2} z x^4 y \bigg) + O \left( y^2 \right),
        \end{aligned} 
    
        \vspace{0.2cm} \\

        \begin{aligned}
            \sum_{n \in \mathbb{Z}} (-1)^n \left( \frac{1}{e^{z \vert x+ny \vert}-1}+\frac{1}{2} \right) \frac{\vert x+ny \vert^{2k+5}}{\Gamma[2-n+k] \Gamma[1+n+k]} = - \delta_{k,0} & \bigg( \left( \frac{1}{e^{z \vert x \vert}-1}+\frac{1}{2} \right) 5 x \vert x \vert^3 y \\
            & - \frac{e^{z \vert x \vert}}{\left( e^{z \vert x \vert}-1 \right)^2} z x^5 y \bigg) + O \left( y^2 \right),
        \end{aligned}
    \end{array}
\end{equation}

\begin{equation}
    \begin{array}{c}
        \displaystyle \sum_{n \in \mathbb{Z}} (-1)^n \left( \frac{1}{e^{z \vert x+ny \vert}-1}+\frac{1}{2} \right) \frac{\vert x+ny \vert^{2k+5}}{\Gamma[3-n+k] \Gamma[1+n+k]} = O \left( y^2 \right),
    
        \vspace{0.2cm} \\

        \displaystyle \sum_{n \in \mathbb{Z}} (-1)^n \left( \frac{1}{e^{z \vert x+ny \vert}-1}+\frac{1}{2} \right) \frac{(x+ny) \vert x+ny \vert^{2k+5}}{\Gamma[3-n+k] \Gamma[1+n+k]} = O \left( y^2 \right),
    \end{array}
\end{equation}

\begin{equation}
    \sum_{n \in \mathbb{Z}} (-1)^n \left( \frac{1}{e^{z \vert x+ny \vert}-1}+\frac{1}{2} \right) \frac{\vert x+ny \vert^{2k+7}}{\Gamma[4-n+k] \Gamma[1+n+k]} = O \left( y^2 \right).
\end{equation}

Using the hypergeometric series expansion of the correlation functions, we write
\begin{equation}\label{Gn0}
    \begin{array}{c}
        \begin{aligned}
            \textstyle{\sum_{n}} & \textstyle{\left( N[\vert \omega_n \vert]+\frac{1}{2} \right) \left( \frac{\vert \omega_n \vert r}{c} \right)^3 G_n^0 \left[ \frac{\vert \omega_n \vert r}{c} \right]} \\
            & = \textstyle{\frac{2}{\sqrt{\pi}} \left( N[\vert \omega \vert]+\frac{1}{2} \right) \bigg( \left( 1 + \left( \frac{\vert \omega \vert r}{c} \right)^2 \right) \sin \left[ \frac{\vert \omega \vert r}{c} \right] - \frac{\vert \omega \vert r}{c} \cos \left[ \frac{\vert \omega \vert r}{c} \right] \bigg) + O \left( \left( \frac{\Omega r}{c} \right)^2 \right)}, 
        \end{aligned}
    
        \vspace{0.2cm} \\

        \sum_{n} (-1)^n \left( N[\vert \omega_n \vert]+\frac{1}{2} \right) \left( \frac{\vert \omega_n \vert r}{c} \right)^3 G_n^0 \left[ \frac{\vert \omega_n \vert r}{c} \right] = \frac{8}{3 \sqrt{\pi}} \left( N[\vert \omega \vert]+\frac{1}{2} \right) \left(\frac{\vert \omega \vert r}{c} \right)^3 + O \left( \left( \frac{\Omega r}{c} \right)^2 \right),
    \end{array}
\end{equation}

\begin{equation}\label{Gnpm}
    \begin{array}{c}
        \begin{aligned}
            \textstyle{\sum_{n}} & \textstyle{\left( N[\vert \omega_n \vert]+\frac{1}{2} \right)\left( \frac{\vert \omega_n \vert r}{c} \right)^3 G_n^\pm \left[ \frac{\vert \omega_n \vert r}{c} \right]} \\
            & \textstyle{ = \frac{1}{\sqrt{\pi}} \bigg( \left( N[\vert \omega \vert]+\frac{1}{2} \right) \bigg( \bigg( 3 - \left( \frac{\vert \omega \vert r}{c} \right)^2 \bigg) \sin \left[ \frac{\vert \omega \vert r}{c} \right] - 3 \frac{\vert \omega \vert r}{c} \cos \left[ \frac{\vert \omega \vert r}{c} \right] \bigg)} \\
            & ~~~~~~~~~ \textstyle{\pm \left( N[\vert \omega \vert]+\frac{1}{2} \right) \frac{\Omega r}{c} \frac{\omega r}{c} \bigg( \sin \left[ \frac{\vert \omega \vert r}{c} \right] - \frac{\vert \omega \vert r}{c} \cos \left[ \frac{\vert \omega \vert r}{c} \right] \bigg)} \\
            & ~~~~~~~~~ \textstyle{\mp \, \frac{\hbar \vert \omega \vert}{k_B T} N[\vert \omega \vert] \left( N[\vert \omega \vert]+1 \right) \frac{\Omega r}{c} \left( \frac{\omega r}{c} \right)^{-1} \bigg( \bigg( 3 - \left( \frac{\vert \omega \vert r}{c} \right)^2 \bigg) \sin \left[ \frac{\vert \omega \vert r}{c} \right]} \\
            & ~~~~~~~~~~~~~~~~~~~~~~~~~~~~~~~~~~~~~~~~~~~~~~~~~~~~~~~~~~~~~~~~~~~~ \textstyle{- \, 3 \frac{\vert \omega \vert r}{c} \cos \left[ \frac{\vert \omega \vert r}{c} \right] \bigg) \bigg)} + O \left( \left( \frac{\Omega r}{c} \right)^2 \right),
        \end{aligned}
    
        \vspace{0.2cm} \\

        \sum_{n} (-1)^n \left( N[\vert \omega_n \vert]+\frac{1}{2} \right) \left( \frac{\vert \omega_n \vert r}{c} \right)^3 G_n^\pm \left[ \frac{\vert \omega_n \vert r}{c} \right] = O \left( \left( \frac{\Omega r}{c} \right)^2 \right),
    \end{array}
\end{equation}

\begin{equation}\label{GnZ}
    \begin{array}{c}
        \begin{aligned}
            \textstyle{\sum_{n}} & \textstyle{ \left( N[\vert \omega_n \vert]+\frac{1}{2} \right) \left( \frac{\vert \omega_n \vert r}{c} \right)^3 G_n^Z \left[ \frac{\vert \omega_n \vert r}{c} \right]} \\
            & \textstyle{= - \frac{4}{\sqrt{\pi}} \left( N[\vert \omega \vert]+\frac{1}{2} \right) \bigg( \bigg( 1 - \left( \frac{\vert \omega \vert r}{c} \right)^2 \bigg) \sin \left[ \frac{\vert \omega \vert r}{c} \right] - \frac{\vert \omega \vert r}{c} \cos \left[ \frac{\vert \omega \vert r}{c} \right] \bigg) + O \left( \left( \frac{\Omega r}{c} \right)^2 \right)},
        \end{aligned}
    
        \vspace{0.2cm} \\

        \sum_{n} (-1)^n \left( N[\vert \omega \vert]+\frac{1}{2} \right) \left( \frac{\vert \omega_n \vert r}{c} \right)^3 G_n^Z \left[ \frac{\vert \omega_n \vert r}{c} \right] = \frac{8}{3 \sqrt{\pi}} \left( N[\vert \omega \vert]+\frac{1}{2} \right) \left( \frac{\vert \omega \vert r}{c} \right)^3 + O \left( \left( \frac{\Omega r}{c} \right)^2 \right),
    \end{array}
\end{equation}

\begin{equation}
    \begin{array}{c}
        \begin{aligned}
            \textstyle{\sum_{n}} & \textstyle{\left( N[\vert \omega_n \vert]+\frac{1}{2} \right) \left( \frac{\omega_n r}{c} \right)^3 H_n^\pm \left[ \frac{\vert \omega_n \vert r}{c} \right]} \\
            & = \textstyle{ \frac{1}{\sqrt{\pi}} \bigg( \left( N[\vert \omega \vert]+\frac{1}{2} \right) 2 \frac{\omega r}{c} \bigg( \sin \left[ \frac{\vert \omega \vert r}{c} \right] - \frac{\vert \omega \vert r}{c} \cos \left[ \frac{\vert \omega \vert r}{c} \right] \bigg)} \\
            & ~~~~~~~~~~ \textstyle{\pm \left( N[\vert \omega \vert]+\frac{1}{2} \right) \frac{\Omega r}{c} \bigg( \bigg( 1 + \left( \frac{\vert \omega \vert r}{c} \right)^2 \bigg) \sin \left[ \frac{\vert \omega \vert r}{c} \right] - \frac{\vert \omega \vert r}{c} \cos \left[ \frac{\vert \omega \vert r}{c} \right] \bigg)} \\
            & ~~~~~~~~~~ \textstyle{\mp \, \frac{\hbar \vert \omega \vert}{k_B T} N[\vert \omega \vert] \left( N[\vert \omega \vert]+1 \right) \frac{\Omega r}{c} \bigg( \sin \left[ \frac{\vert \omega \vert r}{c} \right] - \frac{\vert \omega \vert r}{c} \cos \left[ \frac{\vert \omega \vert r}{c} \right] \bigg) \bigg) + O \left( \left( \frac{\Omega r}{c} \right)^2 \right)},
        \end{aligned}
    
        \vspace{0.2cm} \\

        \begin{aligned}
            \textstyle{\sum_{n}} & \textstyle{ (-1)^n \left( N[\vert \omega_n \vert]+\frac{1}{2} \right) \left( \frac{\omega_n r}{c} \right)^3 H_n^\pm \left[ \frac{\vert \omega_n \vert r}{c} \right]} \\
            & = - \textstyle{\frac{1}{3 \sqrt{\pi}} \bigg( \pm \left( N[\vert \omega \vert]+\frac{1}{2} \right) 4 \frac{\Omega r}{c} \left( \frac{\vert \omega \vert r}{c} \right)^3 \mp \, \frac{\hbar \vert \omega \vert}{k_B T} N[\vert \omega \vert] \left( N[\vert \omega \vert]+1 \right) \frac{\Omega r}{c} \left( \frac{\vert \omega \vert r}{c} \right)^3 \bigg) + O \left( \left( \frac{\Omega r}{c} \right)^2 \right)},
        \end{aligned} 
    \end{array}
\end{equation}

\begin{equation}
    \begin{array}{c}
        \begin{aligned}
            \textstyle{\sum_{n}} & \textstyle{ \left( N[\vert \omega_n \vert]+\frac{1}{2} \right)\left( \frac{\omega_n r}{c} \right)^4 P_n^{\times \pm} \left[ \frac{\vert \omega_n \vert r}{c} \right]} \\
            & = \textstyle{ - \frac{1}{4 \sqrt{\pi}}  \bigg( \left( N[\vert \omega \vert]+\frac{1}{2} \right) 2 \bigg( \bigg( 3 - 2 \left( \frac{\vert \omega \vert r}{c} \right)^2 \bigg) \sin \left[ \frac{\vert \omega \vert r}{c} \right] - \bigg( 3 - \left( \frac{\vert \omega \vert r}{c} \right)^2 \bigg) \frac{\vert \omega \vert r}{c} \cos \left[ \frac{\vert \omega \vert r}{c} \right] \bigg)} \\
            & ~~~~~~~~~~~~~~ \textstyle{\mp \left( N[\vert \omega \vert]+\frac{1}{2} \right) \frac{\Omega r}{c} \frac{\omega r}{c} \bigg( \bigg( 1 + \left( \frac{\vert \omega \vert r}{c} \right)^2 \bigg) \sin \left[ \frac{\vert \omega \vert r}{c} \right] - \frac{\vert \omega \vert r}{c} \cos \left[ \frac{\vert \omega \vert r}{c} \right] \bigg)} \\
            & ~~~~~~~~~~~~~~ \textstyle{\mp \, \frac{\hbar \vert \omega \vert}{k_B T} N[\vert \omega \vert] \left( N[\vert \omega \vert]+1 \right) \frac{\Omega r}{c} \left( \frac{\omega r}{c} \right)^{-1} \bigg( \bigg( 3 - 2 \left( \frac{\vert \omega \vert r}{c} \right)^2 \bigg) \sin \left[ \frac{\vert \omega \vert r}{c} \right]} \\
            & ~~~~~~~~~~~~~~~~~~~~~~~~~~~~~~~~~~~~~~~~~~~~~~~~~~~~~~~~~~~~~~~~~~~~~~~~ \textstyle{- \, \bigg( 3 - \left( \frac{\vert \omega \vert r}{c} \right)^2 \bigg) \frac{\vert \omega \vert r}{c} \cos \left[ \frac{\vert \omega \vert r}{c} \right] \bigg) \bigg) + O \left( \left( \frac{\Omega r}{c} \right)^2 \right)},
        \end{aligned}
    
        \vspace{0.2cm} \\

        \begin{aligned}
            \textstyle{\sum_{n}} & \textstyle{ (-1)^n \left( N[\vert \omega_n \vert]+\frac{1}{2} \right) \left( \frac{\omega_n r}{c}\right)^4 P_n^{\times \pm} \left[ \frac{\vert \omega_n \vert r}{c} \right]} \\
            & = - \textstyle{\frac{1}{15 \sqrt{\pi}} \bigg( \pm \left( N[\vert \omega \vert]+\frac{1}{2} \right) 5 \frac{\Omega r}{c} \frac{\omega r}{c} \left( \frac{\vert \omega \vert r}{c} \right)^3} \\
            & ~~~~~~~~~~~~~~~~ \textstyle{\mp \, \frac{\hbar \vert \omega \vert}{k_B T} N[\vert \omega \vert] \left( N[\vert \omega \vert]+1 \right) \frac{\Omega r}{c} \frac{\omega r}{c} \left( \frac{\vert \omega \vert r}{c} \right)^3 \bigg) + O \left( \left( \frac{\Omega r}{c} \right)^2 \right)},
        \end{aligned}
    \end{array}
\end{equation}

\begin{equation}
    \begin{array}{c}
        \begin{aligned}
            \textstyle{\sum_{n}} & \textstyle{ \left( N[\vert \omega_n \vert]+\frac{1}{2} \right)} \textstyle{\left( \frac{\omega_n r}{c} \right)^4 P_n^{\div \pm} \left[ \frac{\vert \omega_n \vert r}{c} \right]} \\
            & \textstyle{ = \frac{1}{4 \sqrt{\pi}} \bigg( \left( N[\vert \omega \vert]+\frac{1}{2} \right) 2 \bigg( 3 \sin \left[ \frac{\vert \omega \vert r}{c} \right] - \bigg( 3 + \left( \frac{\vert \omega \vert r}{c} \right)^2 \bigg) \frac{\vert \omega \vert r}{c} \cos \left[ \frac{\vert \omega \vert r}{c} \right] \bigg)} \\
            & ~~~~~~~~~~ \textstyle{\pm \left( N[\vert \omega \vert]+\frac{1}{2} \right) \frac{\Omega r}{c} \frac{\omega r}{c} \bigg( \bigg( 3 + \bigg( \frac{\vert \omega \vert r}{c} \bigg)^2 \bigg) \sin \left[ \frac{\vert \omega \vert r}{c} \right] - 3 \frac{\vert \omega \vert r}{c} \cos \left[ \frac{\vert \omega \vert r}{c} \right] \bigg)} \\
            & ~~~~~~~~~~ \textstyle{\mp \, \frac{\hbar \vert \omega \vert}{k_B T} N[\vert \omega \vert] \left(N[\vert \omega \vert]+1 \right) \frac{\Omega r}{c} \left( \frac{\omega r}{c} \right)^{-1} \bigg( 3 \sin \left[ \frac{\vert \omega \vert r}{c} \right]} \\
            & ~~~~~~~~~~~~~~~~~~~~~~~~~~~~~~~~~~~~~~~~~~~~~~~~~~~~~~~ \textstyle{- \, \bigg( 3 + \left( \frac{\vert \omega \vert r}{c} \right)^2 \bigg) \frac{\vert \omega \vert r}{c} \cos \left[ \frac{\vert \omega \vert r}{c} \right] \bigg) \bigg) + O \left( \left( \frac{\Omega r}{c} \right)^2 \right)},
        \end{aligned}
    
        \vspace{0.2cm} \\

        \begin{aligned}
            \textstyle{\sum_{n}} &  \textstyle{(-1)^n \left( N[\vert \omega_n \vert]+\frac{1}{2} \right) \left( \frac{\omega_n r}{c} \right)^4 P_n^{\div \pm} \left[ \frac{\vert \omega_n \vert r}{c} \right]} \\
            & = - \textstyle{\frac{1}{10 \sqrt{\pi}} \bigg( \pm \left( N[\vert \omega \vert]+\frac{1}{2} \right) 5 \frac{\Omega r}{c} \frac{\omega r}{c } \left( \frac{\vert \omega \vert r}{c} \right)^3} \\
            & ~~~~~~~~~~~~~~~~ \textstyle{\mp \, \frac{\hbar \vert \omega \vert}{k_B T} N[\vert \omega \vert] \left( N[\vert \omega \vert]+1 \right) \frac{\Omega r}{c} \frac{\omega r}{c} \left( \frac{\vert \omega \vert r}{c} \right)^3 \bigg) + O \left( \left( \frac{\Omega r}{c} \right)^2 \right)},
        \end{aligned}
    \end{array}
\end{equation}

\begin{equation}
    \begin{array}{c}
        \begin{aligned}
            \textstyle{\sum_{n}} & \textstyle{\left( N[\vert \omega_n \vert]+\frac{1}{2} \right) \left( \frac{\omega_n r}{c} \right)^4 P_n^{3\pm} \left[ \frac{\vert \omega_n \vert r}{c} \right]} \\
            & = \textstyle{\frac{1}{4 \sqrt{\pi}} \bigg( \left( N[\vert \omega \vert]+\frac{1}{2} \right) 2 \bigg( 3 \bigg( 5 - 2 \left( \frac{\vert \omega \vert r}{c} \right)^2 \bigg) \sin \left[ \frac{\vert \omega \vert r}{c} \right] - \bigg( 15 - \left( \frac{\vert \omega \vert r}{c} \right)^2 \bigg) \frac{\vert \omega \vert r}{c} \cos \left[ \frac{\vert \omega \vert r}{c} \right] \bigg)} \\
            & ~~~~~~~~~~~~ \textstyle{\pm \left( N[\vert \omega \vert]+\frac{1}{2} \right) 3 \frac{\Omega r}{c} \frac{\omega r}{c} \bigg( \bigg( 3-\left( \frac{\vert \omega \vert r}{c} \right)^2 \bigg) \sin \left[ \frac{\vert \omega \vert r}{c} \right] - 3 \frac{\vert \omega \vert r}{c} \cos \left[ \frac{\vert \omega \vert r}{c} \right] \bigg)} \\
            & ~~~~~~~~~~~~ \textstyle{\mp \, \frac{\hbar \vert \omega \vert}{k_B T} N[\vert \omega \vert] \left( N[\vert \omega \vert]+1 \right) 3 \frac{\Omega r}{c} \left( \frac{\omega r}{c} \right)^{-1} \bigg( 3 \bigg( 5 - 2 \left( \frac{\vert \omega \vert r}{c} \right)^2 \bigg) \sin \left[ \frac{\vert \omega \vert r}{c} \right]} \\
            & ~~~~~~~~~~~~~~~~~~~~~~~~~~~~~~~~~~~~~~~~~~~~~~~~~~~~~~~~~~~~~~~~~~~~ \textstyle{- \, \bigg( 15 - \left( \frac{\vert \omega \vert r}{c} \right)^2 \bigg) \frac{\vert \omega \vert r}{c} \cos \left[ \frac{\vert \omega \vert r}{c} \right] \bigg) \bigg) + O \left( \left( \frac{\Omega r}{c} \right)^2 \right)},
        \end{aligned}
    
        \vspace{0.2cm} \\

        \sum_{n} (-1)^n \left( N[\vert \omega_n \vert]+\frac{1}{2} \right) \left( \frac{\omega_n r}{c} \right)^4 P_n^{3 \pm} \left[ \frac{\vert \omega_n \vert r}{c} \right] = O \left( \left( \frac{\Omega r}{c} \right)^2 \right),
    \end{array}
\end{equation}

\begin{equation}
    \begin{array}{c}
        \begin{aligned}
            \textstyle{\sum_{n}} & \textstyle{ \left( N[\vert \omega_n \vert]+\frac{1}{2} \right) \left( \frac{\omega_n r}{c} \right)^4 P_n^{Z \pm} \left[ \frac{\vert \omega_n \vert r}{c} \right]} \\
            & \textstyle{ = \frac{1}{\sqrt{\pi}} \bigg( \left( N[\vert \omega \vert]+\frac{1}{2} \right) 2 \bigg( \bigg( 3 - \left( \frac{\vert \omega \vert r}{c} \right)^2 \bigg) \sin \left[ \frac{\vert \omega \vert r}{c} \right] - 3 \frac{\vert \omega \vert r}{c} \cos \left[ \frac{\vert \omega \vert r}{c} \right] \bigg)} \\
            & ~~~~~~~~~ \textstyle{\pm \left( N[\vert \omega \vert]+\frac{1}{2} \right) \frac{\Omega r}{c} \frac{\omega r}{c} \bigg( \sin \left[ \frac{\vert \omega \vert r}{c} \right] - \frac{\vert \omega \vert r}{c} \cos \left[ \frac{\vert \omega \vert r}{c} \right] \bigg)} \\
            & ~~~~~~~~~ \textstyle{\mp \, \frac{\hbar \vert \omega \vert}{k_B T} N[\vert \omega \vert] \left( N[\vert \omega \vert]+1 \right) \frac{\Omega r}{c} \left( \frac{\omega r}{c} \right)^{-1} \bigg( \bigg( 3 - \left( \frac{\vert \omega \vert r}{c} \right)^2 \bigg) \sin \left[ \frac{\vert \omega \vert r}{c} \right]} \\
            & ~~~~~~~~~~~~~~~~~~~~~~~~~~~~~~~~~~~~~~~~~~~~~~~~~~~~~~~~~~~~~~~~~~~ - \textstyle{3 \frac{\vert \omega \vert r}{c} \cos \left[ \frac{\vert \omega \vert r}{c} \right] \bigg) \bigg) + O \left( \left( \frac{\Omega r}{c} \right)^2 \right)},
        \end{aligned}
    
        \vspace{0.2cm} \\

        \begin{aligned}
            \textstyle{\sum_{n}} & \textstyle{(-1)^n \left( N[\vert \omega \vert]+\frac{1}{2} \right) \left( \frac{\omega_n r}{c} \right)^4 P_n^{Z \pm} \left[ \frac{\vert \omega_n \vert r}{c} \right]} \\
            & = - \textstyle{\frac{1}{15 \sqrt{\pi}} \bigg( \pm \left( N[\vert \omega \vert]+\frac{1}{2} \right) 5 \frac{\Omega r}{c} \frac{\omega r}{c} \left( \frac{\vert \omega \vert r}{c} \right)^3} \\
            & ~~~~~~~~~~~~~~~~ \textstyle{\mp \, \frac{\hbar \vert \omega \vert}{k_B T} N[\vert \omega \vert] \left( N[\vert \omega \vert]+1 \right) \frac{\Omega r}{c} \frac{\omega r}{c} \left( \frac{\vert \omega \vert r}{c} \right)^3 \bigg) + O \left( \left( \frac{\Omega r}{c} \right)^2 \right)},
        \end{aligned}
    \end{array}
\end{equation}

\begin{equation}
    \begin{array}{c}
        \begin{aligned}
            \textstyle{\sum_{n}} & \textstyle{\left( N[\vert \omega \vert]+\frac{1}{2} \right) \frac{\omega_n r}{c} \left( \frac{\vert \omega_n \vert r}{c} \right)^3 G_n^Z \left[ \frac{\vert \omega_n \vert r}{c} \right]} \\
            & = \textstyle{ -\frac{4}{\sqrt{\pi}} \left( N[\vert \omega \vert]+\frac{1}{2} \right) \frac{\omega r}{c} \bigg( \bigg( 1 - \left( \frac{\vert \omega \vert r}{c} \right)^2 \bigg) \sin \left[ \frac{\vert \omega \vert r}{c} \right] - \frac{\vert \omega \vert r}{c} \cos \left[ \frac{\vert \omega \vert r}{c} \right] \bigg) + O \left( \left( \frac{\Omega r}{c} \right)^2 \right)},
        \end{aligned}
    
        \vspace{0.2cm} \\

        \sum_{n} (-1)^n \left( N[\vert \omega_n \vert]+\frac{1}{2} \right) \frac{\omega_n r}{c} \left( \frac{\vert \omega_n \vert r}{c} \right)^3 G_n^Z \left[ \frac{\vert \omega_n \vert r}{c} \right] = \frac{8}{3 \sqrt{\pi}} \left( N[\vert \omega \vert]+\frac{1}{2} \right) \frac{\omega r}{c } \left( \frac{\vert \omega \vert r}{c} \right)^3 + O \left( \left( \frac{\Omega r}{c} \right)^2 \right),
    \end{array}
\end{equation}

\begin{equation}
    \begin{array}{c}
        \begin{aligned}
            \textstyle{\sum_{n}} & \textstyle{\left( N[\vert \omega_n \vert]+\frac{1}{2} \right) \frac{\omega_n r}{c} \left( \frac{\vert \omega_n \vert r}{c} \right)^3 G_n^\pm \left[ \frac{\vert \omega_n \vert r}{c} \right]} \\
            & = \textstyle{\frac{1}{\sqrt{\pi}} \bigg( \left( N[\vert \omega \vert]+\frac{1}{2} \right) \frac{\omega r}{c} \bigg( \bigg( 3 - \left( \frac{\vert \omega \vert r}{c} \right)^2 \bigg) \sin \left[ \frac{\vert \omega \vert r}{c} \right] - 3 \frac{\vert \omega \vert r}{c} \cos \left[ \frac{\vert \omega \vert r}{c} \right] \bigg)} \\
            & ~~~~~~~~~~ \textstyle{\pm \left( N[\vert \omega \vert]+\frac{1}{2} \right) \frac{\Omega r}{c} \bigg( 3 \sin \left[ \frac{\vert \omega \vert r}{c} \right] - \bigg( 3 + \left( \frac{\vert \omega \vert r}{c} \right)^2 \bigg) \frac{\vert \omega \vert r}{c} \cos \left[ \frac{\vert \omega \vert r}{c} \right] \bigg)} \\
            & ~~~~~~~~~~ \textstyle{\mp \, \frac{\hbar \vert \omega \vert}{k_B T} N[\vert \omega \vert] \left( N[\vert \omega \vert]+1 \right) \frac{\Omega r}{c} \bigg( \bigg( 3 - \left( \frac{\vert \omega \vert r}{c} \right)^2 \bigg) \sin \left[ \frac{\vert \omega \vert r}{c} \right]} \\
            & ~~~~~~~~~~~~~~~~~~~~~~~~~~~~~~~~~~~~~~~~~~~~~~~~~~~~~~ - \textstyle{3 \frac{\vert \omega \vert r}{c} \cos \left[ \frac{\vert \omega \vert r}{c} \right] \bigg) \bigg) + O \left( \left( \frac{\Omega r}{c} \right)^2 \right)},
        \end{aligned}
    
        \vspace{0.2cm} \\

        \sum_{n} (-1)^n \left( N[\vert \omega_n \vert]+\frac{1}{2} \right) \frac{\omega_n r}{c} \left( \frac{\vert \omega_n \vert r}{c} \right)^3 G_n^\pm \left[ \frac{\vert \omega_n \vert r}{c} \right] = O \left( \left( \frac{\Omega r}{c} \right)^2 \right),
    \end{array}
\end{equation}

\begin{equation}
    \begin{array}{c}
        \begin{aligned}
            \textstyle{\sum_{n}} & \textstyle{\left( N[\vert \omega \vert]+\frac{1}{2} \right) \frac{\omega_n r}{c} \left( \frac{\vert \omega_n \vert r}{c} \right)^3 Q_n^Z \left[ \frac{\vert \omega_n \vert r}{c} \right]} \\
            & = \textstyle{\frac{4}{\sqrt{\pi}} \left( N[\vert \omega \vert]+\frac{1}{2} \right) \frac{\omega r}{c} \bigg( \sin \left[ \frac{\vert \omega \vert r}{c} \right] - \frac{\vert \omega \vert r}{c} \cos \left[ \frac{\vert \omega \vert r}{c} \right] \bigg) + O \left( \left( \frac{\Omega r}{c} \right)^2 \right)},
        \end{aligned}
    
        \vspace{0.2cm} \\

        \sum_{n} (-1)^n \left( N[\vert \omega_n \vert]+\frac{1}{2} \right) \frac{\omega_n r}{c} \left( \frac{\vert \omega_n \vert r}{c} \right)^3 Q_n^Z \left[ \frac{\vert \omega_n \vert r}{c} \right] = \frac{4}{3 \sqrt{\pi}} \left( N[\vert \omega \vert]+\frac{1}{2} \right) \frac{\omega r}{c} \left( \frac{\vert \omega \vert r}{c} \right)^3 + O \left( \left( \frac{\Omega r}{c} \right)^2 \right).
    \end{array}
\end{equation}

\section{Expressions of the correlations involving spatial derivatives to first order in the velocity}\label{approx_EdEB}

With the expressions presented in Appendix \ref{formulas_corr}, one can also derive the following expressions to first order in $\Omega r/c$:

\begin{equation}
    \begin{array}{c}
        \begin{aligned}
            & \langle \Tilde{E}_\pm^A [\omega] \partial_\pm \Tilde{E}_\pm^B [\omega'] \rangle \\
            & \approx -\textstyle{\frac{\hbar}{64 \epsilon_0 \pi^2 r^4} \bigg( \left( N[\vert \omega \vert]+\frac{1}{2} \right) 2 \bigg( 3 \bigg( 5 - 2 \left( \frac{\vert \omega \vert r}{c} \right)^2 \bigg) \sin \left[ \frac{\vert \omega \vert r}{c} \right] -\bigg( 15 - \left( \frac{\vert \omega \vert r}{c} \right)^2 \bigg) \frac{\vert \omega \vert r}{c} \cos \left[ \frac{\vert \omega \vert r}{c} \right] \bigg)} \\
            & ~~~~~~~~~~~~~~~~~~~~ \textstyle{\pm \left( N[\vert \omega \vert]+\frac{1}{2} \right) 3 \frac{\Omega r}{c} \frac{\omega r}{c} \bigg( \bigg( 3-\left( \frac{\vert \omega \vert r}{c} \right)^2 \bigg) \sin \left[ \frac{\vert \omega \vert r}{c} \right] - 3 \frac{\vert \omega \vert r}{c} \cos \left[ \frac{\vert \omega \vert r}{c} \right] \bigg)} \\
            & ~~~~~~~~~~~~~~~~~~~~ \textstyle{\mp \, \frac{\hbar \vert \omega \vert}{k_B T} N[\vert \omega \vert] \left( N[\vert \omega \vert]+1 \right) 3 \frac{\Omega r}{c} \left( \frac{\omega r}{c} \right)^{-1} \bigg( 3 \bigg( 5 - 2 \left( \frac{\vert \omega \vert r}{c} \right)^2 \bigg) \sin \left[ \frac{\vert \omega \vert r}{c} \right]} \\
            & ~~~~~~~~~~~~~~~~~~~~~~~~~~~~~~~~~~~~~~~~~~~~~~~~~~~~~~~~~~~~~~~~~~~~~~~~~~~~~~~~ \textstyle{- \, \bigg( 15 - \left( \frac{\vert \omega \vert r}{c} \right)^2 \bigg) \frac{\vert \omega \vert r}{c} \cos \left[ \frac{\vert \omega \vert r}{c} \right] \bigg) \bigg)} \\
            & ~~~~~~~~~~~~~~~~~~~~~~~~~~~~~~~~~~~~~~~~~~~~~~~~~~~~~~~~~~~~~~~~~~~~~~~~~~~~~~~~~~~~~~~~~~~~~~~~~~~~~~~~~~~~~~~~~~~~~~~~~~~~ \times \textstyle{\delta[\omega+\omega' \pm 3 \Omega]},
        \end{aligned}
    
        \vspace{0.2cm} \\

        \langle \Tilde{E}_\pm^A [\omega] \partial_\pm \Tilde{E}_\pm^A [\omega'] \rangle \approx 0,
    \end{array}
\end{equation}

\begin{equation}
    \begin{array}{c}
        \begin{aligned}
            & \langle \Tilde{E}_\pm^A [\omega] \partial_\pm \Tilde{E}_\mp^B [\omega'] \rangle = \langle \Tilde{E}_\mp^A [\omega] \partial_\pm \Tilde{E}_\pm^B [\omega'] \rangle \\
            & \textstyle{~\approx -\frac{\hbar}{64 \epsilon_0 \pi^2 r^4}} \textstyle{\bigg( \left( N[\vert \omega \vert]+\frac{1}{2} \right) 2 \bigg( 3 \sin \left[ \frac{\vert \omega \vert r}{c} \right] - \bigg( 3 + \left( \frac{\vert \omega \vert r}{c} \right)^2 \bigg) \frac{\vert \omega \vert r}{c} \cos \left[ \frac{\vert \omega \vert r}{c} \right] \bigg)} \\
            & ~~~~~~~~~~~~~~~~~~~~ \textstyle{\pm \left( N[\vert \omega \vert]+\frac{1}{2} \right) \frac{\Omega r}{c} \frac{\omega r}{c} \bigg( \bigg( 3 + \bigg( \frac{\vert \omega \vert r}{c} \bigg)^2 \bigg) \sin \left[ \frac{\vert \omega \vert r}{c} \right] - 3 \frac{\vert \omega \vert r}{c} \cos \left[ \frac{\vert \omega \vert r}{c} \right] \bigg)} \\
            & ~~~~~~~~~~~~~~~~~~~~ \textstyle{\mp \, \frac{\hbar \vert \omega \vert}{k_B T} N[\vert \omega \vert] \left(N[\vert \omega \vert]+1 \right) \frac{\Omega r}{c} \left( \frac{\omega r}{c} \right)^{-1} \bigg( 3 \sin \left[ \frac{\vert \omega \vert r}{c} \right]} \\
            & ~~~~~~~~~~~~~~~~~~~~~~~~~~~~~~~~~~~~~~~~~~~~~~~~~~~~~~~~~~~~~~~~~~~~~~~~~~~~~~ \textstyle{- \, \bigg( 3 + \left( \frac{\vert \omega \vert r}{c} \right)^2 \bigg) \frac{\vert \omega \vert r}{c} \cos \left[ \frac{\vert \omega \vert r}{c} \right] \bigg) \bigg)} \\
            & ~~~~~~~~~~~~~~~~~~~~~~~~~~~~~~~~~~~~~~~~~~~~~~~~~~~~~~~~~~~~~~~~~~~~~~~~~~~~~~~~~~~~~~~~~~~~~~~~~~~~~~~~~~~~~~~~~~~~~~~~~ \times \delta[\omega+\omega' \pm \Omega],
        \end{aligned}
    
        \vspace{0.2cm} \\

        \begin{aligned}
            \langle \Tilde{E}_\pm^A [\omega] \partial_\pm \Tilde{E}_\mp^A [\omega'] \rangle & = \langle \Tilde{E}_\mp^A [\omega] \partial_\pm \Tilde{E}_\pm^A [\omega'] \rangle \\
            & \approx \textstyle{\mp \frac{\hbar \omega \vert \omega \vert^3}{160 \epsilon_0 \pi^2 c^4}} \textstyle{ \bigg( \left( N[\vert \omega \vert]+\frac{1}{2} \right) 5 \frac{\Omega r}{c} - \frac{\hbar \vert \omega \vert}{k_B T} N[\vert \omega \vert] \left( N[\vert \omega \vert]+1 \right) \frac{\Omega r}{c} \bigg) \delta[\omega+\omega' \pm \Omega]}, 
        \end{aligned}
    \end{array}
\end{equation}

\begin{equation}
    \begin{array}{c}
        \begin{aligned}
            & \langle \Tilde{E}_\pm^A [\omega] \partial_\mp \Tilde{E}_\pm^B [\omega'] \rangle \\
            & \approx - \textstyle{\frac{\hbar}{64 \epsilon_0 \pi^2 r^4} \bigg( \left( N[\vert \omega \vert]+\frac{1}{2} \right) 2 \bigg( \bigg( 3 - 2 \left( \frac{\vert \omega \vert r}{c} \right)^2 \bigg) \sin \left[ \frac{\vert \omega \vert r}{c} \right] - \bigg( 3 - \left( \frac{\vert \omega \vert r}{c} \right)^2 \bigg) \frac{\vert \omega \vert r}{c} \cos \left[ \frac{\vert \omega \vert r}{c} \right] \bigg)} \\
            & ~~~~~~~~~~~~~~~~~~~ \textstyle{\mp \left( N[\vert \omega \vert]+\frac{1}{2} \right) \frac{\Omega r}{c} \frac{\omega r}{c} \bigg( \bigg( 1 + \left( \frac{\vert \omega \vert r}{c} \right)^2 \bigg) \sin \left[ \frac{\vert \omega \vert r}{c} \right] - \frac{\vert \omega \vert r}{c} \cos \left[ \frac{\vert \omega \vert r}{c} \right] \bigg)} \\
            & ~~~~~~~~~~~~~~~~~~~ \textstyle{\mp \, \frac{\hbar \vert \omega \vert}{k_B T} N[\vert \omega \vert] \left( N[\vert \omega \vert]+1 \right) \frac{\Omega r}{c} \left( \frac{\omega r}{c} \right)^{-1} \bigg( \bigg( 3 - 2 \left( \frac{\vert \omega \vert r}{c} \right)^2 \bigg) \sin \left[ \frac{\vert \omega \vert r}{c} \right]} \\
            & ~~~~~~~~~~~~~~~~~~~~~~~~~~~~~~~~~~~~~~~~~~~~~~~~~~~~~~~~~~~~~~~~~~~~~~~~~~~~~ \textstyle{- \, \bigg( 3 - \left( \frac{\vert \omega \vert r}{c} \right)^2 \bigg) \frac{\vert \omega \vert r}{c} \cos \left[ \frac{\vert \omega \vert r}{c} \right] \bigg) \bigg)} \\
            & ~~~~~~~~~~~~~~~~~~~~~~~~~~~~~~~~~~~~~~~~~~~~~~~~~~~~~~~~~~~~~~~~~~~~~~~~~~~~~~~~~~~~~~~~~~~~~~~~~~~~~~~~~~~~~~~~~~~~~~~~ \times \delta[\omega+\omega' \pm \Omega],
        \end{aligned}
    
        \vspace{0.2cm} \\

        \begin{aligned}
            \textstyle{ \langle \Tilde{E}^A_\pm [\omega] \partial_\mp \Tilde{E}^A_\pm [\omega'] \rangle \approx \pm \frac{\hbar \omega \vert \omega \vert^3}{240 \epsilon_0 \pi^2 c^4}}\textstyle{ \bigg( \left( N[\vert \omega \vert]+\frac{1}{2} \right) 5 \frac{\Omega r}{c} - \frac{\hbar \vert \omega \vert}{k_B T} N[\vert \omega \vert] \left( N[\vert \omega \vert]+1 \right) \frac{\Omega r}{c} \bigg) \delta[\omega+\omega' \pm \Omega]},
        \end{aligned}
    \end{array}
\end{equation}

\begin{equation}
    \begin{array}{c}
        \begin{aligned}
            &\langle \Tilde{E}_\pm^A [\omega] \partial_Z \Tilde{E}_Z^B [\omega'] \rangle = \langle \Tilde{E}_Z^A [\omega] \partial_Z \Tilde{E}_\pm^B [\omega'] \rangle \\
            & \textstyle{ \approx \frac{\hbar}{16 \epsilon_0 \pi^2 r^4} \bigg( \left( N[\vert \omega \vert]+\frac{1}{2} \right) 2 \bigg( \bigg( 3 - \left( \frac{\vert \omega \vert r}{c} \right)^2 \bigg) \sin \left[ \frac{\vert \omega \vert r}{c} \right] - 3 \frac{\vert \omega \vert r}{c} \cos \left[ \frac{\vert \omega \vert r}{c} \right] \bigg)} \\
            & ~~~~~~~~~~~~~~~~ \textstyle{\pm \left( N[\vert \omega \vert]+\frac{1}{2} \right) \frac{\Omega r}{c} \frac{\omega r}{c} \bigg( \sin \left[ \frac{\vert \omega \vert r}{c} \right] - \frac{\vert \omega \vert r}{c} \cos \left[ \frac{\vert \omega \vert r}{c} \right] \bigg)} \\
            & ~~~~~~~~~~~~~~~~ \textstyle{\mp \, \frac{\hbar \vert \omega \vert}{k_B T} N[\vert \omega \vert] \left( N[\vert \omega \vert]+1 \right) \frac{\Omega r}{c} \left( \frac{\omega r}{c} \right)^{-1} \bigg( \bigg( 3 - \left( \frac{\vert \omega \vert r}{c} \right)^2 \bigg) \sin \left[ \frac{\vert \omega \vert r}{c} \right]} \\
            & ~~~~~~~~~~~~~~~~~~~~~~~~~~~~~~~~~~~~~~~~~~~~~~~~~~~~~~~~~~~~~~~~~~~~~~~~~ - \textstyle{3 \frac{\vert \omega \vert r}{c} \cos \left[ \frac{\vert \omega \vert r}{c} \right] \bigg) \bigg) \delta[\omega+\omega' \pm \Omega]}, 
        \end{aligned}
    
        \vspace{0.2cm} \\

        \begin{aligned}
            \textstyle{\langle \Tilde{E}_\pm^A [\omega] \partial_Z \Tilde{E}_Z^A [\omega'] \rangle} & = \langle \Tilde{E}_Z^A [\omega] \partial_Z \Tilde{E}_\pm^A [\omega'] \rangle \\
            & \approx \textstyle{\pm \frac{\hbar \omega \vert \omega \vert^3}{240 \epsilon_0 \pi^2 c^4}} \textstyle{ \bigg( \left( N[\vert \omega \vert]+\frac{1}{2} \right) 5 \frac{\Omega r}{c} -\frac{\hbar \vert \omega \vert}{k_B T} N[\vert \omega \vert] \left( N[\vert \omega \vert]+1 \right) \frac{\Omega r}{c} \bigg) \delta[\omega+\omega' \pm \Omega] },
        \end{aligned}
    \end{array}
\end{equation}

\begin{equation}
    \begin{array}{c}
        \begin{aligned}
            & \langle \Tilde{E}_Z^A [\omega] \partial_\pm \Tilde{E}_Z^B [\omega'] \rangle \\
            & \textstyle{ \approx \frac{\hbar}{16 \epsilon_0 \pi^2 r^4}  \bigg( \left( N[\vert \omega \vert]+\frac{1}{2} \right) 2 \bigg( \bigg( 3 - 2 \left( \frac{\vert \omega \vert r}{c} \right)^2 \bigg) \sin \left[ \frac{\vert \omega \vert r}{c} \right] - \bigg( 3 - \left( \frac{\vert \omega \vert r}{c} \right)^2 \bigg) \frac{\vert \omega \vert r}{c} \cos \left[ \frac{\vert \omega \vert r}{c} \right] \bigg)} \\
            & ~~~~~~~~~~~~~~~~ \textstyle{\mp \left( N[\vert \omega \vert]+\frac{1}{2} \right) \frac{\Omega r}{c} \frac{\omega r}{c} \bigg( \bigg( 1 + \left( \frac{\vert \omega \vert r}{c} \right)^2 \bigg) \sin \left[ \frac{\vert \omega \vert r}{c} \right] - \frac{\vert \omega \vert r}{c} \cos \left[ \frac{\vert \omega \vert r}{c} \right] \bigg)} \\
            & ~~~~~~~~~~~~~~~~ \textstyle{\mp \, \frac{\hbar \vert \omega \vert}{k_B T} N[\vert \omega \vert] \left( N[\vert \omega \vert]+1 \right) \frac{\Omega r}{c} \left( \frac{\omega r}{c} \right)^{-1} \bigg( \bigg( 3 - 2 \left( \frac{\vert \omega \vert r}{c} \right)^2 \bigg) \sin \left[ \frac{\vert \omega \vert r}{c} \right]} \\
            & ~~~~~~~~~~~~~~~~~~~~~~~~~~~~~~~~~~~~~~~~~~~~~~~~~~~~~~~~~~~~~~~~~~~~~~~~~~ \textstyle{- \, \bigg( 3 - \left( \frac{\vert \omega \vert r}{c} \right)^2 \bigg) \frac{\vert \omega \vert r}{c} \cos \left[ \frac{\vert \omega \vert r}{c} \right] \bigg) \bigg) \delta[\omega+\omega' \pm \Omega]},
        \end{aligned}
    
        \vspace{0.2cm} \\

        \begin{aligned}
            \textstyle{\langle \Tilde{E}_Z^A [\omega] \partial_\pm \Tilde{E}_Z^A [\omega'] \rangle \approx \mp \frac{\hbar \omega \vert \omega \vert^3}{60 \epsilon_0 \pi^2 c^4}} \textstyle{ \bigg( \left( N[\vert \omega \vert]+\frac{1}{2} \right) 5 \frac{\Omega r}{c} - \frac{\hbar \vert \omega \vert}{k_B T} N[\vert \omega \vert] \left( N[\vert \omega \vert]+1 \right) \frac{\Omega r}{c} \bigg) \delta[\omega+\omega' \pm \Omega]},
        \end{aligned}
    \end{array}
\end{equation}

\begin{equation}
    \begin{array}{c}
        \begin{aligned}
            & \langle \Tilde{B}_\pm^A [\omega] \partial_\pm \Tilde{E}_Z^B [\omega'] \rangle = - \langle \Tilde{B}_Z^A [\omega] \partial_\pm \Tilde{E}_\pm^B [\omega'] \rangle 
            \\
            & \textstyle{ \approx \pm \frac{\hbar}{16 \epsilon_0 \pi^2 c r^4} \bigg( \left( N[\vert \omega \vert]+\frac{1}{2} \right) \frac{\omega r}{c} \bigg( \bigg( 3 - \left( \frac{\vert \omega \vert r}{c} \right)^2 \bigg) \sin \left[ \frac{\vert \omega \vert r}{c} \right] - 3 \frac{\vert \omega \vert r}{c} \cos \left[ \frac{\vert \omega \vert r}{c} \right] \bigg)} \\
            & ~~~~~~~~~~~~~~~~~~ \textstyle{\pm \left( N[\vert \omega \vert]+\frac{1}{2} \right) \frac{\Omega r}{c} \bigg( 3 \sin \left[ \frac{\vert \omega \vert r}{c} \right] - \bigg( 3 + \left( \frac{\vert \omega \vert r}{c} \right)^2 \bigg) \frac{\vert \omega \vert r}{c} \cos \left[ \frac{\vert \omega \vert r}{c} \right] \bigg)} \\
            & ~~~~~~~~~~~~~~~~~~ \textstyle{\mp \, \frac{\hbar \vert \omega \vert}{k_B T} N[\vert \omega \vert] \left( N[\vert \omega \vert]+1 \right) \frac{\Omega r}{c} \bigg( \bigg( 3 - \left( \frac{\vert \omega \vert r}{c} \right)^2 \bigg) \sin \left[ \frac{\vert \omega \vert r}{c} \right] - 3 \frac{\vert \omega \vert r}{c} \cos \left[ \frac{\vert \omega \vert r}{c} \right] \bigg) \bigg)} \\
            & ~~~~~~~~~~~~~~~~~~~~~~~~~~~~~~~~~~~~~~~~~~~~~~~~~~~~~~~~~~~~~~~~~~~~~~~~~~~~~~~~~~~~~~~~~~~~~~~~~~~~~~~~~~~~~~~~~~~~~~~~~~~~ \times \delta[\omega+\omega' \pm 2 \Omega],
        \end{aligned}
    
        \vspace{0.2cm} \\

        \langle \Tilde{B}_\pm^A [\omega] \partial_\pm \Tilde{E}_Z^A [\omega'] \rangle = - \langle \Tilde{B}_Z^A [\omega] \partial_\pm \Tilde{E}_\pm^A [\omega'] \rangle \approx 0, 
    \end{array}
\end{equation}

\begin{equation}
    \begin{array}{c}
        \begin{aligned}
            \langle \Tilde{B}_\pm^A [\omega] \partial_\mp \Tilde{E}_Z^B [\omega'] \rangle & = - \langle \Tilde{B}_Z^A [\omega] \partial_\mp \Tilde{E}_\pm^B [\omega'] \rangle 
            \\
            & \textstyle{ ~\approx \pm \frac{\hbar}{16 \epsilon_0 \pi^2 c r^4} \left( N[\vert \omega \vert]+\frac{1}{2} \right) \frac{\omega r}{c} \bigg( \bigg( 1 - \left( \frac{\vert \omega \vert r}{c} \right)^2 \bigg) \sin \left[ \frac{\vert \omega \vert r}{c} \right]} \\
            & ~~~~~~~~~~~~~~~~~~~~~~~~~~~~~~~~~~~~~~~~~~~~~~~ - \textstyle{\frac{\vert \omega \vert r}{c} \cos \left[ \frac{\vert \omega \vert r}{c} \right] \bigg) \delta[\omega+\omega']},
        \end{aligned}
    
        \vspace{0.2cm} \\

        \begin{aligned}
            \langle \Tilde{B}_\pm^A [\omega] \partial_\mp \Tilde{E}_Z^A [\omega'] \rangle = - \langle \Tilde{B}_Z^A [\omega] \partial_\mp \Tilde{E}_\pm^A [\omega'] \rangle 
            \textstyle{\approx \mp \frac{\hbar \omega \vert \omega \vert^3}{24 \epsilon_0 \pi^2 c^5} \left( N[\vert \omega \vert]+\frac{1}{2} \right) \delta[\omega+\omega']},
        \end{aligned}
    \end{array}
\end{equation}

\begin{equation}
    \begin{array}{c}
        \begin{aligned}
            \textstyle{ \langle \Tilde{B}_\pm^A [\omega] \partial_Z \Tilde{E}_\mp^B [\omega'] \rangle \approx \pm \frac{\hbar}{8 \epsilon_0 \pi^2 c r^4} \left( N[\vert \omega \vert]+\frac{1}{2} \right) \frac{\omega r}{c} \bigg( \sin \left[ \frac{\vert \omega \vert r}{c} \right] - \frac{\vert \omega \vert r}{c} \cos \left[ \frac{\vert \omega \vert r}{c} \right] \bigg) \delta[\omega+\omega']},
        \end{aligned}
    
        \vspace{0.2cm} \\

        \langle \Tilde{B}_\pm^A [\omega] \partial_Z \Tilde{E}_\mp^B [\omega'] \rangle \approx \pm \frac{\hbar \omega \vert \omega \vert^3}{24 \epsilon_0 \pi^2 c^5} \left( N[\vert \omega \vert]+\frac{1}{2} \right) \delta[\omega+\omega'].
    \end{array}
\end{equation}

\bibliography{article_bib}

\end{document}